\documentclass[aps,pra,twocolumn,noeprint,superscriptaddress,amsmath,amssymb,nofootinbib]{revtex4-2}

\usepackage{graphicx}
\usepackage{dcolumn}
\usepackage{amsmath}
\usepackage{csquotes}
\usepackage{rotating}
\usepackage{braket}
\usepackage{bm}
\usepackage{multirow}
\usepackage{comment}
\usepackage{booktabs}
\usepackage{tabularx}
\usepackage{enumerate}
\usepackage{float}
\usepackage{array}

\usepackage{siunitx}
\DeclareSIUnit\gauss{G}
\DeclareSIUnit\electron{\textit{e}}

\renewcommand{\arraystretch}{1.4} 

\graphicspath{ {./images/} }

\usepackage[e]{esvect}

\begin{document}

\title{Systematic and statistical uncertainty evaluation of the HfF$^+$ electron electric dipole moment experiment}

\author{Luke Caldwell}
\thanks{These authors contributed equally to this work.}
\affiliation{JILA, NIST and University of Colorado, Boulder, Colorado 80309, USA}
\affiliation{Department of Physics, University of Colorado, Boulder, Colorado 80309, USA}

\author{Tanya S. Roussy}
\thanks{These authors contributed equally to this work.}
\affiliation{JILA, NIST and University of Colorado, Boulder, Colorado 80309, USA}
\affiliation{Department of Physics, University of Colorado, Boulder, Colorado 80309, USA}

\author{Trevor Wright}
\affiliation{JILA, NIST and University of Colorado, Boulder, Colorado 80309, USA}
\affiliation{Department of Physics, University of Colorado, Boulder, Colorado 80309, USA}

\author{William B. Cairncross}
\altaffiliation[Present address: ]{Atom Computing, 918 Parker Street, suite A-13, Berkeley California, 94710}
\affiliation{JILA, NIST and University of Colorado, Boulder, Colorado 80309, USA}
\affiliation{Department of Physics, University of Colorado, Boulder, Colorado 80309, USA}

\author{Yuval Shagam}
\altaffiliation[Present address: ]{Schulich Faculty of Chemistry, Technion - Israel Institute of Technology, Haifa 3200003, Israel}
\affiliation{JILA, NIST and University of Colorado, Boulder, Colorado 80309, USA}
\affiliation{Department of Physics, University of Colorado, Boulder, Colorado 80309, USA}

\author{Kia Boon Ng}
\affiliation{JILA, NIST and University of Colorado, Boulder, Colorado 80309, USA}
\affiliation{Department of Physics, University of Colorado, Boulder, Colorado 80309, USA}

\author{Noah Schlossberger}
\affiliation{JILA, NIST and University of Colorado, Boulder, Colorado 80309, USA}
\affiliation{Department of Physics, University of Colorado, Boulder, Colorado 80309, USA}

\author{Sun Yool Park}
\affiliation{JILA, NIST and University of Colorado, Boulder, Colorado 80309, USA}
\affiliation{Department of Physics, University of Colorado, Boulder, Colorado 80309, USA}

\author{Anzhou Wang}
\affiliation{JILA, NIST and University of Colorado, Boulder, Colorado 80309, USA}
\affiliation{Department of Physics, University of Colorado, Boulder, Colorado 80309, USA}

\author{Jun Ye}
\affiliation{JILA, NIST and University of Colorado, Boulder, Colorado 80309, USA}
\affiliation{Department of Physics, University of Colorado, Boulder, Colorado 80309, USA}

\author{Eric A. Cornell}
\affiliation{JILA, NIST and University of Colorado, Boulder, Colorado 80309, USA}
\affiliation{Department of Physics, University of Colorado, Boulder, Colorado 80309, USA}


\begin{abstract} 
We have completed a new precision measurement of the electron's electric dipole moment using trapped  HfF$^+$ in rotating bias fields. We report on the accuracy evaluation of this measurement, describing the mechanisms behind our systematic shifts. Our systematic uncertainty is reduced by a factor of 30 compared to the first generation of this measurement \cite{Cairncross2017}. Our combined statistical and systematic accuracy is improved by a factor of 2 relative to any previous measurement \cite{ACME2018}.
\end{abstract}
 

\newcommand{\Cdoub}{{\mathcal C}_{\rm d}}

\newcommand{\numblocks}{1329}

\newcommand{\tPz}{{}^3\Pi_{0^+}}
\newcommand{\tPzm}{{}^3\Pi_{0^-}}
\newcommand{\sSp}{{}^1\Sigma^+}
\newcommand{\tDo}{{}^3\Delta_1}
\newcommand{\tSm}{{}^3\Sigma_{0^+}^-}

\newcommand{\ltrans}{$\mathcal{L}_{\rm trans}^{961}$}
\newcommand{\lvc}{$\mathcal{L}_{\rm vc}^{818}$}
\newcommand{\lop}{$\mathcal{L}_{\rm op}^{1082}$}
\newcommand{\ldepl}{$\mathcal{L}_{\rm depl}^{814}$}

\newcommand{\gFul}{g_F^{u/l}}
\newcommand{\gFbar}{\bar{g}_F}
\newcommand{\dgF}{{\delta g_{F}}}
\newcommand{\dgeff}{{\delta g}_{\rm eff}}
\newcommand{\dmf}{d_{{\rm mf}}}
\newcommand{\wef}{\omega_{ef}}
\newcommand{\fef}{f_{ef}}
\newcommand{\Ehf}{E_{\rm hf}}
\newcommand{\Apar}{A_\parallel}
\newcommand{\Gpar}{G_\parallel}
\newcommand{\Eeff}{{\mathcal E}_{\rm eff}}
\newcommand{\dDelta}{{\delta \Delta}}
\newcommand{\Deltabar}{\bar{\Delta}}

\newcommand{\fn}{f^0}
\newcommand{\fB}{f^B}
\newcommand{\fD}{f^D}
\newcommand{\fR}{f^R}
\newcommand{\fBD}{f^{DB}}
\newcommand{\fBR}{f^{BR}}
\newcommand{\fDR}{f^{DR}}
\newcommand{\fBDR}{f^{DBR}}
\newcommand{\fI}{f^I}
\newcommand{\fBI}{f^{BI}}
\newcommand{\fDI}{f^{DI}}
\newcommand{\fRI}{f^{RI}}
\newcommand{\fDBI}{f^{DBI}}
\newcommand{\fBRI}{f^{BRI}}
\newcommand{\fDRI}{f^{DRI}}
\newcommand{\fBDRI}{f^{DBRI}}

\newcommand{\Erot}{{\mathcal E}_{\rm rot}}
\newcommand{\hatErot}{\hat{\mathcal E}_{\rm rot}}
\newcommand{\Vrot}{V_{\rm rot}}
\newcommand{\Vrf}{V_{\rm rf}}
\newcommand{\vecErot}{\vec{\mathcal E}_{\rm rot}}
\newcommand{\rrot}{r_{\rm rot}}

\newcommand{\Brot}{{\mathcal B}_{\rm rot}}
\newcommand{\vecBrot}{\bm{\mathcal B}_{\rm rot}}
\newcommand{\Bperp}{{\mathcal B}_{\perp}}
\newcommand{\vecBperp}{\bm{\mathcal B}_{\perp}}
\newcommand{\BX}{{\mathcal B}_X}
\newcommand{\BY}{{\mathcal B}_Y}
\newcommand{\BZ}{{\mathcal B}_Z}
\newcommand{\Bx}{{\mathcal B}_x}
\newcommand{\By}{{\mathcal B}_y}
\newcommand{\Bz}{{\mathcal B}_z}
\newcommand{\EX}{{\mathcal E}_X}
\newcommand{\EY}{{\mathcal E}_Y}
\newcommand{\EZ}{{\mathcal E}_Z}
\newcommand{\Ex}{{\mathcal E}_x}
\newcommand{\Ey}{{\mathcal E}_y}
\newcommand{\Ez}{{\mathcal E}_z}
\newcommand{\Eperp}{{\mathcal E}_\perp}
\newcommand{\Baxgrad}{{\mathcal B}'_{\rm axgrad}}
\newcommand{\Btrans}{{\mathcal B}'_{\rm trans}}
\newcommand{\Bone}{{\mathcal B}'_{1}}
\newcommand{\Btwo}{{\mathcal B}'_{2}}
\newcommand{\Bthree}{{\mathcal B}'_{3}}
\newcommand{\Baxgradnr}{{\mathcal B}_{\rm axgrad}^{\prime{\rm nr}}}
\newcommand{\Btransnr}{{\mathcal B}_{\rm trans}^{\prime{\rm nr}}}
\newcommand{\Bonenr}{{\mathcal B}_{1}^{\prime{\rm nr}}}
\newcommand{\Btwonr}{{\mathcal B}_{2}^{\prime{\rm nr}}}
\newcommand{\Bthreenr}{{\mathcal B}_{3}^{\prime{\rm nr}}}
\newcommand{\Brotnr}{{\mathcal B}_{\rm rot}^{\rm nr}}
\newcommand{\Bxnr}{{\mathcal B}_x^{\rm nr}}
\newcommand{\Bynr}{{\mathcal B}_y^{\rm nr}}
\newcommand{\Bznr}{{\mathcal B}_z^{\rm nr}}
\newcommand{\BXnr}{{\mathcal B}_X^{\rm nr}}
\newcommand{\BYnr}{{\mathcal B}_Y^{\rm nr}}
\newcommand{\BZnr}{{\mathcal B}_Z^{\rm nr}}

\newcommand{\wrot}{\omega_{\rm rot}}
\newcommand{\vecwrot}{\bm{\omega}_{\rm rot}}
\newcommand{\wrf}{\omega_{\rm RF}}
\newcommand{\frot}{f_{\rm rot}}
\newcommand{\Trot}{T_{\rm rot}}
\newcommand{\vecfrot}{\bm{f}_{\rm rot}}
\newcommand{\frf}{f_{\rm rf}}
\newcommand{\wsec}{\omega_{\rm sec}}

\newcommand{\Bm}{{\mathcal B}}
\newcommand{\Em}{{\mathcal E}}
\newcommand{\hffp}{HfF$^+$}
\newcommand{\hfp}{Hf$^+$}
\newcommand{\td}{{^3\Delta_1}}
\newcommand{\orderof}{{\mathcal O}}
\newcommand{\sgn}{{\rm sgn}}
\newcommand{\stat}{{\rm stat}}
\newcommand{\syst}{{\rm syst}}
\newcommand{\mus}{\,\mu{\rm s}}
\newcommand{\percm}{\,{\rm cm}^{-1}}
\newcommand{\mVcm}{\,{\rm mV/cm}}
\newcommand{\Vcm}{\,{\rm V/cm}}
\newcommand{\Volt}{\, {\rm V}}
\newcommand{\musmm}{\,\mu{\rm s/mm}}
\newcommand{\Kelvin}{\,{\rm K}}

\newcommand{\tilB}{{\tilde{B}}}
\newcommand{\tilD}{{\tilde{D}}}
\newcommand{\tilR}{{\tilde{R}}}
\newcommand{\tilI}{{\tilde{I}}}
\newcommand{\tilP}{{\tilde{P}}}
\newcommand{\tilBD}{\widetilde{BD}}
\newcommand{\tilBR}{\widetilde{BR}}
\newcommand{\tilDR}{\widetilde{DR}}
\newcommand{\tilBDR}{\widetilde{BDR}}
\newcommand{\tilS}{{\tilde{S}}}
\newcommand{\tils}{{\tilde{s}}}

\newcommand{\Heff}{H_{\rm eff}}
\newcommand{\Htum}{H_{\rm tum}}
\newcommand{\Hhf}{H_{\rm hf}}
\newcommand{\HS}{H_{\rm S}}
\newcommand{\HZe}{H_{{\rm Z},e}}
\newcommand{\HZN}{H_{{\rm Z},N}}
\newcommand{\Hrot}{H_{\rm rot}}
\newcommand{\HOD}{H_{\Omega}}
\newcommand{\Hedm}{H_{\rm edm}}


\newcommand{\pdet}{p_{\rm det}}

\newcommand{\Expect}[1]{{\rm E}\left( #1 \right)}
\newcommand{\Var}[1]{{\rm Var}\left( #1 \right)}
\newcommand{\Std}[1]{{\rm Std}\left( #1 \right)}
\newcommand{\Cov}[1]{{\rm Cov}\left( #1 \right)}
\newcommand{\Kurt}[1]{{\rm Kurt}\left( #1 \right)}
\newcommand{\threej}[6]{\begin{pmatrix} 
    #1 & #3 & #5 \\ 
    #2 & #4 & #6 
    \end{pmatrix}}
\newcommand{\sixj}[6]{\begin{Bmatrix} 
    #1 & #2 & #3 \\ 
    #4 & #5 & #6 
    \end{Bmatrix}}
\newcommand{\ME}[3]{\left\langle #1 \left| #2  \right| #3 \right\rangle}
\newcommand{\RME}[3]{\left\langle #1 \left\| #2 \right\| #3 \right\rangle}
\newcommand{\xmark}{$\bm{\times}$}

\newcommand{\wbc}[1]{\textcolor{red!50!black}{$^{\textrm{wbc}}${#1}}}
\newcommand{\wbcx}[1]{\todo[author=wbc,inline,color=red!30]{#1}}

\newcommand{\fcorrA}{-1}
\newcommand{\sfcorrA}{5}
\newcommand{\sftotA}{5}
\newcommand{\fcorrB}{-3}
\newcommand{\sfcorrB}{2}
\newcommand{\sftotB}{4}
\newcommand{\fcorrC}{-1}
\newcommand{\sfcorrC}{1}
\newcommand{\sftotC}{2}
\newcommand{\fcorrD}{0}
\newcommand{\sfcorrD}{0}
\newcommand{\sftotD}{0}
\newcommand{\fcorrE}{6}
\newcommand{\sfcorrE}{13}
\newcommand{\sftotE}{14}
\newcommand{\fcorrF}{-2}
\newcommand{\sfcorrF}{8}
\newcommand{\sftotF}{8}
\newcommand{\fcorrG}{-107}
\newcommand{\sfcorrG}{163}
\newcommand{\sftotG}{195}
\newcommand{\feedmmHz}{0.10}
\newcommand{\sstatmHz}{0.87}
\newcommand{\ssystmHz}{0.20}
\newcommand{\sstatuHz}{868}
\newcommand{\ssystuHz}{195}
\newcommand{\stotuHz}{890}
\newcommand{\deecm}{0.9}
\newcommand{\sstatecm}{7.7}
\newcommand{\ssystecm}{1.7}
\newcommand{\ubecm}{1.3}
\newcommand{\mhztoecm}{1.13}

\newcommand{\vecpi}[0]{{\bm \pi}}
\newcommand{\vecPi}[0]{{\bm \Pi}}
\newcommand{\vecsigma}[0]{{\bm \sigma}}
\newcommand{\veca}[0]{{\bf a}}
\newcommand{\vecA}[0]{{\bf A}}
\newcommand{\vecx}[0]{{\bf x}}
\newcommand{\vecB}[0]{{\bf B}}
\newcommand{\vecalpha}[0]{{\bm \alpha}}
\newcommand{\vecomega}[0]{{\bm \omega}}
\newcommand{\vecL}[0]{{\bf L}}
\newcommand{\vecJ}[0]{{\bf J}}
\newcommand{\vecE}[0]{{\bm{\mathcal E}}}

\maketitle


\section{\label{sec:introduction}Introduction} 

Symmetries are fundamental in physics and exploring them has been vital to revealing much of what we understand about nature today. In 1967 Sakharov showed that violation of combined charge and parity (CP) symmetry is required to explain the observed baryon asymmetry of the universe—an imbalance between the amount of matter and anti-matter \cite{Sakharov1967}. CP symmetry is not a good symmetry of nature but is only weakly violated by the Standard Model. Explaining the baryon asymmetry requires new physics with additional sources of CP violation and many extensions to the Standard Model have been proposed \cite{Canetti2012}.

Electric dipole moments of fundamental particles such as the electron violate time-reversal (T) symmetry, equivalent to CP-violation assuming CPT invariance. The Standard Model predicts an electron electric dipole moment (eEDM) which is well below current experimental sensitivity \cite{Yamaguchi2020, Ema2022}. However many proposed extensions predict eEDMs which are several orders of magnitude larger, bringing its observation within experimental reach. Measurements of the eEDM thus constitute sensitive probes for physics beyond the Standard Model \cite{Chupp2019}.

We recently completed a new measurement of the eEDM, providing the most precise measurement yet of its size. This paper is intended to accompany the result paper \cite{Roussy2022} and explains the details of our experimental procedure and analysis. Section~\ref{sec:experimental_overview} describes our apparatus and experimental sequence, Section~\ref{sec:data_analysis} the protocol used for data analysis, and Section~\ref{sec:effective-hamiltonian} details the effective Hamiltonian we use to model the results. In Sections~\ref{sec:accuracy_evaluation_overview}--\ref{sec:conclusion} we report the measures we have taken to identify, characterize and mitigate sources of systematic error.


\section{\label{sec:experimental_overview}Experiment} 

\begin{figure*}
    \centering
        \includegraphics[width=\textwidth]{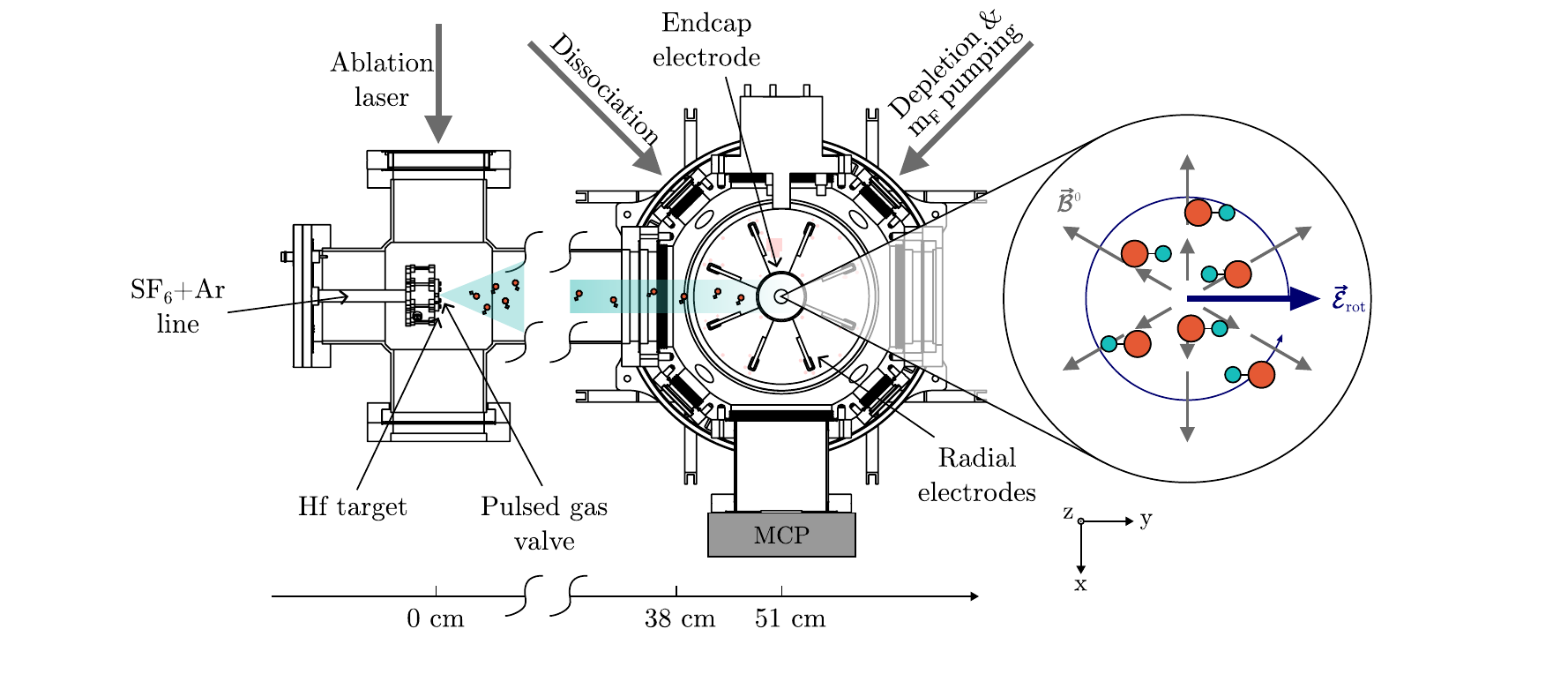}
    \caption{Schematic of experimental apparatus. On the left is the source chamber, where we produce neutral molecules. On the right is the main experimental chamber, containing the ion trap. The two chambers are connected by a differential pumping chamber with two small apertures at either end. The endcap electrodes of the ion trap have a hole in the center to allow optical access along the $z$ direction (vertical in the lab). Inset shows fields applied during experimental sequence: the rotating electric bias field $\vecErot$, and the quadrupole magnetic field $\vec{\mathcal{B}^0}$. The molecular axis of each of the ions is either aligned or anti-aligned with $\vecErot$.}
    \label{fig:apparatus}
\end{figure*}

\begin{table*}
    \caption{Spectroscopic constants for $\tDo$ state of \hffp{} used throughout this document. The total magnetic g-factor of $J=1, F=3/2$ states $g_F \equiv (\Gpar + g_N \mu_N/\mu_{\rm B})/3$ results from the combination of nuclear and electronic Zeeman effects.}
    \centering
    \begin{ruledtabular}
    \begin{tabular}{@{}l l l r@{}}
            Constant \hspace{10pt}  &   Value   & Description    & Reference  \\ \hline
            $B_e/h$   &   \SI{8.983+-0.001}{\giga\hertz} & Rotational constant & \cite{Cossel2012} \\
            $\Apar/h$ &   \SI{-62.0+-0.2}{\mega\hertz} & Hyperfine constant & \cite{Cairncross2017} \\ 
            $\dmf/h$ &   \SI{1.79+-0.01}{\mega\hertz\per\volt\centi\meter} & Molecule-frame electric dipole moment  & \cite{Cairncross2017} \\
            $\wef/(2\pi)$ &   \SI{0.74+-0.04}{\mega\hertz}   & $\Omega$-doubling constant  & \cite{Cossel2012} \\
            $g_N$  &   \num{5.25774+-0.00002} & Nuclear magnetic g-factor of $^{19}$F  & \cite{Stone2005} \\
            $G_\parallel$ & \num{-0.0122+-0.0003} & Electronic g-factor & \cite{Loh2013} \\
            $g_F$ &   \num{-0.0031+-0.0001} & $F=3/2$ state g-factor  & \cite{Loh2013} \\
            $|\Eeff|/h$  &   \SI{5.63e24}{\hertz\per\electron\per\centi\meter}  & Effective electric field & \cite{Fleig2013} \\
    \end{tabular}
    \label{tab:mol_constants}
    \end{ruledtabular}
\end{table*}
        
\begin{table*}
	\caption{Example experimental parameters and associated derived parameters from our 2022 data.}
	\centering
    \begin{ruledtabular}
	\begin{tabular}{@{}l r r@{}}
		Parameter & Value & Description \\ \hline
		$\Erot$ & \SI{58}{\volt\per\centi\meter} & Magnitude of rotating electric field during free evolution \\
        $\Erot^{\pi/2}$ & \SI{7}{\volt\per\centi\meter} & Magnitude of rotating electric field during $\pi/2$ pulses\\
		$\wrot$ &  $2\pi\times\SI{375}{\kilo\hertz}$ & Angular frequency of $\Erot$ \\
		$\Brot$ & \SI{10}{\milli\gauss} (typ.) & Effective rotating magnetic field \\ 
		$\mathcal{B}_{2,0}^{\rm rev}$ &  \SI{200}{\milli\gauss\per\centi\meter} (typ.) & Applied magnetic quadrupole gradient \\
		$r_{\rm rot}$ & \SI{0.5}{\milli\meter} & Radius of ion circular motion \\ 
		$\frac{\dgF}{g_F}$ & \num{-0.002146+-0.000002} & Stark doublet-odd magnetic g-factor ratio (see Figure \ref{fig:dgF}) \\ 
		$\Delta^0$ & $\sim\SI{1}{\hertz}$ & Rotation induced $m_F$ coupling \\ 
		$\Delta^D$ & $\sim\SI{-0.6}{\hertz}$ & Doublet-odd correction to $\Delta$  \\ 
		$V_{\rm RF}$ & \SI{23.5}{\volt} & RF radial confinement voltage during free evolution \\ 
        $\mathcal{E}_{\rm RF}$ & \SI{0.5}{\volt\per\centi\meter} & RF electric field amplitude at typical ion during free evolution \\ 
		$\omega_{\rm RF}$ & $2\pi\times\SI{50}{\kilo\hertz}$ & Radial-confinement RF frequency \\ 
		$V_{\rm DC}$ & \SI{3.7}{\volt} & DC axial confinement voltage during free evolution \\ 
  		$\mathcal{E}_{\rm DC}$ & \SI{10}{\milli\volt\per\centi\meter} & DC axial confinement electric field at typical ion \\
        $\omega_x$ & $2\pi\times\SI{0.95}{\kilo\hertz}$ & $x$ secular frequency during free evolution \\ 
        $\omega_y$ & $2\pi\times\SI{1.51}{\kilo\hertz}$ & $y$ secular frequency during free evolution \\ 
        $\omega_z$ & $2\pi\times\SI{1.60}{\kilo\hertz}$ & $z$ secular frequency during free evolution \\ 
	\end{tabular}
	\label{tab:typ_values}
	\end{ruledtabular}
\end{table*}

Our experiment uses \hffp{} ions, confined in an ion trap and prepared in the metastable $\tDo$ state. Relevant molecular properties are given in Table~\ref{tab:mol_constants}. In the $\tDo$ state, one of the valence electrons is subject to a large intramolecular effective electric field $\Eeff=\SI{23}{\giga\volt\per\centi\meter}$ \cite{Fleig2013}, along the internuclear axis of the molecule. We orient this molecular axis in the lab frame by applying an external electric field which rotates to maintain confinement of the ions. We then prepare the electron spin of the molecule in a coherent superposition of states oriented parallel and antiparallel to $\Eeff$, and measure the energy difference between them using Ramsey spectroscopy. The eEDM will give a contribution to this energy proportional to $d_e \Eeff$. To reject other unwanted contributions, we perform this measurement simultaneously on two spatially overlapping clouds of ions with their molecular axes aligned and anti-aligned with the externally applied field. The difference between the measured energies in each case is our science signal.

This section describes the apparatus and each of the steps used in state preparation and measurement of the ions. A summary of typical experimental parameters is given in Table~\ref{tab:typ_values}.

\subsection{Lasers}

\begin{table*}
	\caption{Photons used in our experimental sequence. All lasers address $v=0$ state of ground and excited levels, except \lvc{} which addresses $v=1$. \ltrans{} and \lvc{} propagate along the trap $z$ axis, all other lasers propagate in the $x$-$y$ plane.}
	\centering
	\begin{ruledtabular}
	\begin{tabular}{@{}l c r r r c r@{}}
		Name & Symbol & Transition & Power/Energy  & $\lambda$ (\si{\nano\meter}) & Polarization & Pulse Width \\ \hline
		Ablation & & -- & \SI{10}{\milli\joule} & $532$ & linear & \SI{10}{\nano\second} \\
		Photoionization 1 & & $\Omega =3/2 \leftarrow  X{}^2\Delta_{3/2}$ & \SI{30}{\micro\joule} & $309.388$ & linear & \SI{10}{\nano\second} \\
		Photoionization 2 & & Rydberg $ \leftarrow \Omega =3/2$ & \SI{1.3}{\milli\joule} & $367.732$ & linear & \SI{10}{\nano\second} \\
		Transfer & \ltrans{} & $P(1)$ $ ^3\Pi_{0^+}\leftarrow X^1\Sigma^+$ & $600$ mW & $961.43495$ & linear & CW \\
		$m_F$ Pumping & \lop{} & $P(1)$ $^3\Pi_{0^-} \leftarrow {}^3\Delta_1$ & $21$ mW  & $1082.4137$ & circular & CW, strobed \\
		$m_F$ Depletion & \ldepl{} & $Q(1) ^3\Sigma^-_{0^+} \leftarrow {}^3\Delta_1$ & $550$ mW  & $814.508$ & circular & CW, strobed \\
		Vibrational Cleanup & \lvc{} & $P(1) ^3\Sigma^-_{0^+}  \leftarrow {}^3\Delta_1$ & $30$ mW & $818.37198$ & linear & CW  \\
		Dissociation 1 & & $\Omega=2 \leftarrow {}^3\Delta_1$ & $1.6$ mJ & $368.494$  & circular & $10$ ns \\
		Dissociation 2 & & $? \leftarrow \Omega=2$ & 25 mJ & 266 & circular & $10$ ns  \\
	\end{tabular}
	\label{tab:lasers}
	\end{ruledtabular}
\end{table*}

\begin{figure}
    \centering
        \includegraphics[width=\columnwidth]{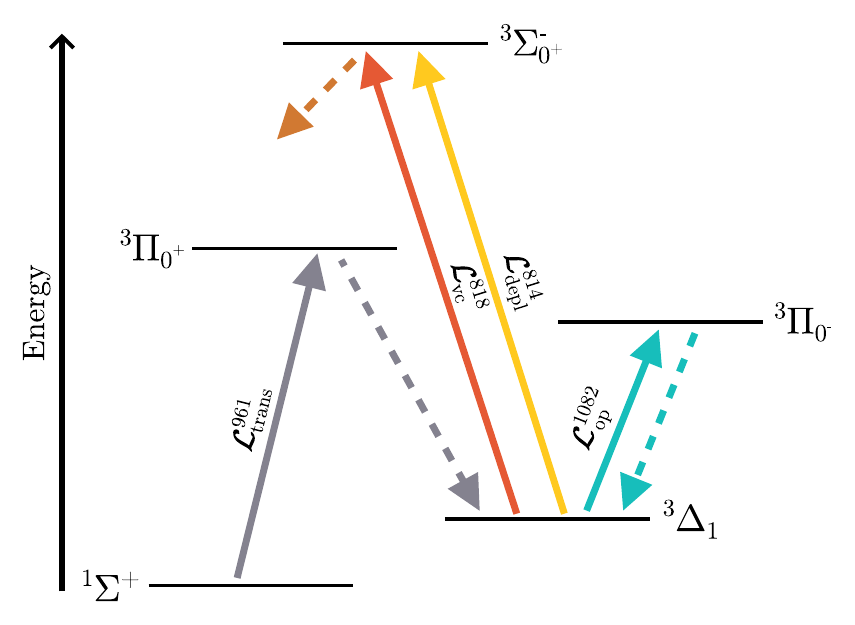}
    \caption{Cartoon depicting the transitions used during our state preparation. The ${}^3\Pi_{0^+}$ and ${}^3\Pi_{0^-}$ states decay preferentially to ${}^3\Delta_1$, while the ${}^3\Sigma_0^-$ state decays preferentially to ${}^1\Sigma^+$.} 
    \label{fig:state_prep_level_diagram}
\end{figure}

The experiment uses a total of 9 lasers; 5 pulsed lasers used for ablation, ionization, and photodissociation, and 4 CW lasers---which we denote \ltrans, \lvc, \lop, \ldepl---used for state-preparation and readout. A summary is given in Table~\ref{tab:lasers} and Fig.~\ref{fig:state_prep_level_diagram}, and each is described in detail in the following sections.  All lasers are locked to wavemeters using simple, $\sim\SI{1}{\hertz}$ servo loops. The CW lasers are locked to within $\sim\pm\SI{30}{\mega\hertz}$, and the pulsed lasers to $\sim\pm\SI{500}{\mega\hertz}$.

\subsection{Molecular beam and ionization}

Our experiment begins with a pulsed beam of neutral molecules. We use a pulsed Nd:YAG laser to ablate a solid Hf rod into a pulsed supersonic expansion of Ar, seeded with 1\% SF$_6$. Chemical reactions between the Hf plasma and the SF$_6$ produce neutral HfF which are entrained in the supersonic expansion and rovibrationally cooled by collisions with the Ar atoms to a temperature of $\sim \SI{10}{\kelvin}$. When they arrive in our main chamber, $\sim 50$ cm away, a pair of pulsed UV lasers at \SI{309}{\nano\meter} and \SI{368}{\nano\meter} excite a two-photon transition to a Rydberg state 54 cm$^{-1}$ above the ionization threshold, from which they autoionize \cite{Loh2011,Loh2013}. The molecular ions are created in the first few rotational levels of $\sSp(v=0)$, the electronic and vibrational ground state of the molecule. The ions are stopped at the center of our RF ion trap by pulsed voltages on the radial trap electrodes, after which the confining potentials are immediately turned on. We typically trap $\sim\num{2e4}$ \hffp{} ions with a lifetime\footnote{We note that the trap lifetime is limited by slow heating of the ions and is strongly dependent on the trapping parameters. The \SI{5}{\second} here is for the very shallow trap used during the Ramsey interrogation time.} of $\sim\SI{5}{\second}$. The trap is described in detail in the next section. 

\subsection{Ion trap} 

Our linear Paul trap has 8 radial electrodes and 2 endcaps. The radial confinement is provided by driving the radial electrodes in a quadrupole configuration producing a field, 
\begin{equation}
    \vec{\mathcal E}_{\rm RF}(\vec{r},t) = \frac{V_{\rm RF}}{R_0^2} \cos{(\wrf t)} (\vec{x} - \vec{y}),
\end{equation}
where $\wrf= 2 \pi\times \SI{50}{\kilo\hertz}$, $V_{\rm RF}$ is the voltage applied on each electrode, $R_0\sim\SI{4.8}{\centi\meter}$ is the effective radius the RF trap, and $\vec{x},\vec{y}$ are the radial position coordinates of the ions in the laboratory frame. Axial confinement is provided by DC voltages $V_{\rm DC}$ on a pair of endcaps, producing a field
\begin{equation}
    \vec{\mathcal E}_{\rm DC}(\vec{r},t) =  \frac{V_{\rm DC}}{Z_0^2} (\vec{x} + \vec{y} - 2 \vec{z}),
\end{equation}
where $Z_0\sim \SI{17}{\centi\meter}$ is the effective height of the RF trap. We choose the values of $V_{\rm RF}$ and $V_{\rm DC}$ immediately after ionization to best match the spatial mode of the initial ion cloud, giving trap frequencies $\sim\SI{5}{\kilo\hertz}$ in all directions. We then linearly ramp the trapping voltages down over \SI{10}{\milli\second} to expand and cool the ion cloud. The ramp takes the trap frequencies to $\sim\SI{2.8}{\kilo\hertz}$ and $\sim\SI{2.0}{\kilo\hertz}$ in the radial and axial directions respectively.

In addition to the confinement fields, we also apply a rotating electric field $\vecErot$, 
\begin{equation}
    \vecErot(t) = \Erot \left[ \hat{x} \cos{(\wrot t)} + \tilR \hat{y} \sin{(\wrot t)} \right],
\end{equation}
where $\wrot= 2 \pi \times \SI{375}{\kilo\hertz}$, $\tilR=\pm1$ indicates the rotation direction and $\Erot=|\vecErot|$ is typically $\sim\SI{58}{\volt\per\centi\meter}$. This field serves to orientate the molecular axis, and thus the effective electric field, of the ions and we do our spectroscopy in this rotating frame. $\vecErot$ causes an additional micromotion of the ions,
\begin{equation}
    -\frac{e}{m\wrot^2}\vecErot=-\rrot \hatErot,
\end{equation}
where $\rrot\sim\SI{0.5}{\milli\meter}$. The shape of the radial electrodes is optimized to minimize inhomogeneities in $\vecErot$ across the ion cloud \cite{Cairncross2019a}. 

\subsection{Magnetic fields}

Measuring the electron EDM also requires orienting the electron spin of the molecules which we do with an applied magnetic field $\vec{\mathcal{B}^0}$. In order for the unpaired electrons to experience a time-averaged interaction with the intramolecular effective electric field, this magnetic field must corotate with $\vecErot$. We achieve this using a pair of coils in anti-helmholtz configuration aligned along the axial direction, giving
\begin{equation}
    \vec{\mathcal{B}^0} = \tilB \mathcal{B}_{2,0}^{\rm rev} (2 \vec{z} - \vec{x} - \vec{y}).
\end{equation}
Here $\mathcal{B}_{2,0}^{\rm rev}$ is typically $\sim\SI{200}{\milli\gauss\per\centi\meter}$ and $\tilB=\pm1$ indicates the direction of the current in the coils, explained in more detail in Sec.~\ref{sec:exp-switch-states}. In the rotating and co-moving frame of the ions, this quadrupole magnetic field appears as a time-averaged magnetic bias, 
\begin{equation}
    \Brot = \langle \vec{\mathcal{B}^0}\cdot\vecErot\rangle = \mathcal{B}_{2,0}^{\rm rev} \rrot.
\end{equation}
The coil pair is driven by a precision current source with \SI{1}{\pico\ampere} resolution. We refer to the pair of coils that produces this field as the $\vec{\mathcal{B}^0}$-coils. 

The apparatus also includes three pairs of coils setup along the lab frame $\hat{x},\hat{y},\hat{z}$ axes in Helmholtz configuration for tuning the magnetic field at the position of the ions. The $z$ coil is driven by the second channel of the precision current supply used for the $\vec{\mathcal{B}^0}$-coils, the $x$ and $y$ coils are driven by a lower precision current supply. The magnetic field around the periphery of the trap is measured by an array of eight, 3-axis fluxgate magnetometers bolted to the outside of the main experimental chamber. We use these measurements to infer the magnetic field at the center of the trap. In contrast to other modern eEDM experiments \cite{ACME2018,Hudson2011,Zhu2013}, the apparatus includes no magnetic shielding as we are principally only sensitive to magnetic fields rotating at $\wrot$, as discussed in detail in Sec.~\ref{subsec:magnetic_effects}.

\subsection{State preparation}

Immediately after ionization, the \hffp{} ions are in the ground electronic and vibrational state ($\sSp(v=0)$), primarily distributed over the lowest 4 rotational levels $J=0$--$3$. We connect these rotational levels using microwaves and perform incoherent transfer to the eEDM-sensitive $\tDo(v=0,J=1)$ science state by using light from \ltrans{} to drive the $\tPz(J=0) \leftarrow \sSp$ transition, the excited state of which decays preferentially to $\tDo$. This light enters the chamber along the $z$-axis and is on for \SI{80}{\milli\second} beginning immediately after ionization. The decay from $\tPz$ puts population in several vibrational levels in $\tDo$, which can decay into the $v=0$ science state if left untreated. We remove the population in higher vibrational levels by illuminating the cloud with \lvc{} light which connects $\tSm (v=1, J=0) \leftarrow  \tDo(v=1, J=1)$ at $\sim\SI{818}{\nano\meter}$, preferentially decaying back to $\sSp$. The \lvc{} laser also enters the chamber along the $z$ axis and remains on for the duration of the experiment.

\begin{figure}
    \centering
        \includegraphics[width=\columnwidth]{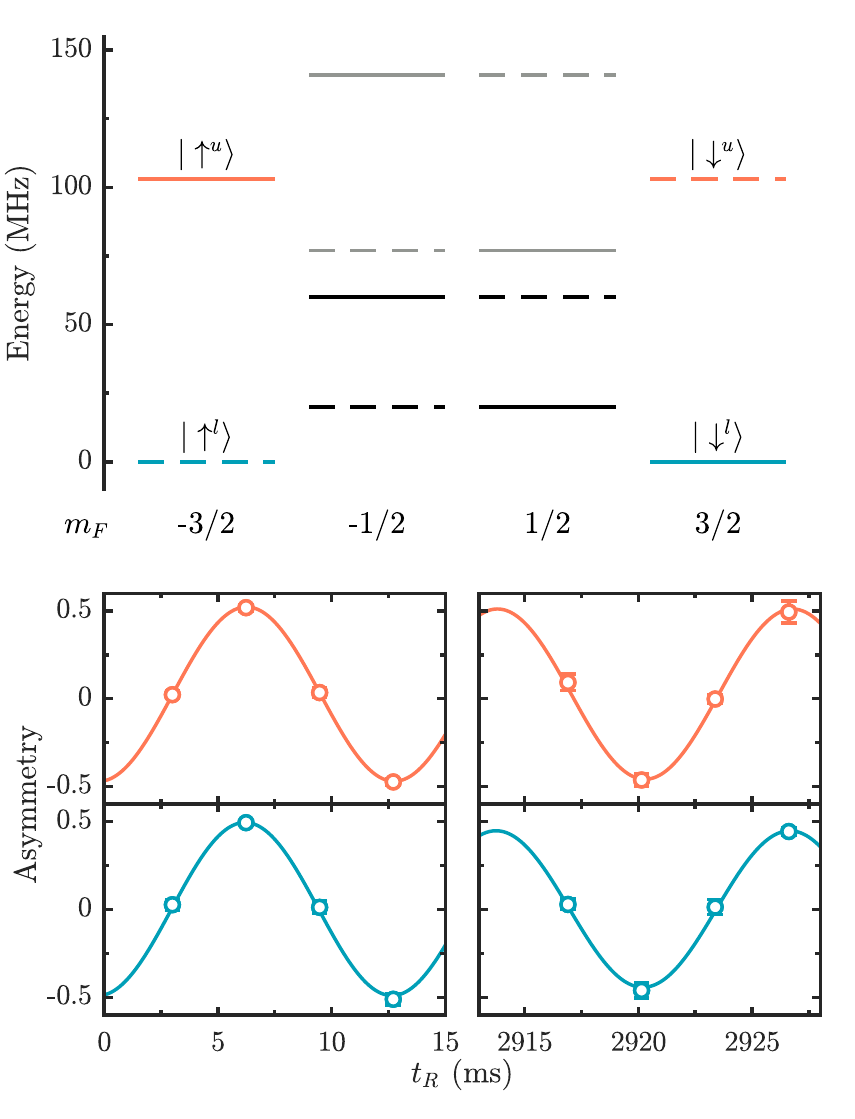}
    \caption{Ramsey spectroscopy in \hffp{}. Top: level structure of the eEDM-sensitive $\tDo(v=0,J=1)$ manifold in external electric $\mathcal{E}_{\rm rot}\sim\SI{58}{\volt\per\centi\meter}$. Solid (dashed) lines correspond to states with $\Omega=+1(-1)$. Gray lines correspond to states which asymptote to $F=1/2$ at zero field, all other states asymptote to $F=3/2$. The upper and lower doublet used for the measurement, highlighted in pink and blue respectively, are separated by $\sim\SI{100}{\mega\hertz}$. The two states in each doublet are further split by Zeeman energy and interaction of the eEDM with $\Eeff$. Bottom: example Ramsey fringes from our dataset. The fringes from the two doublets are collected simultaneously.}
    \label{fig:ramsey_fringe_figure}
\end{figure}

After transferring the ions to the science state, we ramp on $\vecErot$ in \SI{5}{\milli\second}. Figure~\ref{fig:ramsey_fringe_figure} shows the structure of the science state at $\Erot=\SI{58}{\volt\per\centi\meter}$. In this field, the stretched states of $\tDo(J=1,F=3/2)$, can be considered states of good orientation. They form two pairs of levels---which we call the upper and lower doublet, highlighted in pink and blue respectively---with their molecular dipole, and thus $\Eeff$, either aligned or anti-aligned with $\Erot$. Each doublet consists of one state with $m_F=3/2$ and one with $m_F=-3/2$.

We polarize the molecules in the rotating frame by optically pumping them using \lop{} light addressing the $\tPzm(v=0,J=0)\leftarrow\tDo(v=0,J=1)$ at \SI{1082}{\nano\meter}. The light is circularly polarized with its $k$-vector in the plane of $\vecErot$. We use an AOM to strobe \lop{} synchronously with the rotation of $\vecErot$ on a 50\% duty cycle\footnote{We note that, although the light is on for 50\% of each cycle, the micromotion-induced Doppler shifts mean it is only resonant with the ions for less than 5\%.} such that it drives either $\sigma^{+}$ or $\sigma^{-}$ transitions to an $F' = 3/2$ manifold in the excited state. This eventually leaves population only in either the $m_F=3/2$ or $m_F=-3/2$ states of $\tDo(v=0,J=1)$. We define the \textit{preparation phase} of the experiment as the orientation of $\vecErot$ relative to the $k$-vector of the light when the light is on; \textit{in} when $\vecErot$ is parallel, and \textit{anti} when it is anti-parallel. This preparation phase can be changed by adjusting the timing of the strobing cycle. \lop{} is on for a total of \SI{80}{\milli\second}, starting \SI{40}{\milli\second} after trapping. 

The final step of state preparation is applying \ldepl{} light at \SI{814}{\nano\meter}. This laser is tuned to address the $\tSm(v=0, J=0) \leftarrow \tDo(v=0, J=1)$, which preferentially decays to $\sSp$ by a 10:1 ratio, having weaker coupling to the $\tDo$ state. \ldepl{} is circularly polarized with the same handedness and $k$-vector as \lop{}. It is again strobed so as to only address and remove any residual population left over in other $m_F$ states after \lop{} is turned off. \ldepl{} light is on for \SI{7}{\milli\second}, beginning \SI{3}{\milli\second} after \lop{} is turned off.

These steps leave the population in an incoherent mixture of one of the stretched states of the two doublets. The key difference from our previous measurement \cite{Cairncross2017} is that the experiment proceeds on \textit{both} doublets simultaneously. Our detection scheme \cite{Zhou2020}, described in Sec.~\ref{sec:measurement}, allows us to read out each independently, enabling us to take advantage of common-mode noise cancellation.

\subsection{Ramsey sequence}
\label{sec:ramsey-sequence}

Immediately prior to the Ramsey sequence, we ramp the radial confinement of the ions down further, to trap frequencies of $\sim\SI{1}{\kilo\hertz}$. This reduces the density of the cloud and improves the coherence time.

We apply a $\pi/2$ pulse to the ensemble of ions by temporarily ramping down the magnitude of $\vecErot$ from $\sim\SI{58}{\volt\per\centi\meter}$ to $\sim\SI{7}{\volt\per\centi\meter}$ in \SI{16}{\micro\second}, holding it there for \SI{1}{\milli\second} and then ramping back up in a further \SI{16}{\micro\second}. Reducing $\Erot$ increases a rotation-induced coupling between $m_F = \pm 3/2$ states in a doublet (see Sec. \ref{sec:effective-hamiltonian}), causing the pure spin states in each doublet to evolve into a coherent superposition. We allow this superposition to evolve for a variable amount of time $t_R$---up to \SI{3}{\second}---and then apply a second $\pi/2$ pulse to map the relative phase onto a population difference between the two states in a doublet. 

\subsection{Measurement}
\label{sec:measurement}

We project the ions into their final state by applying \ldepl{} again to remove population from one of the stretched states in each doublet. The \textit{readout phase} is defined in the same way as the \textit{preparation phase}; \textit{in} for $\vecErot$ is parallel with the $k$-vector of the light when it is on and \textit{anti} for antiparallel. 

Finally, we detect and count the number of ions in the remaining stretched states via resonance-enhanced multiphoton dissociation \cite{Ni2014}, driven by two pulsed UV lasers at \SI{368}{\nano\meter} and \SI{266}{\nano\meter}. Immediately prior to the dissociation pulse, we ramp up both radial and axial confinement to compress the cloud and improve the dissociation efficiency. The dissociation pulse is timed so that $\vecErot$, is along $\tilI \hat{y}$, parallel to the plane of a microchannel plate (MCP) and phosphor screen assembly. Here $\tilI=\pm1$ and $\hat{y}$ is defined by Fig.~\ref{fig:apparatus}. Because the dissociation lasers enter at an angle to $\hat{y}$, there is considerable Doppler shift from the micromotion of the ions at \SI{45}{\degree} to the $k$-vector of the light. To account for this we adjust the frequency of the \SI{368}{\nano\meter} light by $\sim\pm\SI{2}{\giga\hertz}$ depending on the product $\tilR\tilI$ which gives the sign of the Doppler shift.

Each of the lasers is circularly polarized to drive transitions which preserve the orientation of the molecules during dissociation \cite{Cairncross2019a}. In this way the resultant \hfp{} ions from each doublet are ejected in opposite directions. The handedness of the dissociation lasers $\tilP$ is determined by a $\lambda/2$ waveplate which can be moved into or out of the beam path on a motorized mount. Immediately after dissociation, we turn off the RF confinement and apply pulsed voltage on the radial electrodes to kick the ions towards the MCP. The \hfp{} ions from each doublet are imaged on opposite sides of the phosphor screen; the side each doublet is imaged on is set by the value of $\tilI$. We time-gate the phosphor screen such that we only image the dissociated \hfp{} and not any remaining \hffp{} ions. We detect both \hfp{} and \hffp{} ions in time of flight. Technical details of our imaging and counting system are described in \cite{Shagam2020a} and \cite{Zhou2020}.

We typically detect $\sim 550$ Hf$^+$ ions on each side of the screen at early time, and $\sim 120$ after $t_R\sim\SI{3}{\second}$. The latter is principally limited by the finite lifetime of the $\tDo$ state but with a contribution from ions being heated out of our shallow trapping potential during $t_R$.

\subsection{Noise}

Instability of the intensity of the pulsed lasers used for ablation, ionization and photo-dissociation means that the fluctuations in the number of Hf$^+$ ions detected at the end of each shot are $\sim30\%$, roughly $3\times$ the quantum projection noise limit for 120 ions. However these sources of noise, and many others, are common mode; the exact same laser pulses address the ions which end up in the upper and lower doublets. If we measure the ion number when the Ramsey oscillations of the two doublets are close to in phase with one another then we can take advantage of excellent noise cancellation in the number difference \cite{Zhou2020} which we use to extract our science signal (see Sec~\ref{sec:data_analysis}). The two doublets oscillate at slightly different frequencies owing to a part in 230 difference in their magnetic moments and so we deliberately take our data at a \textit{beat}; our early time data is taken when the two doublets are in phase\footnote{Due to the finite length of the $\pi/2$ pulses, the doublets are already slightly out of phase at the earliest Ramsey times accessible to us. Systematic effects associated with this imperfection are discussed in Sec~\ref{sec:phase_shifts}.} and our late time data $\sim 230$ oscillations later when they come back into phase again. The time of the second \textit{beat} can be controlled by varying the oscillation frequency via the $\vec{\mathcal{B}^0}$-coils. In our final dataset, the noise in the science signal is roughly 30\% above the shot noise limit.

\subsection{Experimental protocol and switch states}
\label{sec:exp-switch-states}

In each shot of the experiment we can choose the preparation phase to be either \textit{in} or \textit{anti}. For a given choice of $\tilI$, the direction of $\vecErot$ at the moment of dissociation, and $\tilP$, the handedness of the dissociation laser polarization, the readout phase is constrained by the need to drive stretched-to-stretched transitions which preserve molecule orientation. We label each shot of the experiment with each of these phases. For example \textit{in}-\textit{anti} labels a shot where the cleanup (and optical pumping) laser are parallel to $\vecErot$ during state preparation but anti-parallel during readout.

To record a Ramsey fringe, we repeat our measurement at different free evolution times. For a given fringe, the phase of readout is kept fixed. At each Ramsey time, we take an even number of shots with each pair consisting of one shot with each phase of state preparation. This set of shots is called a point and a Ramsey fringe consists of 8 of these points taken at different $t_R$; we take 4 points at early Ramsey time, and 4 points at late Ramsey time, each consisting of two points on the sides of fringes and one point each on the top and bottom as shown in Fig.~\ref{fig:ramsey_fringe_figure}. The points on the sides of the fringes consist of 20 shots each, while the points on top and bottom consist of 8 shots each.

We record our data in `blocks'. Each block is constructed from a set of $2^3=8$ fringes recorded in each possible combination of 3 experimental switches. Each switch corresponds to an experimental parameter whose sign can be reversed: $\tilB$ the direction of the current in the $\vec{\mathcal{B}^0}$-coils, $\tilR$ the direction of rotation of $\vecErot$, and $\tilI$ the direction of $\vecErot$ relative to the $y$ axis at the time of dissociation, corresponding to which side of the phosphor screen each of the doublets are imaged onto. Note that in our implementation of the $\tilI$ switch, the direction of $\vecErot$ is reversed at all points in time so that the opposite switch is prepared and read out. A fourth experimental switch, $\tilP$ the polarization of the dissociation light, is alternated every block.

We refer to each experimental configuration with $\{\tilB,\tilR,\tilI,\tilP\}=\pm1$ as a switch state. In each block, the first Ramsey time is recorded for all switch states before moving onto the second Ramsey time for each switch state etc. The order of the switch states at each point is $\{\tilB,\tilR,\tilI\}=$\{1,1,1\}, \{-1,1,1\}, \{1,-1,1\}, \{-1,-1,1\}, \{1,1,-1\}, \{-1,1,-1\}, \{1,-1,-1\}, \{-1,-1,-1\}. Every other block, the order of switch states is reversed. In each switch state, we simultaneously collect data for molecules in each doublets, corresponding to orientation of the molecular axis with respect to the applied electric field which we represent by another switch $\tilD=\pm1$. The Ramsey times for each switch state are adjusted independently based on the data from the previous block to ensure that the 20-shot points are as close as possible to both the sides of the fringes, where our sensitivity is highest, and to the beat, where our noise cancellation is best and where various systematic shifts are minimized.

For the eEDM dataset, we collected $1370$ blocks or $\sim600$ hours of data over a $\sim2$ month period of April--June 2022. During the data run, we took data with 3 different values of the $\mathcal{B}_{2,0}^{\rm rev}$, corresponding to fringe frequencies of $\sim75$, 105 and \SI{151}{\hertz}. About halfway through the dataset, we rotated the waveplates of \lop{}, \ldepl{} and the dissociation lasers to reverse the handedness of the light from each.

\subsection{Images to determine doublet positions}\label{sec:images}

To determine where the dissociated \hfp{} ions from each doublet fall on the phosphor screen in each switch state, we take a series of images with no Ramsey sequence. For these images we prepare the stretched states as described in Sec.~\ref{sec:state-prep} but apply no $\pi/2$ pulses before removing the population in one of the doublets using low-power \ldepl{} light, tuned to resonance with the doublet to be depleted, and propagating along the $z$ direction to avoid micromotion-induced Doppler shifts. We take 3 types of image per switch state: one where we deplete the lower Stark doublet, one where we deplete the upper Stark doublet, and one where the laser is tuned between the doublets to deplete both symmetrically. We use these three images to determine the center line between the two blobs for each switch state as described in Section~\ref{sec:ion-counting}. We repeated this imaging routine roughly every 10 blocks during the dataset.

\section{\label{sec:data_analysis}Data Analysis} 

\subsection{Ion counting and asymmetry} 
\label{sec:ion-counting}

\begin{figure}
    \centering
    \includegraphics{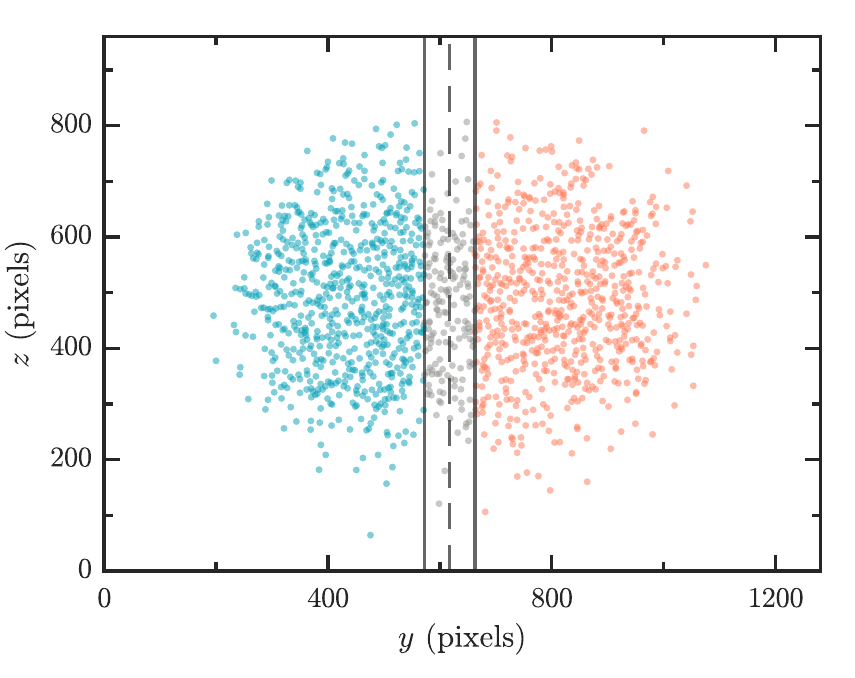}
    \caption{Example ion-detection data for a single shot of the experiment. Swatch (solid lines), from which ion counts are discarded, is defined by region $\pm45$ pixels from center line (dashed line). Ions assigned to the upper and lower doublets are shown in pink and blue respectively.}
    \label{fig:example-image}
\end{figure}

Our experimental signal is dissociated Hf$^+$ ions read out via phosphorescence on an imaging microchannel plate (MCP). The images are processed asynchronously and we save a file which contains the locations of each bright spot which was determined to be an ion according to our smoothing and noise-reducing processing algorithm. The data from a tyical shot is shown in Fig~\ref{fig:example-image}. The full eEDM dataset contains $\sim10^8$ ion detection events.

We use this same algorithm to analyze the test images described in Sec.~\ref{sec:images} and find a \textit{center line} for each switch state. We use this center line when analyzing the Ramsey data to define a \textit{swatch} which is a rectangular area, of fixed width, at the center of the image from which ion counts are discarded, as shown in Fig.~\ref{fig:example-image}. We do this because the doublets are not entirely separated on the screen, so in this area we cannot be sure that we will assign ions to the correct doublet. The extent to which we are able to isolate the two doublets is given by the imaging contrast $C_{\rm I}$. As the swatch width increases, the imaging contrast improves but total ion number $N$ decreases as we throw out more ions. For the final analysis of the dataset, we used a swatch of 90 pixels---to be compared with the total width covered by the detected ions, about 900 pixels---which roughly maximizes $C_{\rm I}\sqrt{N}$, proportional to our sensitivity. 

Once we have our center line for each switch state we can properly count Hf$^+$ ions in our Ramsey data images and assign them to the correct doublet. For every image (which corresponds to a single run of the experiment), we end up with a number of ions in the upper doublet $N^{\rm u}$, and number of ions in the lower doublet $N^{\rm l}$.

For every switch state, we take data in both the in-in and anti-in (or the anti-anti and in-anti) combinations of preparation and readout phase. If the preparation and readout phase are the same (i.e. in-in and anti-anti), then the fringe formed as we vary $t_R$ will have a $\pi$ phase shift from the fringe formed when they are different (anti-in or in-anti). We will refer to in-in and anti-anti as ``In" and in-anti and anti-in as ``Anti". Now for each pair of shots, we can form our spin asymmetry,
\begin{equation}
\mathcal{A}_{\rm u/l} = \frac{N_{\rm In}^{\rm u/l}-N_{\rm Anti}^{\rm u/l}}{N_{\rm In}^{\rm u/l}+N_{\rm Anti}^{\rm u/l}},
\label{eq:asymmetry}
\end{equation}
where the subscript refers to the preparation and readout phase combination. For each Ramsey time and switch state we take $n$ shots and so can form $n/2$ asymmetries for each of the upper and lower doublet. From these $n/2$ asymmetries we construct a mean $\mathcal{A}^{\rm m}_{\rm u/l}$, and from their scatter, a standard error on the mean $\delta\mathcal{A}_{\rm u/l}$.

We then form two `meta' asymmetries by taking the difference ($D$) and sum ($S$) of the upper and lower  asymmetries,
\begin{equation}
\begin{split}
\mathcal{A}_{\rm D} &= \mathcal{A}_{\rm u} - \mathcal{A}_{\rm l}, \\
\mathcal{A}_{\rm S} &= \mathcal{A}_{\rm u} + \mathcal{A}_{\rm l}, \\
\end{split}
\label{eq:asymmetry-meta}
\end{equation}
with means $\mathcal{A}^{\rm m}_{\rm D},\mathcal{A}^{\rm m}_{\rm S}$ and standard errors $\delta\mathcal{A}_{\rm D},\delta\mathcal{A}_{\rm S}$. The $\delta\mathcal{A}_{\rm D}$ are reduced compared to $\delta\mathcal{A}_{\rm S}$ (and $\delta\mathcal{A}_{\rm u/l})$ due to common-mode cancellation of many sources of noise.

\subsection{Least squares fitting}

As mentioned previously, we perform our Ramsey experiment simultaneously on both doublets and use their opposing orientations at the time of dissociation to read them out on opposing sides on the imaging MCP screen. Because the data are acquired simultaneously, the difference asymmetry allows us to cancel much of the common-mode noise, leaving us with doublet-odd data with less scatter than the raw data. Unfortunately, the doublets are not fully separated on the screen, so we must be careful with how we fit our data. 

For an ideal Ramsey fringe, with no leakage from the other doublet, we can define a functional form for the asymmetry,
\begin{equation}
h_{\rm u/l}(t_R) = C_{\rm u/l} e^{-\gamma_{\rm u/l} t_R} \cos(2 \pi f_{\rm u/l} t_R + \phi_{\rm u/l}) + O_{\rm u/l}. \label{eq:decaying-sine-wave}
\end{equation}
Here $C$ is the initial fringe contrast, $\gamma$ the contrast decay rate, $f$ the fringe frequency, $t_R$ the free evolution time, $\phi$ the initial phase, $O$ the offset, and the subscripts indicate the upper or lower Stark doublet. In our fitting routine, we initially fit each fringe separately to this function. From the fit parameters we define the mean and difference parameters as 
\begin{equation}
    \begin{split}
        \alpha_{\rm m} &= \frac{\alpha_{\rm u}+\alpha_{\rm l}}{2},\\
        \alpha_{\rm d} &= \frac{\alpha_{\rm u}-\alpha_{\rm l}}{2},
    \end{split}
\end{equation}
with $\alpha\in\{C,\gamma,f,\phi,O\}$. Due to imperfect imaging contrast $C_{\rm I}$, in reality each doublet's signal has a contribution from the other doublet. In this case, the measured asymmetries are
\begin{equation} 
\begin{split}
    A_{\rm u} & = (\frac{1+C_{\rm I}}{2})h_{\rm u}+ (\frac{1-C_{\rm I}}{2})h_{\rm l},\\
    A_{\rm l} & = (\frac{1-C_{\rm I}}{2})h_{\rm u}+ (\frac{1+C_{\rm I}}{2})h_{\rm l}.\label{eq:asymmetries}
\end{split}
\end{equation} 
Now our sum and difference asymmetries are
\begin{equation} 
\begin{split}
    A_{\rm S} & = h_{\rm u}+h_{\rm l}\\
    A_{\rm D} & = C_{\rm I}(h_{\rm u}-h_{\rm l}),
\end{split}
\end{equation} 
which we can express in terms of the mean and difference fitting parameters,
\begin{widetext}
\begin{equation} 
\begin{split}
    A_{\rm S} = {}& (C_{\rm m}+C_{\rm d}) e^{-2\gamma_{\rm m} t_R} \cos(2 \pi (f_{\rm m}+f_{\rm d}) t_R + (\phi_{\rm m}+\phi_{\rm d})) + (O_{\rm m}+O_{\rm d})  ...\\
    &+ (C_{\rm m}-C_{\rm d}) e^{-2\gamma_{\rm d} t_R} \cos(2 \pi (f_{\rm m}-f_{\rm d}) t_R + (\phi_{\rm m}-\phi_{\rm d})) + (O_{\rm m}-O_{\rm d}), \\
    A_{\rm D}  ={}& C_{\rm I}((C_{\rm m}+C_{\rm d}) e^{-2\gamma_{\rm m} t_R} \cos(2 \pi (f_{\rm m}+f_{\rm d}) t_R + (\phi_{\rm m}+\phi_{\rm d})) + (O_{\rm m}+O_{\rm d}) ...\\
    &-(C_{\rm m}-C_{\rm d}) e^{-2\gamma_{\rm d} t_R} \cos(2 \pi (f_{\rm m}-f_{\rm d}) t_R + (\phi_{\rm m}-\phi_{\rm d})) - (O_{\rm m}-O_{\rm d}) ).\label{eq:asymmetry-simultaneous}
\end{split}
\end{equation} 
\end{widetext}
We use these two expressions to perform a simultaneous least-squares fit for the sum and difference asymmetries in each experimental switch state. The value of $C_{\rm I}$, the imaging contrast, is fixed at 0.89 for this fit---determined as described in Sec.~\ref{sec:systematic_fDpulling}. The parameter uncertainties are extracted based solely on the standard errors of the asymmetries used in the fits. The resultant uncertainties on the fitted values of $f_{\rm d}$ and $\phi_{\rm d}$ are close to the shot noise limit and much smaller than those on $f_{\rm m}$ and $\phi_{\rm m}$ thanks to our simultaneous data collection and fitting routine, which cancels most of the common-mode noise. The outputs of these simultaneous fits are used for all further analysis.

\subsection{Switch-parity channels}

After fitting to each Ramsey fringe in a block to extract the 10 fitting parameters, we use the resulting 8 values of each parameter to form 8 linear combinations which we call switch-parity channels. The switch-parity channel for the mean and difference parameters $\alpha_{\rm m}$ and $\alpha_{\rm d}$ which are odd under the product of switches $[\tilde{\mathcal{S}}_a\tilde{\mathcal{S}}_b...]$ are given by
\begin{equation}
\begin{split}
    \alpha^{\mathcal{S}_a\mathcal{S}_b...} ={}& \frac{1}{8}\sum_{\tilB,\tilR,\tilI=\pm 1}[\tilde{\mathcal{S}}_a\tilde{\mathcal{S}}_b...] \alpha_{\rm m}(\tilB,\tilR,\tilI),\\
    \alpha^{D\mathcal{S}_a\mathcal{S}_b...} ={}& \frac{1}{8}\sum_{\tilB,\tilR,\tilI=\pm1}[\tilde{\mathcal{S}}_a\tilde{\mathcal{S}}_b...] \alpha_{\rm d}(\tilB,\tilR,\tilI).
\end{split}
\end{equation}
So for example, $C^{DBRI}$ is formed from adding together the measured $C_{\rm d}$ in all switch states for which the product $\tilB\tilR\tilI=1$, subtracting all switch states for which the product $\tilB\tilR\tilI=-1$ and dividing by the number of switch states. $\fBR$ is formed by adding together the $f_{\rm m}$ measured in all switch states for which the product $\tilB\tilR=1$, subtracting the $f_{\rm m}$ measured in all switch states for which the product $\tilB\tilR=-1$ and, again, dividing by the total number of switch states. We note that, because the measured $f_{\rm u/l}$ are defined as positive quantities (see Sec.~\ref{sec:effective-hamiltonian}), the $\tilB$ switch is anomalous in that frequency contributions which change sign with $\tilB$ appear in $\tilB$-even channels while contributions which are independent of the $\tilB$ switch appear in $\tilB$-odd channels. The eEDM channel is $\fBD$. All other parity channels allow us to to diagnose experimental issues and identify sources of systematic error.

\subsection{Blinding}

We blinded the dataset by programming the fitting routine to add an unknown constant offset to the $\fBD$ channel. This offset was stored in an encrypted file and was not removed until all statistical and systematic checks on the dataset had been completed, and the uncertainties finalized. 

\subsection{Data cuts}

\begin{figure}
    \centering
        \includegraphics[]{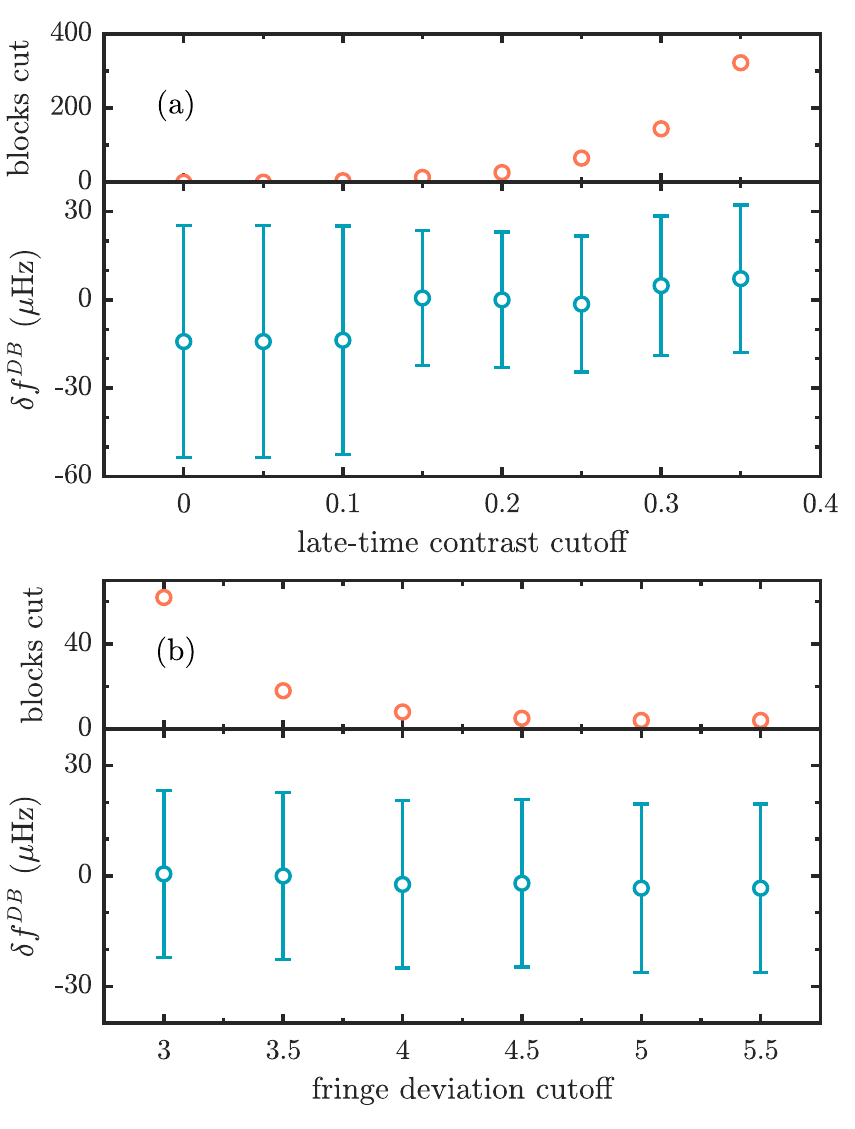}
    \caption{Change in $\fBD$ over whole dataset as a function of cuts explained in text. (a) Late-time contrast cut. (b) Individual fringe outlier cut. In each plot, only one cut is applied. The error bars on $\fBD$ are corrected by a factor of $\sqrt{\chi^2}$.}
    \label{fig:fBD_vs_cuts}
\end{figure}

After completing the dataset, we applied cuts to the blinded data based on non-EDM channels indicating signal quality. Blocks with any individual fringe with late-time contrast below $C_{\rm late}=0.2$ were cut due to low signal to noise. By inspection of least-squares fits of individual fringes, this cut served as a good proxy for pathological fitting results. Blocks were also cut if they contained a fringe with a fitted difference frequency $f_{\rm d}$ in any switch state which was more than $3.5\sigma$ different from the mean fringe frequency for that switch state. The mean fringe frequencies were calculated over all blocks not removed by the late-time-contrast cut and which had the same value of $\vec{\mathcal{B}^0}$. They were constructed from the linear combinations, including the blinded offset on $\fBD$. This cut helped remove blocks where an experiment malfunction, e.g. laser unlocking, affected just one or two shots in a fringe. If our data were perfectly normally distributed, with no outliers, this would be expected to remove $\sim 5$ blocks, and decrease $\chi^2$ by $\sim0.7\%$. Figure~\ref{fig:fBD_vs_cuts} shows the shift in the center value, and the error bar of the eEDM channel, as a function of each of these two cuts. The first cut removed 26 blocks and the second 15, leaving \numblocks{} blocks in the final dataset, with $\chi^2=\num{1.07+-0.04}$ for $\fBD$. Our final $1\sigma$ statistical error of \SI{23}{\micro\hertz} has been relaxed by a factor $\sqrt{\chi^2}=1.035$.

\section{Effective Hamiltonian for doublets}
\label{sec:effective-hamiltonian}

To an excellent approximation, we can model the evolution of either of the Stark doublets, shown in pink and blue in Fig.~\ref{fig:ramsey_fringe_figure}, as a two-level system. The effective Hamiltonian for each pair can be parameterized
\begin{equation}
    H_{\rm eff}=\frac{h}{2}\begin{pmatrix}
        f_0 & \Delta \\
        \Delta & -f_0
    \end{pmatrix}.
\end{equation}
The diagonal components $f_0$ contain all terms which directly shift the energies of the two states relative to one another, whilst the off-diagonal components $\Delta$ contain all terms which mix the two states. We can expand both $f_0$ and $\Delta$ in terms of their leading contributions,
\begin{align}
    f_0 &= \tilB f_0^0 + \delta \mathcal{F}_0,\\
    \Delta &= \tilR \Delta^0 + \tilR \tilD \Delta^D + \delta\mathcal{D},
\end{align}
Here the quantities with tildes are equal to $\pm1$ and are determined by the experimental switch state, discussed in Sec.~\ref{sec:exp-switch-states}. The remaining parameters are defined in the next two paragraphs. 

The principal contribution to $f_0$ is the Zeeman splitting $f^0_0=3 g_F \mu_{\rm B} \Brot\sim \SI{100}{\hertz}$. The off-diagonal component is dominated by two terms with similar magnitude, but different switch state dependence; $\Delta^0$ and $\Delta^D\sim \SI{1}{\hertz}$ represent a slight mixing of the two $m_F=\pm3/2$ states in each doublet and arise at fourth order in perturbation theory in the full molecular Hamiltonian from the combined perturbations of rotation and $\Omega$-doubling, breaking the degeneracy of either Stark doublet at $\Brot=0$ \cite{Leanhardt2011,Meyer2009}. $\Delta^0$ and $\Delta^D$ are given by
\begin{equation}
\begin{split}
    h \Delta^0 &= \frac{3 \hbar \wef}{2} \left( \frac{\hbar \wrot}{\dmf \Erot} \right)^3
        \left( \frac{18 \Apar^2 - 19 \dmf^2 \Erot^2}{\Apar^2 - \dmf^2 \Erot^2} \right),\\
    h \Delta^D &= \frac{3 \hbar \wef}{2} \left( \frac{\hbar^3 \wrot^3}{\dmf^2 \Erot^2 \Apar} \right)
        \left( \frac{9 \Apar^2 - 8 \dmf^2 \Erot^2}{\Apar^2 - \dmf^2 \Erot^2} \right),
        \label{eq:Delta_eq}
        \end{split}
\end{equation}
where the various constants are defined in Table~\ref{tab:mol_constants}.
These expressions are valid as long as $\dmf\Erot \gg  \hbar \wef$ and $\dmf \Erot \gg  \hbar \wrot$ or, in other words, if the molecule is fully polarized. The strong scaling of $\Delta$ with $\Erot$ allows us to perform off-resonant $\pi/2$ pulses by modulating the magnitude of $\Erot$ as described in Sec.~\ref{sec:ramsey-sequence}.

The additional perturbations are given by
\begin{align}
    \delta \mathcal{F}_0(\tilB,\tilR,\tilI,\tilD) &= \sum_{\tilS\in\mathcal{W}}\tilS \delta f_0^s,\\
    \delta \mathcal{D}(\tilB,\tilR,\tilI,\tilD) &= \sum_{\tilS\in\mathcal{W}}\tilS \delta \Delta^s,
\end{align}
where both summations are over $\mathcal{W}$, the set of all possible products of $\{\tilB, \tilR, \tilI\, \tilD \}$, and the superscript $s$ on the $\delta f_0^s$ and $\delta \Delta^s$ denote the switch state dependence of the perturbation \textit{relative} to the largest term in each matrix element, $f_0^0$ and $\Delta^0$ respectively. For our purposes, the most important perturbation is that due to the eEDM which contributes a diagonal term, $\tilD \delta f^{DB}_0=2 \tilD d_e \Eeff$. Others which are important for our determination of $d_e$ are discussed in detail in Sec.~\ref{sec:frequency_shifts}.

For each experimental switch state $(\tilB,\tilR,\tilI)$, and doublet $\tilD$, we measure a frequency $f(\tilB,\tilR,\tilI,\tilD)$ corresponding to the energy difference between the two eigenstates, which we define to be always positive. For typical experimental parameters, $f^0_0$ is roughly two orders of magnitude larger than any other term in the Hamiltonian, and so we can expand $f$ about $f^0_0$,
\begin{widetext}
\begin{equation}
    \begin{split}
        f(\tilB,\tilR,\tilI,\tilD) ={}& \Big|\tilB f^0_0 + \delta \mathcal{F}_0+ \frac{{(\Delta^0)}^2+{(\Delta^D)}^2+2\tilD\Delta^0\Delta^D+2\tilR(\Delta^0+\tilD\Delta^D)\delta\mathcal{D}+{\delta\mathcal{D}}^2}{2\tilB f_0^0}\\
        &-\delta\mathcal{F}_0\frac{{(\Delta^0)}^2+{(\Delta^D)}^2+2\tilD\Delta^0\Delta^D+2\tilR(\Delta^0+\tilD\Delta^D)\delta\mathcal{D}+{\delta\mathcal{D}}^2}{2{f_0^0}^2} + ...\Big|\\
        ={}& |f^0_0| + \tilB\delta\mathcal{F}_0+ \frac{{(\Delta^0)}^2+{(\Delta^D)}^2+2\tilD\Delta^0\Delta^D+2\tilR(\Delta^0+\tilD\Delta^D)\delta\mathcal{D}+{\delta\mathcal{D}}^2}{2|f_0^0|}\\
        &-\tilB\delta\mathcal{F}_0\frac{{(\Delta^0)}^2+{(\Delta^D)}^2+2\tilD\Delta^0\Delta^D+2\tilR(\Delta^0+\tilD\Delta^D)\delta\mathcal{D}+{\delta\mathcal{D}}^2}{2|f_0^0|^2} + ...\label{eq:freq-expansion}
    \end{split}
\end{equation}
\end{widetext}


\section{\label{sec:accuracy_evaluation_overview}Evaluation of systematic uncertainty}

Accurate determination of the eEDM-induced energy shift in \hffp{} requires a careful evaluation of all significant systematic shifts. There are several methods we employ to search for and understand systematics. First is to change experimental parameters in a controlled way and look for shifts in the eEDM channel. Section \ref{sec:baxgradnr} is a good example of doing this. To see unambiguous shifts with this method, on a timescale which is short compared to the full dataset, one typically needs to change the parameter in question by a factor of $10$ or more, which is not always possible. Second, we can \textquote{break} the data analysis by making extreme assumptions during fitting. This can be very precise since it can be performed on our entire dataset before unblinding, and runs no risk of actually breaking the physical experiment. Section \ref{sec:internal_state_measurement} contains several examples of this approach. Third, we can use other parity channels for insight into the eEDM channel as they can at times magnify certain effects. Section \ref{sec:axial2f_plus_radial_ellipticity} is a good example of this approach. Finally, one can perform detailed analytical calculations or simulations to understand how experimental parameters might affect our measurement. Section \ref{subsec:magnetic_effects} is an example of this approach.

Our systematics themselves can be divided into $2$ categories: shifts in the actual frequency of phase evolution between the two states in a doublet, and shifts in the phase at early or late free evolution time. Frequency shifts are the dominant class of systematics in our experiment, and are covered in Section \ref{sec:frequency_shifts}. The effects of phase shifts---which tend to be smaller, because they are suppressed by our long coherence times and ability to measure both the early-time and late-time phase (not possible in beam experiments)---are covered in Section \ref{sec:phase_shifts}.

\section{\label{sec:frequency_shifts}Frequency shifts} 

\subsection{\label{subsec:magnetic_effects}Magnetic effects} 

Many experiments to measure EDMs are plagued by the interaction of the atom or molecule with stray magnetic fields. In our experiment, three features make us comparatively immune to these effects. First, the $^3\Delta_1$ state we use has an extremely small magnetic moment, roughly 200 times smaller than the magnetic moment of a bare electron. Second, our measurement is a differential one taken simultaneously between two pairs of states with almost identical magnetic moments. This allows us to measure magnetic effects by looking at the $\fB$ channel whilst they cancel to a high degree in our measurement channel $\fBD$. Finally, the rotating quantization axis in our experiment means that most effects considered in other experiments tend to average to zero over an integer number of rotation cycles. However, there are still important magnetic effects that can shift our measured value of $\fBD$ and we consider those in this section.

The interaction of the science states with a magnetic field $\vec{\mathcal{B}}$ is well described by the effective Zeeman Hamiltonian,
\begin{equation}
    \begin{aligned}
        H_Z &= (\vec{\mu_0} + \tilD \vec{\mu_{\rm D}})\cdot\vec{\mathcal{B}},\\
        &= m_F \mu_{\rm B}(g_F\pm\dgF)\hatErot\cdot\vec{\mathcal{B}}, \label{eqn:zeeman-ham}
    \end{aligned}
\end{equation}
where $\tilD=\pm1$ corresponds to the upper and lower doublet respectively. Here $\vec{\mu_0}$ is the part of the magnetic moment that is common to both doublets and $\vec{\mu_{\rm D}}$ is the differential part, typically 460 times smaller for our choice of experimental values. In the second line, we have made explicit the fact that both magnetic moments track the quantization axis defined by the unit vector pointing along $\Erot$\footnote{While strictly an approximation, this is true to high precision for the fields used in our experiment. The typical electric fields of $\sim\SI{50}{\volt\per\centi\meter}$ cause shifts between adjacent $m_F$ levels of $\sim\SI{30}{\mega\hertz}$ while typical magnetic fields of $\sim\SI{10}{\milli\gauss}$ cause shifts of $\sim\SI{50}{\hertz}$. This means that deflection angles caused by magnetic fields are $\sim10^{-6}$}. We choose to write the magnetic field as a sum of two parts, $\vec{\mathcal{B}} = \tilB\vec{\mathcal{B}^0} +  \vec{\delta\mathcal{B}}$ where the first part is due to the idealized applied quadrupole magnetic field gradient which reverses perfectly with the $\tilB$ switch, and the second part $\vec{\delta\mathcal{B}}$ represents any additional magnetic field experienced by the molecules. In general, $\vec{\delta\mathcal{B}}$ will be composed of many components with different dependence on the switch state and can be written
\begin{equation}
    \vec{\delta\mathcal{B}} = \sum_{\tilS\in\mathcal{W}}\tilS \,\vec{\delta\mathcal{B}^s},
\end{equation}
where the summation is over $\mathcal{W}$, the set of all possible products of $\{\tilB,\tilD,\tilR,\tilI\}$, and the superscript $s$ denotes the switch state dependence of the field relative to $\vec{\mathcal{B}^0}$. For example, the largest magnetic shifts measured in the experiment are caused by charging currents induced by the oscillating voltages on the radial electrodes. The charging currents produce a uniform magnetic field which rotates with $\vecErot$ and whose sign depends on the rotation direction,
\begin{equation}
    \tilR \,\vec{\delta\mathcal{B}^{BR}} = \tilR  \frac{\mu_0 \wrot C_{\rm eff} V_{\rm rot}}{2\pi R_{\rm eff}}\hatErot.
\end{equation}
Here $V_{\rm rot}$ is the amplitude of the oscillating voltages on each of the fins, $C_{\rm eff}$ is the effective capacitance, and $R_{\rm eff}$ the effective radius of the trap. In our experiment $|\vec{\delta\mathcal{B}^{BR}}|\sim\SI{15}{\micro\gauss}$, causing $\fBR\sim\SI{200}{\milli\hertz}$.

We will be particularly interested in $\vec{\delta\mathcal{B}^B}$, the magnetic field which does not chop sign with $\tilB$, causing shifts in $\fB$ and $\fBD$. To first order\footnote{We note that there are higher-order corrections to these shifts arising from mixing of the two states in a doublet described in Section~\ref{sec:effective-hamiltonian}. These effects are well understood but, under the experimental parameters used in the dataset, contribute less than 10\,\% corrections to the shifts discussed and so are not included in our systematic uncertainty budget.} these are
\begin{align}
    h \delta \fB & = 3 g_F \mu_{\rm B} \hat{\mathcal{E}}\cdot\vec{\delta\mathcal{B}{}^B},\\
    h \delta \fBD & = 3 \dgF \mu_{\rm B} \hat{\mathcal{E}}\cdot\vec{\delta\mathcal{B}^B},\label{eqn:zeeman-shifts}
\end{align}
where $\hat{\mathcal{E}}$ is the unit vector pointing along the total electric field $\vec{\mathcal{E}}$.

The differential magnetic moment $\vec{\mu_{\rm D}}$ arises principally from two effects---mixing of adjacent $m_F$ levels in the rotating frame, and mixing of the $J=1$ levels with $J=2$ induced by the electric field. For the experimental parameters used throughout our dataset, the latter dominates and so we neglect the former here. The differential g-factor $\dgF$ can then be calculated from second-order perturbation theory as,
\begin{equation}
    \dgF \simeq \frac{3 \dmf\Gpar |\vec{\mathcal{E}}|}{20 B},
\end{equation}
where the values of molecular constants are given in Table~\ref{tab:mol_constants}. We see that, as expected, the mixing is linear in the size of the electric field. We can use this, in combination with Eq.~\ref{eqn:zeeman-shifts} to give expressions for the time-averaged frequency shifts we measure in the experiment,
\begin{align}
    h \langle\delta \fB\rangle & = 3 g_F \mu_{\rm B} \langle\hat{\mathcal{E}}\cdot\vec{\delta\mathcal{B}^B}\rangle,\\
    h \langle\delta \fBD\rangle & = 3\frac{\dgF}{|\vec{\mathcal{E}}|} \mu_{\rm B} \langle\vec{\mathcal{E}}\cdot\vec{\delta\mathcal{B}^B}\rangle,
\end{align}
where $\dgF/|\vec{\mathcal{E}}|$ is independent of the electric field. We see that the common mode part of the Zeeman interaction is proportional to $\langle\hat{\mathcal{E}}\cdot\vec{\delta\mathcal{B}^B}\rangle$, while the differential part is proportional to $\langle\vec{\mathcal{E}}\cdot\vec{\delta\mathcal{B}^B}\rangle$. This subtle difference will be the source of most of the systematics described in this section.

\begin{table}
	\caption{Magnetic-field expansion}
	\centering
	\begin{ruledtabular}
	\begin{tabular}{@{}c c@{}}
	    Coefficient & $\vec{\nabla} r^l \mathcal{Y}_{l,m}$\\
	    \midrule
	    \multicolumn{2}{c}{Uniform fields}\\
	    \midrule
	    $\mathcal{}{B}_{1,-1}$ & $\hat{y}$\\
	    $\mathcal{B}_{1,0}$ & $\hat{z}$\\
	    $\mathcal{B}_{1,1}$ & $\hat{x}$\\
	    \midrule
	    \multicolumn{2}{c}{First-order gradients}\\
	    \midrule
	    $\mathcal{B}_{2,-2}$ & $\sqrt{3}(y\hat{x} + x\hat{y})$\\
	    $\mathcal{B}_{2,-1}$ & $\sqrt{3}(z\hat{y} + y\hat{z})$\\
	    $\mathcal{B}_{2,0}$ & $-x\hat{x}-y\hat{y}+2z\hat{z}$\\
	    $\mathcal{B}_{2,1}$ & $\sqrt{3}(z\hat{x}+x\hat{z})$\\
	    $\mathcal{B}_{2,2}$ & $\sqrt{3}(x\hat{x}-y\hat{y})$\\
	    \midrule
	    \multicolumn{2}{c}{Second-order gradients}\\
	    \midrule
	    $\mathcal{B}_{3,-3}$ & $\frac{3}{2} \sqrt{\frac{5}{2}} (2 x y\hat{x} + (x-y)(x+y)\hat{y})$ \\
	    $\mathcal{B}_{3,-2}$ & $\sqrt{15}(yz\hat{x} + xz\hat{y} + xy\hat{z})$\\
	    $\mathcal{B}_{3,-1}$ & $\sqrt{\frac{3}{2}}(-xy\hat{x} + \frac{1}{2}(-x^2-3y^2+4z^2)\hat{y}+4yz\hat{z}$\\
	    $\mathcal{B}_{3,0}$ & $-3(xz\hat{x} + yz\hat{y} +\frac{1}{2}(x^2+y^2-2z^2)\hat{z})$\\
	    $\mathcal{B}_{3,1}$ & $\sqrt{\frac{3}{2}}(\frac{1}{2}(-3x^2-y^2+4z^2)\hat{x}-xy\hat{y}+4xz\hat{z})$\\
	    $\mathcal{B}_{3,2}$ & $\sqrt{15}(xz\hat{x}-yz\hat{y}+\frac{1}{2}(x-y)(x+y)\hat{z})$\\
	    $\mathcal{B}_{3,3}$ & $\frac{3}{2} \sqrt{\frac{5}{2}} ((x-y)(x+y)\hat{x}-2xy\hat{y})$
	\end{tabular}
	\label{tab:b-field-imperfections}
	\end{ruledtabular}
\end{table}

Any possible $\vec{\delta\mathcal{B}^B}$ can be expanded as
\begin{equation}
    \vec{\delta\mathcal{B}^B}(\vec{x},\vec{y},\vec{z}) = \sum_{l=1,2,...}\sum_{m=-l}^l \mathcal{B}_{l,m} \vec{\nabla} r^l \mathcal{Y}_{l,m},\label{eq:mag-field-imperfections}
\end{equation}
where $\vec{x},\vec{y},\vec{z}$ are the ions position relative to the trap center, $\mathcal{B}_{l,m}$ is the coefficient of each component and $\mathcal{Y}_{l,m}$ are the semi-normalized real spherical harmonics
\begin{equation}
    \mathcal{Y}_{l,m} = \sqrt{\frac{4\pi}{2l+1}}\times
    \begin{cases}
    \frac{i}{\sqrt{2}}(Y_l^m-(-1)^mY_l^{-m})&\text{if } m<0\\
    Y_l^0&\text{if } m=0\\
    \frac{1}{\sqrt{2}}(Y_l^{-m}+(-1)^mY_l^m)&\text{if } m>0
    \end{cases}.
\end{equation}
The $\vec{\nabla} r^l \mathcal{Y}_{l,m}$ for $l\le3$ are given in Table~\ref{tab:b-field-imperfections}. 

\subsubsection{Non-reversing $\Baxgrad$}
\label{sec:baxgradnr}
\begin{figure}
    \centering
        \includegraphics[]{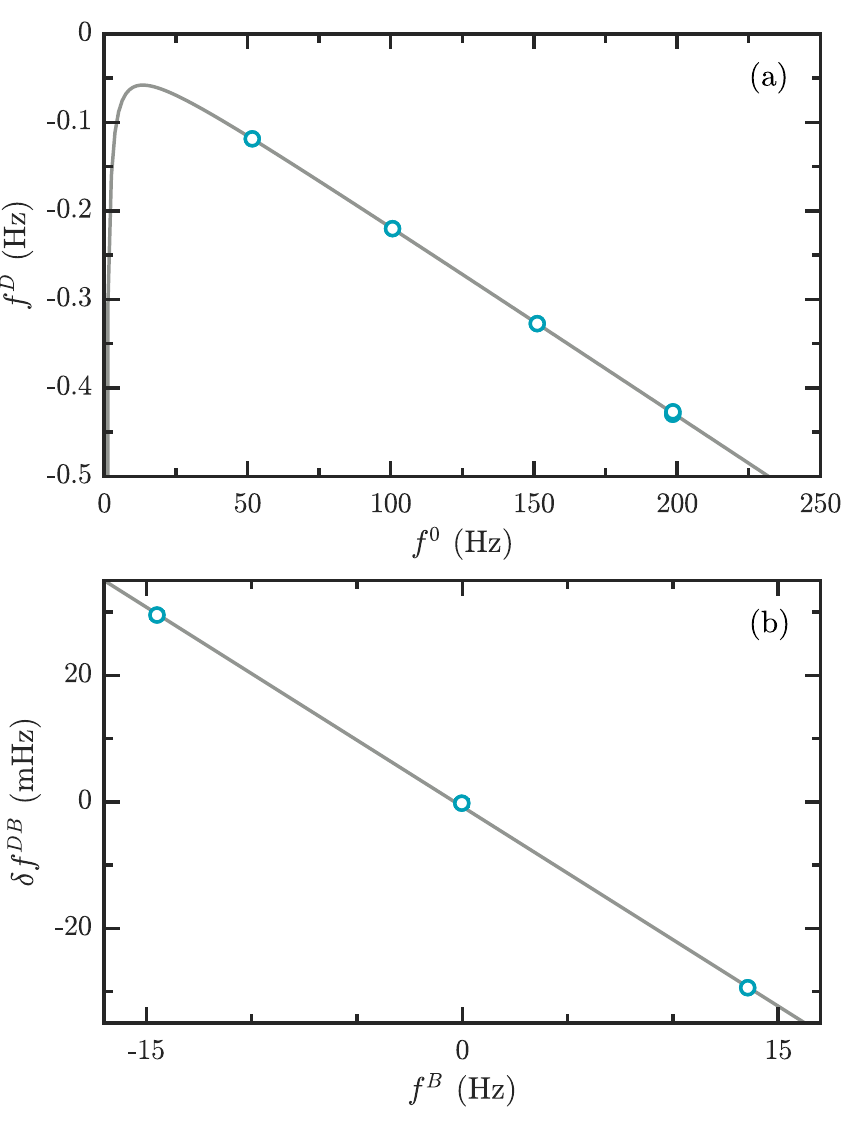}
    \caption{Measurement of $\tfrac{\dgF}{g_F}$. (a) $\fD$ vs $\fn$ for various values of the applied quadrupole magnetic field, $\vec{\mathcal{B}^0}$. Data taken at $\Erot \sim \SI{58}{\volt\per\centi\meter}$. Fit is to $f^D = \tfrac{\dgF}{g_F}f^0+\tfrac{\Delta^0\Delta^D}{f^0}$, giving $\tfrac{\dgF}{g_F}=\num{-0.002149+-0.000003}$, $\Delta^0\Delta^D=\SI{-0.39+-0.04}{\hertz\squared}$. (b) Change in $\fBD$ vs $\fB$ induced by introducing a large non-reversing $\mathcal{B}_{2,0}$. Fitted gradient is equal to $\tfrac{\dgF}{g_F}-\tfrac{\Delta^0\Delta^D}{{f^0}^2}$. Combined with value of $\Delta^0\Delta^D$ from (a) gives $\tfrac{\dgF}{g_F}=\num{-0.00213+-0.00002}$, in excellent agreement. In both plots, error bars are smaller than points.}
    \label{fig:dgF}
\end{figure}

In an idealized version of the experiment where the ions are subject only to a perfect rotating electric field $\vecErot=\Erot(\cos(\wrot t)\hat{x} + \sin(\wrot t) \hat{y})$, the micromotion of the ions is $-\tfrac{e}{m \wrot^2}\vecErot=-\rrot\hatErot$. In this case the only component of the magnetic field with $l\le3$ which causes a non-zero time-averaged shift in any frequency channel is the quadrupole magnetic field $\vec{\delta\mathcal{B}^B}=\mathcal{B}_{2,0}(-x\hat{x}-y\hat{y}+2z\hat{z})$. As described in Section~\ref{sec:experimental_overview}, we deliberately apply such a quadrupole magnetic field $\vec{\mathcal{B}^0}$ whose sign reverses in the $\tilB$ chop and which causes the main contribution to $f$. However there can be an additional contribution which does not reverse sign with $\tilB$, arising either from a background field present in the lab, or from imperfect reversal of the applied field. In this case we have 
\begin{align}
    h \langle\delta \fB\rangle & = -3 g_F \mu_{\rm B} \mathcal{B}_{2,0} \rrot,\\
    h \langle\delta \fBD\rangle & = -3 \dgF \mu_{\rm B} \mathcal{B}_{2,0} \rrot.
\end{align}
Both shifts are proportional to $\mathcal{B}_{2,0}$ but with the $\tilD$-even shift being $\sim 460$ times larger. There are two possible approaches to removing the effect of this systematic from our measurement. The first is to make a correction to the measured value of $\fBD$ based on the measured value of $\fB$,
\begin{equation}
\label{eq:fB-corr}
    \delta \fBD_{\rm corr} = \fB \frac{\dgF}{g_F}.
\end{equation}
The second is to shim out $\mathcal{B}_{2,0}$ by deliberately applying slightly different currents through the $\vec{\mathcal{B}^0}$ coil in each direction. Both approaches are equivalent, we choose to combine them. We minimize $\fB$ by shimming $\mathcal{B}_{2,0}$ on each block based on the measured value of $\fB$ in the previous block and then correct the measured value of $\fBD$ on each block by the measured value of $\fB$ on that block. The quantity $\tfrac{\dgF}{g_F}$ in Eq.~\ref{eq:fB-corr} is determined directly from experiment as shown in Fig.~\ref{fig:dgF}. The largest $\fB$ measured on any single block is \SI{35}{\milli\hertz}, corresponding to a correction to $\fBD$ of \SI{75}{\micro\hertz}, while the median correction size is \SI{5}{\micro\hertz}. The average correction over the whole dataset is just \SI{90}{\nano\hertz}.

This approach is effective in reducing the contribution of this systematic shift to levels well below our statistical error bar, but only holds if there are no other shifts in $\fB$ that scale with $\fBD$ with a constant of proportionality different from $\tfrac{\dgF}{g_F}$---or even no shift in $\fBD$ at all. The remainder of this subsection explores and places limits on such shifts arising from stray magnetic fields. Section \ref{subsec:berrys_phase_effects} explores possible contributions from Berry's phase effects.

\subsubsection{General principles in determining other magnetic effects}
\label{subsubsecsec:magnetic_shifts_general_principles}

Although no other magnetic field components couple directly to an idealized rotating electric field to give significant shifts, a number of potentially problematic shifts can arise when other contributions to the electric field experienced by the ions are taken into account. In particular, when the magnitude of the total electric field experienced by the ions is not constant in time, there can be effects for which $\langle\vec{\mathcal{E}}\cdot\vec{\delta\mathcal{B}^B}\rangle\neq\langle|\vec{\mathcal{E}}|\rangle\langle\hat{\mathcal{E}}\cdot\vec{\delta\mathcal{B}^B}\rangle$. 

To enumerate possible shifts in the experiment, we analytically calculate the micromotion of a classical charged particle subject to an electric field $\vec{\mathcal{E}} = \vecErot + \kappa\vec{\delta \mathcal{E}}$ where $\vecErot$ is the idealized rotating electric field, and $\vec{\delta \mathcal{E}}$ represents perturbations to this, discussed in the sections below. This micromotion allows us to write down the time-dependent magnetic field experienced by a molecule moving through magnetic field imperfections $\kappa\vec{\delta\mathcal{B}}$. Finally we find the frequency shifts $\langle\fB\rangle$ and $\langle\fBD\rangle$ by calculating $\langle\hat{\mathcal{E}}\cdot\vec{\delta\mathcal{B}}\rangle$ and $\langle\vec{\mathcal{E}}\cdot\vec{\delta\mathcal{B}}\rangle$. Here the time average is over an integer number of periods of all relevant frequencies and we keep terms up to $\mathcal{O}(\kappa^2)$, before setting $\kappa=1$. 

For those effects which give non-zero shifts in either channel, we constrain the maximum size of the shifts by direct measurement using the ions, or by auxiliary measurements of the size of the imperfections. We note that this analytical approach, by necessity, does not include the effects of spatially dependent electric fields and so misses the effects of ponderomotive forces such as those from the RF confining fields. To validate this approximation, we performed comprehensive numerical simulations including the full motion of the ions in the trap while varying the size of the imperfections listed over ranges well above their expected size in the experiment and observed no additional systematic effects. 

\subsubsection{Second harmonic of $\Erot$ and transverse magnetic field}
\label{sec:2h+B}

The rotating electric field in our experiment is generated by sinusoidally varying voltages on each of the radial electrodes. These voltages are driven by op-amps which inevitably suffer from harmonic distortion, adding higher harmonics of the input signal. Consider the effect of an additional electric field oscillating at the second-harmonic frequency. We have
\begin{align}
\begin{split}
    \vec{\mathcal{E}} &= \Erot \begin{pmatrix}
        \cos(\wrot t)\\
        \sin(\wrot t)\\
        0
    \end{pmatrix}\\
    &+\begin{pmatrix}
        \mathcal{E}_{\rm 2hX}\cos(2\wrot t + \phi_{\rm 2hX})\\
        \mathcal{E}_{\rm 2hY}\cos(2\wrot t + \phi_{\rm 2hY})\\
        0
    \end{pmatrix},
\end{split}
\end{align}
\begin{align}
\begin{split}
\hat{\mathcal{E}} &\simeq \begin{pmatrix}
        \cos(\wrot t)\\
        \sin(\wrot t)\\
        0
    \end{pmatrix}
    +\frac{1}{2\Erot}\begin{pmatrix}
        \mathcal{E}_{\rm 2hX}\cos(2\wrot t + \phi_{\rm 2hX})\\
        \mathcal{E}_{\rm 2hY}\cos(2\wrot t + \phi_{\rm 2hY})\\
        0
    \end{pmatrix}\\
    &+\frac{\mathcal{E}_{\rm 2hX}}{4\Erot}\begin{pmatrix}
        -\cos(\phi_{\rm 2hX})\\
        \sin(\phi_{\rm 2hX})\\
        0
    \end{pmatrix}
    +\frac{\mathcal{E}_{\rm 2hY}}{4\Erot}\begin{pmatrix}
        \sin(\phi_{\rm 2hY})\\
        \cos(\phi_{\rm 2hY})\\
        0
    \end{pmatrix}\\
    &+\frac{\mathcal{E}_{\rm 2hX}}{4\Erot}\begin{pmatrix}
        -\cos(4\wrot t + \phi_{\rm 2hX})\\
        -\sin(4\wrot t +\phi_{\rm 2hX})\\
        0
    \end{pmatrix}\\
    &+\frac{\mathcal{E}_{\rm 2hY}}{4\Erot}\begin{pmatrix}
        -\sin(4\wrot t +\phi_{\rm 2hY})\\
        \cos(4\wrot t +\phi_{\rm 2hY})\\
        0
    \end{pmatrix}
\end{split}
\end{align}
where $\hat{\mathcal{E}}$ has been expanded to first order in $1/\Erot$. Alongside the components oscillating at $\wrot$, and the corrections oscillating at $2\wrot$, $\hat{\mathcal{E}}$ also has terms that are time-independent, and terms that oscillate at $4\wrot$. The time-independent terms can couple to a uniform transverse magnetic field $\vec{\delta\mathcal{B}^B}=\mathcal{B}_{1,1}\hat{x} + \mathcal{B}_{1,-1}\hat{y}$ to give time averaged shifts,
\begin{align}
\begin{split}
    h\langle\delta \fB\rangle ={}& \frac{3 g_F \mu_{\rm B}}{4\Erot}\big(-\mathcal{B}_{1,1}\mathcal{E}_{\rm 2hX}\cos(\phi_{\rm 2hX})\\
    &+\mathcal{B}_{1,-1}\mathcal{E}_{\rm 2hY}\cos(\phi_{\rm 2hY})\\
    &+\mathcal{B}_{1,-1}\mathcal{E}_{\rm 2hX}\sin(\phi_{\rm 2hX})\\
    &+\mathcal{B}_{1,1}\mathcal{E}_{\rm 2hY}\sin(\phi_{\rm 2hY})\big),
\end{split}\\
    h\langle\delta \fBD\rangle ={}& 0.
\end{align}
We note that, depending on the nature of the origin of the second harmonic, the phases $\phi_{\rm 2hX}$ and $\phi_{\rm 2hY}$ can change sign under rotation. In this way, some of this shift can (and usually does) show up in $\fBR$ rather than $\fB$. We have verified these shifts experimentally by intentionally applying a large second harmonic to our electrodes along with large transverse magnetic fields. Although the effect causes no direct shift in $\fBD$, any shift in $\fB$ will cause our applied correction---described in Section~\ref{sec:baxgradnr}---to include a systematic $\delta\fBD=\tfrac{\dgF}{g_F}\langle\delta\fB\rangle$.

During the data run, we measure the transverse field using an array of magnetometers positioned on the vacuum chamber and tune it to zero using three pairs of shim coils arranged around the apparatus. The magnetometers can be subject to small offsets and are calibrated at the beginning of the data run by intentionally applying large second-harmonic electric fields and measuring shifts in $\fB$ and $\fBR$. We used the same technique at regular (roughly weekly) intervals throughout the dataset to check the size of the magnetic fields and guard against any drifts in the magnetometer offsets. Based on these measurements, we conservatively estimate the maximum size of the transverse fields to be $<\SI{10}{\milli\gauss}$ in $x$ and $y$, limited principally by the precision of the current supplies used to drive the shim coils.

To reduce the size of the second-harmonic electric fields present in the experiment, we intentionally apply a feedforward voltage at $2\wrot$ with the opposite phase, canceling the inherent voltage at $2\wrot$ down to about a part in 40,000 of the voltage at $\wrot$. In the worst case, where the relative phases of the residual $2\wrot$ signals are such that the corresponding electric field at the position of the ions is maximized, this corresponds to residual shifts in $\fB$ of about \SI{80}{\milli\hertz\per\gauss}. After this procedure, we deliberately apply large magnetic fields $\sim\SI{2}{\gauss}$ to the ions and observe the residual shifts in $\fB$ to be \SI{103+-1}{\milli\hertz\per\gauss} and \SI{46+-9}{\milli\hertz\per\gauss} for fields along $x$ and $y$ respectively, in good agreement with our direct voltage measurements. The amplitude of the second harmonic was checked periodically throughout the dataset to guard against any drifts, for example in the characteristics of the power operation amplifiers. Combined with the uncertainty on the transverse fields above, the measurement of these shifts limit the contribution of this effect to $\fB$ to less than \SI{1.0}{\milli\hertz} with a corresponding systematic uncertainty in $\fBD$ of \SI{2.2}{\micro\hertz}.

\subsubsection{Higher-harmonics of $\Erot$}
\label{sec:nh-magnetic}

We can generalize the arguments of the previous section to higher harmonics of $\Erot$. For the $n$th harmonic, the expansion of $\hat{\mathcal{E}}$ in powers of $1/\Erot$ will contain components which oscillate at $(n-2)\wrot$ and $(n+2)\wrot$. The $\vecErot$-induced micromotion of the ions combined with an $m$th order magnetic field gradient, produces a magnetic field in the frame of the ions which oscillates at $m\wrot $. So in general an $n$th harmonic of $\Erot$ can couple to $(n-2)$th and $(n+2)$th order magnetic field gradients ($l=n-1$ or $n+3$) to give non-zero time-averaged frequency shifts. After the feedforward is applied to reduce the second harmonic as described above, the next largest shift from harmonic distortion is due to $3\wrot$, with an amplitude on each fin of about a part in 1500 of the amplitude at $\wrot$. The 3rd harmonic can couple to first-order field gradients of the form $\vec{\delta\mathcal{B}^B}=\mathcal{B}_{2,-2}\sqrt{3}(y\hat{x}+x\hat{y}) + \mathcal{B}_{2,2}\sqrt{3}(x\hat{x}-y\hat{y})$ to give shifts,
\begin{align}
\begin{split}
    h\langle\delta \fB\rangle ={}& \frac{3\sqrt{3} g_F \mu_{\rm B} e}{4m \wrot^2}\big(\mathcal{B}_{2,2}\mathcal{E}_{\rm 3hX}\cos(\phi_{\rm 3hX})\\
    &-\mathcal{B}_{2,-2}\mathcal{E}_{\rm 3hY}\cos(\phi_{\rm 3hY})\\
    &-\mathcal{B}_{2,-2}\mathcal{E}_{\rm 3hX}\cos(\phi_{\rm 3hX})\\
    &-\mathcal{B}_{2,2}\mathcal{E}_{\rm 3hY}\cos(\phi_{\rm 3hY})\big),
\end{split}\\
    h\langle\delta \fBD\rangle ={}& 0.
\end{align}
We intentionally applied a large third-harmonic electric field with various phases and observed shifts of $\fB\sim \SI{20}{\milli\hertz}$. The applied third harmonic field was 28 times larger than that present in the dataset and so we conservatively set a limit on the maximum size of any possible shift in $\fB$ at \SI{0.7}{\milli\hertz}, with a corresponding systematic uncertainty in $\fBD$ of \SI{1.5}{\micro\hertz}.

Higher harmonics are comparable to, or lower in amplitude (the fourth harmonic is down about a factor of $2/3$ compared to the third, whilst the fifth is up a factor of about $4/3$; all higher harmonics are smaller) and couple to higher-order magnetic field gradients. To set limits on their possible contributions to shifts in $\fB$ we made measurements of the magnetic field gradients. These measurements were made outside of, but within a few \si{\centi\meter} of, the main vacuum chamber. The electrodes in the experiment are machined from non-magnetic titanium and there are no other sources of magnetic fields closer to the ions than the steel vacuum chamber which is $\sim \SI{10}{\centi\meter}$ away. Because of this, we expect that the field gradients immediately outside the chamber are similar to, or greater than, those inside the chamber. From these measurements we estimate the maximum size of first-order field gradients to be \SI{10}{\milli\gauss\per\centi\meter} and second-order gradients to be \SI{10}{\milli\gauss\per\centi\meter\squared}. To determine the size of the magnetic field oscillating at $n\wrot$, the $n$th order magnetic field gradient should be multiplied by $\rrot^n$ where $\rrot=\SI{0.05}{\centi\meter}$. This means we expect any higher-order harmonic contributions to shifts in $\fB$ to be reduced by a factor of $\sim20$ from those for the third harmonic, corresponding to systematic uncertainties in $\fBD<\SI{100}{\nano\hertz}$. Therefore, we do not include any higher-order effects in our systematic uncertainty budget.

\subsubsection{Ellipticity of $\Erot$}
\label{sec:ellipticity-mag}

Another possible electric field imperfection is an ellipticity of $\Erot$ (i.e. the oscillating voltage on different electrodes in the trap having slightly different amplitudes). A general rotating electric field having an ellipticity with its major axis orientated at angle $\theta$ to the $x$ axis can be written
\begin{equation}
\begin{aligned}
    \vec{\mathcal{E}} ={}& \Erot \begin{pmatrix}
        \cos(\wrot t)\\
        \sin(\wrot t)\\
        0
    \end{pmatrix}
    +\mathcal{E}_\epsilon\begin{pmatrix}
        \cos(2\theta - \wrot t)\\
        \sin(2\theta - \wrot t)\\
        0
    \end{pmatrix}.\label{eq:ellipticity-no-rot}
\end{aligned}
\end{equation}
The corresponding $\hat{\mathcal{E}}$ is
\begin{equation}
    \begin{aligned}
        \hat{\mathcal{E}} = & \begin{pmatrix}
        \cos(\wrot t)\\
        \sin(\wrot t)\\
        0
    \end{pmatrix}
    + \frac{\mathcal{E}_\epsilon}{2\Erot} \begin{pmatrix}
        \cos(2\theta - \wrot t)\\
        \sin(2\theta - \wrot t)\\
        0
    \end{pmatrix}\\
    &+ \frac{\mathcal{E}_\epsilon}{2\Erot} \begin{pmatrix}
        -\cos(2\theta - 3\wrot t)\\
        \sin(2\theta - 3\wrot t)\\
        0
    \end{pmatrix}.
    \end{aligned}
\end{equation}
This field can once again couple to first-order magnetic field gradients of the form $\vec{\delta\mathcal{B}^B}=\mathcal{B}_{2,-2}\sqrt{3}(y\hat{x}+x\hat{y}) + \mathcal{B}_{2,2}\sqrt{3}(x\hat{x}-y\hat{y})$. After taking into account the modification to the ion micromotion caused by this perturbation, the resulting shifts are
\begin{align}
    h\langle\delta \fB\rangle =& -\frac{9\sqrt{3} g_F \mu_{\rm B} e \mathcal{E}_\epsilon}{2m \wrot^2}\big(\mathcal{B}_{2,2}\cos(2\theta) + \mathcal{B}_{2,-2}\sin(2\theta)\big),\\
    h\langle\delta \fBD\rangle =& -\frac{6\sqrt{3} \dgF \mu_{\rm B} e \mathcal{E}_\epsilon}{m \wrot^2}\big(\mathcal{B}_{2,2}\cos(2\theta) \nonumber\\
    &\quad\quad+ \mathcal{B}_{2,-2}\sin(2\theta)\big).
\end{align}
Note that in contrast to the effects of higher-harmonics discussed above, the ellipticity does cause a shift in $\fBD$, with $\delta\fBD/\delta\fB=\tfrac{4}{3}\tfrac{\dgF}{g_F}$. After accounting for the correction described in Sec.~\ref{sec:baxgradnr}, which assumes $\delta\fBD/\delta \fB=\tfrac{\dgF}{g_F}$, the systematic associated with the shift described in this section will be $1/3$ of the shift induced in $\fB$, or a quarter of the shift induced in $\fBD$. The size of the ellipticity in our experiment was measured as described in Sec.~\ref{sec:axial2f_plus_radial_ellipticity}. To set limits on the size of the shift, we applied an ellipticity 5 times larger than this, limited by our shallow ion trap during the Ramsey sequence, and varied its angle. The one-sigma upper limit on the shifts seen in this way was \SI{12}{\milli\hertz} and so we set an upper limit on the size of any shift in $\fB$ due to this effect of \SI{2.4}{\milli\hertz}, corresponding to a systematic uncertainty in $\fBD$ of \SI{1.7}{\micro\hertz}.

\subsubsection{RF micromotion}

In addition to the rotating electric field, the ions are subject to the RF confining field oscillating at $\omega_{\rm rf}$. This RF field induces its own micromotion of the ions and, through $\mathcal{B}_{2,0}$, a magnetic field oscillating at $\omega_{\rm rf}$. The corresponding time-averaged shifts are 
\begin{align}
        h\langle\delta \fB\rangle ={}& 3 g_F \mu_{\rm B} \frac{e \mathcal{E}_{\rm rf}^2\mathcal{B}_{2,0}}{4\Erot m \omega_{\rm rf}^2},\\
        h\langle\delta \fBD\rangle ={}& 3 \dgF \mu_{\rm B} \frac{e \mathcal{E}_{\rm rf}^2\mathcal{B}_{2,0}}{2\Erot m \omega_{\rm rf}^2}.
\end{align}
In this case we have $\langle\delta \fB\rangle/\langle\delta \fBD\rangle=2\dgF/g_F$ and so there is potential for a systematic, but both shifts are proportional to $\mathcal{B}_{2,0}$ which we are shimming to zero. The ratio of the magnitude of the shift caused by the RF field to that caused by $\Erot$ is $\langle\mathcal{E}_{\rm rf}^2 \rangle\wrot^2/\Erot^2\omega_{\rm rf}^2\sim10^{-2}$ so that this shift will just provide a $\sim1\%$ correction to the---already very small---systematic uncertainty in Sec.~\ref{sec:baxgradnr}. In addition, because our measurement of $\dgF/g_F$ was performed with the RF fields present, it already includes this correction. Therefore, we do not include any contribution from this effect in our systematic uncertainty budget. Other electric fields which oscillate at frequencies other than integer multiples of $\wrot$ (e.g. time-averaged electric field due to secular motion) behave similarly but with even smaller contributions.

\subsubsection{Out of plane electric fields}

All the electric field imperfections discussed thus far are additional fields contained within the $x,y$-plane, the plane in which $\Erot$ rotates. However, it's also possible for the ions to experience oscillating fields in the $z$ direction. This can occur, for example, due to thermal motion of the ions or patch charges displacing the center of the ion cloud from the geometrical center of the trapping electrodes. The largest shifts from fields in the $z$ direction are those due to the motion of the ions induced by these fields coupling to $\mathcal{B}_{2,0}$. The shifts have $\delta\fBD/\delta\fB=\dgF/g_F$ and so produce no systematic effects not already accounted for by the approach described in Sec.~\ref{sec:baxgradnr}.

\subsection{Berry's phase effects} 
\label{subsec:berrys_phase_effects}

The rotating electric bias field $\Erot$ defines our quantization axis, and we do our spectroscopy in the rotating frame. Working in this rotating frame has two important effects: a mixing of the eigenstates of the system, which depends on the rotation rate; and an additional phase accumulation between eigenstates with different total angular momentum projections onto the quantization axis, which does not. This latter part, a geometric or Berry’s phase---depending only on the path traced out by the quantization axis in time---is discussed in this section, whilst the former is discussed in Section \ref{subsec:rotation-induced-mixing-effects}.

As the electric field vector in the experiment rotates, it traces out closed loops in phase space. In each loop, the two states in a doublet accrue a differential geometric phase given by 
\begin{equation}
    \phi_{\rm geo} = \Delta m_F \Omega = 3 \Omega,
\end{equation}
where $\Delta m_F$ is the difference in the angular momentum projection of the two states and $\Omega$ is the solid angle traced out by the electric field. In the idealized experiment, the electric field rotates strictly in the $X,Y$-plane, subtending a solid angle of $2\pi$ every $\Trot=1/\frot$. Phase shifts of integer multiples of $2\pi$ are indistinguishable from zero and so not observable in the experiment. However effects that tilt the quantization axis out the $X,Y$-plane can cause a non-zero frequency shift. For small tilts, the shift is given by
\begin{equation}
    \delta f \simeq -\frac{3\frot}{2\pi} \int_0^{\Trot} \alpha(t) \dot{\phi}(t)\mathrm{d}t\label{eq:berrys-phase}
\end{equation}
where $\alpha(t)$ is the tilt of the quantization axis out of the equatorial plane, $\dot{\phi}(t)$ is its azimuthal angular velocity, and the integral is over one period of the rotating field, $\Trot$.

This frequency shift affects both doublets identically but is independent of the $\tilB$ chop and so appears in $\tilB$-odd frequency channels. We note that the main contribution to $\dot{\phi}(t)\sim\tilR\wrot$ is rotation odd and so the most natural channel for the shift to appear in is $\fBR$. However rotation-odd components of $\alpha(t)$ can  cause effects in $\fB$ as well. While the shifts do not appear directly in our measurement channel, any shifts in $\fB$ will cause a systematic shift---suppressed by a factor of $\dgF/g$---due to the correction we make based on the measured value of $\fB$, described in Sec.~\ref{sec:baxgradnr}.

The most straightforward way to tilt the quantization axis is to add a  time-independent electric field in the axial direction. Our measurement using trapped ions is well protected from this possibility; any electric field in the axial direction which does \textit{not} time average to zero, \textit{and} which is not balanced by another force on the ions, will cause the mean position of the ions to shift to where the time-averaged electric field \textit{is} zero. However, there are a number of mechanisms that can cause non-zero time-averaged geometric phases in the experiment. In this section, we discuss these mechanisms and place constraints on their possible size during our measurement.

\subsubsection{Gravity}
\label{subsubsec:gravity-berrys-phase}

The gravitational force on the ions causes them to sit at a position in the trap where the axial electric field is slightly different from zero. This field is $\mathcal{E}_{\rm grav}=mg/e\sim\SI{20}{\micro\volt\per\meter}$, causing a shift $\delta f=3 \frot \mathcal{E}_{\rm grav}/\Erot\sim\SI{4}{\milli\hertz}$. This shift appears rigorously in $\fBR$ which we do not use for correction of $\fBD$ and so we do not include any contribution from this effect in our systematic uncertainty budget.

\subsubsection{ac Stark shift from vibrational cleanup light}
\label{subsubsec:ac_stark_shift}

The vibrational cleanup light is tuned to resonance with the ${^3\Delta_1(v=1)}\leftarrow {^3\Sigma_{0+}}(v=0)$ transition to remove any population in ${^3\Delta_1(v=1)}$ and prevent it decaying into the science state. The light is left on throughout the free evolution time and propagates in the axial direction through the chamber. To estimate the effect of this light, we assume that the polarizability of the $^3\Delta_1(v=0)$ state is dominated by interaction with $^3\Sigma_{0+}(v=0)$, from which the laser is detuned $\delta_L\sim\SI{1.7}{\tera\hertz}$, and the nearby $^3\Sigma_{1}(v=0)$, a further $\sim \SI{30}{\tera\hertz}$ away. Conservatively assuming transition matrix elements of $d_{\rm trans}\sim0.1\,ea_0$ for each, we obtain a polarizability of $d_{\rm trans}^2/(h^2\epsilon_0 c\delta_L)\sim\SI{4e-4}{\hertz\per\watt\meter\squared}$. The light intensity at the position of the ions is $\sim \SI{300}{\watt\per\meter\squared}$, giving an ac Stark shift of order $\sim \SI{0.1}{\hertz}$. 

The scalar ac Stark shift affects all states in $^3\Delta_1(v=0)$ equally and so has no effect on our measurement. However, a gradient of the intensity could exert a force on the ions, pushing them to a position where the electric field is non-zero in a similar way to the gravitational effect described above. The light is roughly collimated through the chamber and we conservatively constrain the intensity gradient to be $<\SI{500}{\watt\per\meter\cubed}$, corresponding to a force more than 10 orders of magnitude smaller than that due to gravity discussed in section \ref{subsubsec:gravity-berrys-phase}.

Effects from possible vector or tensor ac stark shifts depend on the polarization of the light which is nominally linear in the $X,Y$-plane in our experiment. The tensor shift causes an ac stark shift which is differential between doublets, but common mode within a doublet, but which oscillates at $2\frot$, imitating an ellipticity of $\Erot$. Conservatively assuming the tensor polarizability is of similar size to the scalar, this effective ellipticity is $\sim h\times\SI{0.1}{\hertz}/\Erot\dmf\sim10^{-9}$, much smaller than our best limit on the true ellipticity in the trap.

For linearly polarized light, the vector Stark shift is zero but we can consider the effect of a residual handedness. In this case, the vector shift can tilt the quantization axis out of the $x,y$-plane. Conservatively assuming vector polarizability similar size to the scalar, and a 2\,\% residual handedness of the light, we find a geometric phase of $\sim\SI{1}{\milli\hertz}$. Like the gravitational effect, this shift is in $\fBR$ and so has no systematic effect on our measurement. We took some auxiliary blocks with the vibrational cleanup light turned off during the free evolution time and observed no shifts, empirically constraining any shift in $\fBR$ to less than \SI{9}{\milli\hertz}.

\subsubsection{Phase modulation due to axial secular motion}  
\label{subsec:systematic_zslosh}

Ions displaced from the center of the trap experience a non-zero axial confining field. Any axial motion due to the axial confining potential is harmonic, so the axial component of the electric field will average out to zero over many cycles. \textit{Instantaneously}, however, the ions will accrue a Berry's phase,
\begin{equation}
\begin{split}
    \delta(t_R) ={}& \int_0^{t_R} e^{-t/\tau_z}3 \wrot \frac{z_{\rm sec}}{\rrot}\cos(\omega_z t+\phi_z)\mathrm{d}t,\\
    \simeq{}&\delta_0(-\sin\phi_z+e^{-t_R/\tau_z}\sin(\omega_z t_R+\phi_z)).
\end{split}
\end{equation}
Here $z_{\rm sec}$, $\omega_z$ and $\phi_z$ are the amplitude, angular frequency and phase of the center-of-mass axial secular motion, $\tau_z$ the coherence time of that motion, and $\delta_0=3\wrot z_{\rm sec}/(\omega_z\rrot)$. The second line holds only for $\tau_z\omega_z\gg1$; we find $\tau_z=\SI{90+-10}{\milli\second}$, while $\omega_z\sim2\pi\times\SI{1.6}{\kilo\hertz}$ and so this is condition is well satisfied. We are sensitive to the difference in this phase evaluated at early and late $t_R$ and so the first term cancels. The late $t_R$ used are long compared to $\tau_z$ and so the maximum size of the difference is $\delta_0$. 

\begin{figure}
    \centering
        \includegraphics[width=\columnwidth]{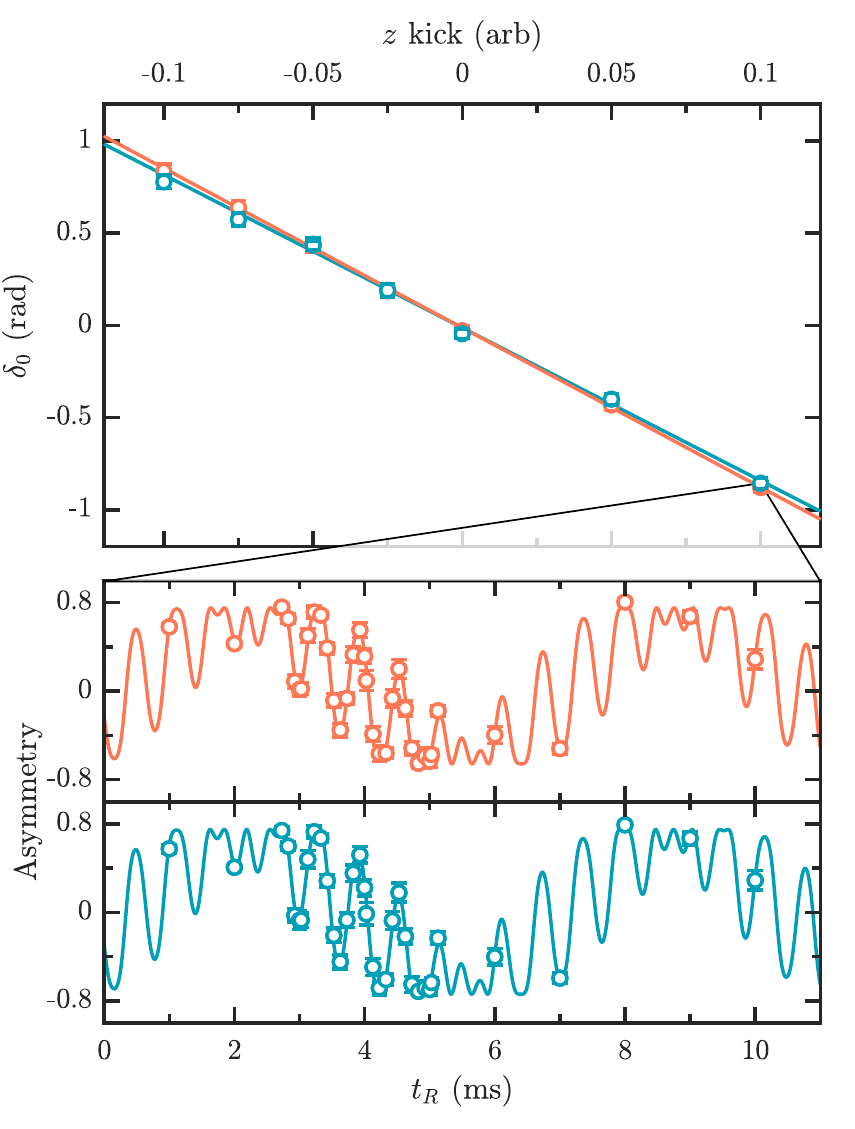}
    \caption{Phase modulation amplitude $\delta$, in each doublet for various sizes of $z$ kick. Upper doublet data is in pink and lower doublet data is in blue. Both fits have intercepts of \SI{-0.013+-0.011}{\radian}. Inset shows example data for one kick value.}
    \label{fig:phase_modulation_linear_fit}
\end{figure}

We investigated this effect by pulsing the endcap electrode to `kick' the ion cloud in the axial direction right before the first $\pi/2$ pulse---causing temporary coherent axial oscillations---then collecting a short-time Ramsey fringe. Example data is shown in the inset of Fig.~\ref{fig:phase_modulation_linear_fit}. We repeated this procedure over a range of kick amplitudes, in each instance fitting the data to the model
\begin{equation}
   A\cos(2\pi ft + \phi + \delta_0 \cos(\omega_z t + \phi_z)).
\end{equation}
Figure~\ref{fig:phase_modulation_linear_fit} shows $\delta_0$, plotted against the amplitude of the applied kick. The data constrain $\delta_0$ for no applied kick to $<0.024$ radians at the $1\sigma$ confidence level, corresponding to $z_{\rm sec}\lesssim\SI{1}{\milli\meter}$. The worst-case frequency shift is then $\delta_0/(2\pi t_R)<\SI{1.6}{\milli\hertz}$ for the effective average late $t_R$ ($\sim 2.4$ seconds) used. This shift is likely to be in $\fBR$ but could leak into $\fB$ if the phase of the axial oscillation were rotation odd. We did not monitor this during our dataset and so include a systematic uncertainty of $\SI{3.4}{\micro\hertz}$ in our uncertainty budget.


\subsubsection{Axial 2nd harmonic + radial ellipticity} 
\label{sec:axial2f_plus_radial_ellipticity}

In addition to static tilts of the quantization axis and aliasing of time-varying tilts, perturbations to the radial electric field can cause variations in the magnitude of the rotating field and its azimuthal angular velocity which can couple to axial displacements to give geometric phase effects which do not time-average to zero. An important example of this type of effect is an ellipticity of $\Erot$ combined with a second-harmonic field in the axial direction. Such an axial field could result either from leakage of the signals that drive the radial electrodes onto the endcaps, or from axial displacement of the ions from the geometric center of the trap combined with fields generated by the radial electrodes themselves. 

\begin{figure}
    \centering
        \includegraphics[width=\columnwidth]{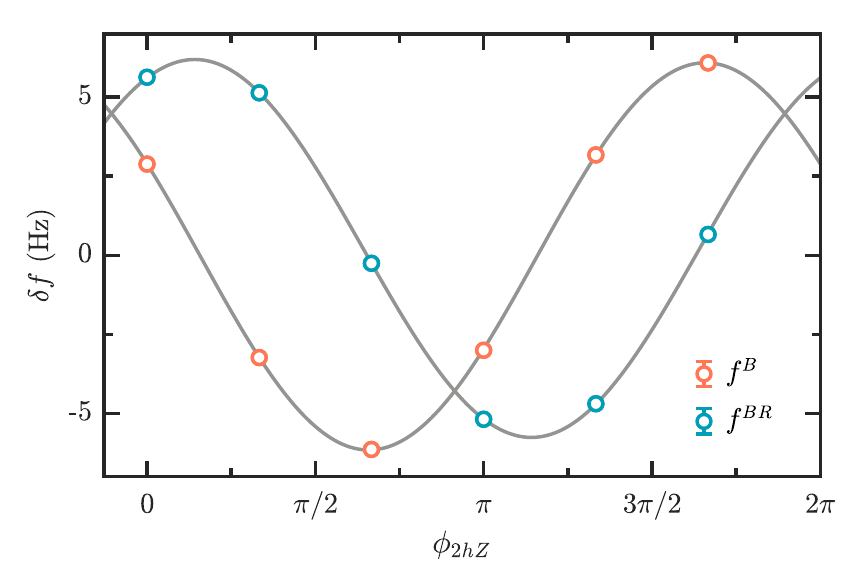}
    \caption{Example data collected with $\mathcal{E}_{2hZ}\sim\SI{4}{\volt\per\meter}$ and an ellipticity $\mathcal{E}_\epsilon/\Erot\sim \num{7e-3}$. As we vary the phase of the axial modulation, $\fB$ and $\fBR$ vary sinusoidally with a $\pi/2$ phase offset, as expected.}
    \label{fig:endcap2f_ellipticity_example_data}
\end{figure}

We can express a radial ellipticity in $\Erot$ as we did before (Eq.~\ref{eq:ellipticity-no-rot}) but now explicitly including rotational dependence,
\begin{equation}
\begin{aligned}
    \vec{\mathcal{E}} ={}& \Erot \begin{pmatrix}
        \cos(\wrot t+\phi_R+\tilR\phi_0)\\
        \tilR\sin(\wrot t+\phi_R+\tilR\phi_0)\\
        0
    \end{pmatrix}\\
    &+\mathcal{E}_\epsilon\begin{pmatrix}
        \cos(2\tilR\theta - \wrot t-\phi_R-\tilR\phi_0)\\
        \tilR\sin(2\tilR\theta - \wrot t-\phi_R-\tilR\phi_0)\\
        0
    \end{pmatrix}.
\end{aligned}
\end{equation}
Here $\phi_0+\tilR\phi_R$ is the angle of the radial electric field to the $X$-axis at $t=0$ and $\theta$ is the angle between the major axis of the ellipse and the $X$-axis. An axial field oscillating at the second harmonic is given by
\begin{equation}
\begin{aligned}
    \vec{\mathcal{E}}_{2hZ} =&  \mathcal{E}_{2hZ}\begin{pmatrix}
        0\\
        0\\
        \cos(2\wrot t+\phi_{2hZ})\\
    \end{pmatrix}.
\end{aligned}
\end{equation}
With these electric fields we find the tilt angle and azimuthal angular velocity are
\begin{align}
    \begin{split}
    \alpha(t) &\simeq  \frac{\mathcal{E}_{2hZ}}{\Erot}\cos(2\wrot t+2\phi_{2hZ})\\
    &\times\left(1-\frac{\mathcal{E}_\epsilon}{\Erot}\cos(2\wrot t+2\phi_R + 2\tilR\phi_0-2\tilR\theta)\right),
    \end{split}\\
    \dot{\phi}(t) &\simeq \tilR\wrot-\frac{2\tilR\mathcal{E}_\epsilon\wrot}{\Erot}\cos(2\wrot t+2\phi_R + 2\tilR\phi_0-2\tilR\theta),
\end{align}
where in each case we have expanded to first order in $\mathcal{E}_\epsilon/\Erot$. Using Eq.~\ref{eq:berrys-phase}, we find
\begin{equation}
    \delta f \simeq \frac{9 \tilR\mathcal{E}_{2hZ}\mathcal{E}_\epsilon\frot}{2\Erot^2}\cos(\phi_{2hZ}+2\tilR\theta-2\tilR\phi_0-2\phi_R).
\end{equation}
Finally we can calculate the expected shifts in $\fB$ and $\fBR$ by taking the sum and difference of the shifts in the two rotation directions,
\begin{align}
    \delta \fB &= -\frac{9 \mathcal{E}_{2hZ}\mathcal{E}_\epsilon\frot}{2\Erot^2}\sin(2\theta-2\phi_0)\sin(\phi_{2hZ}-2\phi_R),\\
    \delta \fBR &= \frac{9 \mathcal{E}_{2hZ}\mathcal{E}_\epsilon\frot}{2\Erot^2}\cos(2\theta-2\phi_0)\cos(\phi_{2hZ}-2\phi_R).
\end{align}
We see that we expect frequency shifts which vary sinusoidally as a function of either $\theta$ or $\phi_{2hZ}$, with a $\pi/2$ phase shift between $\fB$ and $\fBR$. Figure~\ref{fig:endcap2f_ellipticity_example_data} shows example data for deliberately applied ellipticity $\mathcal{E}_\epsilon/\Erot\sim\num{7e-3}$ and an axial second-harmonic $\mathcal{E}_{2hZ}\sim\SI{4}{\volt\per\meter}$.

We used data like this to manually shim out any residual ellipticity by applying $\mathcal{E}_{2hZ}$, and adjusting the $\Erot$ multiplying DAC on each electrode until we saw the effect was minimized. We then measured the residual ellipticity in the same way and found $\mathcal{E}_\epsilon/\Erot = \num{3e-4}$, suppressed by a factor of 28 relative to our applied ellipticity. We constrain the axial second harmonic field by deliberately applying an ellipticity and varying $\theta$ to constrain $|\mathcal{E}_{2hZ}|<\SI{10}{\milli\volt\per\meter}$ at the $1\sigma$ confidence level, suppressed by a factor 400 relative to our deliberately applied second harmonic. Combining these suppression factors allows us to place a limit on the maximum size of any shift in $\fB$, $\delta\fB<\SI{0.8}{\milli\hertz}$. We include a corresponding contribution to our systematic uncertainty of $\SI{1.7}{\micro\hertz}$.

\subsubsection{Effects of higher harmonics}
\label{sec:berrys-phase-higher-harmonics}

We can generalize the discussion of the previous section to a pair of perturbations $\vec{\delta\mathcal{E}}_{X,Y}$ and $\vec{\delta\mathcal{E}}_Z$ which act in radial and axial direction respectively, and which are both phase-locked to $\Erot$. We can write general expressions for these perturbations,
\begin{align}
\begin{split}
    \vec{\delta\mathcal{E}}_{X,Y}={}&\mathcal{E}_{nhX}\cos(n\wrot t + \phi_{nhX})\hat{x}\\
    {}&+ \mathcal{E}_{nhY}\cos(n\wrot t + \phi_{nhY})\hat{y},
\end{split}\\
    \vec{\delta\mathcal{E}}_Z={}&\mathcal{E}_{mhZ}\cos(m\wrot t+\phi_{mhZ})\hat{z}
\end{align}
where $m,n$ are both integer. The combination of these perturbations will result in a non-zero time-averaged frequency shift at first order provided $m$ and $n$ differ by $\pm1$. The shift is given by
\begin{equation}
\begin{split}
    \delta f={}&\mp \frac{3 \mathcal{E}_{mhZ} \frot (n\pm2)}{4\Erot^2}\big(\mathcal{E}_{nhX}\cos(\phi_{nhX}-\phi_{mhZ})\\
    {}&\quad\pm\mathcal{E}_{nhY}\sin(\phi_{nhY}-\phi_{mhZ})\big),\label{eq:higher-harmonic-berrys-phase}
\end{split}
\end{equation}
where the $\pm$ correspond to $m=n\pm1$. Depending on the $\tilR$ dependence of the various phases, this shift could appear in either $\fB$ or $\fBR$. We note that the frequency shift corresponding to the first harmonic on the endcap and second harmonic radially, $m=1$, $n=2$, is zero, and so the next largest shifts are expected to come from effects involving the third harmonic or higher.

To place constraints on the possible size of these effects, we measured the Fourier transform of the signal on each of the radial electrodes and the two endcaps%
\footnote{To characterize higher harmonics with power as low as -80\,dB relative to the fundamental, we suppressed the carrier with a home-built notch filter with well-characterized linearity.}
The measured amplitudes of each harmonic are shown in the second column of Table~\ref{tab:higher-harmonic-fields}. We first consider the possible radial electric fields caused by these signals. The radial electric fields at the center of the trap generated by an $n\wrot$ voltage signal on each electrode are
\begin{align}
    \delta\mathcal{E}_{Xnh}={}& -\sum_{k=1}^8 \frac{V_{nhk}}{R_{X,Y}} \cos((9-2k)\frac{\pi}{8})\cos(n\wrot t + \phi_{nhk}), \label{eq:higher-harmonic-field-x}\\
    \delta\mathcal{E}_{Ynh}={}& \sum_{k=1}^8 \frac{V_{nhk}}{R_{X,Y}} \sin((9-2k)\frac{\pi}{8})\cos(n\wrot t + \phi_{nhk}),\label{eq:higher-harmonic-field-y}
\end{align}
where $V_{nhk}$ and $\phi_{nhk}$ are amplitude and phase of the voltage on each of the 8 electrodes and $R_{X,Y}\simeq\SI{23}{\centi\meter}$ is a constant relating to the geometry of the trap. The amplitude of these fields is given by
\begin{align}
\begin{split}
    |\delta\mathcal{E}_{Xnh}|={}&\Bigg[\left(\sum_{k=1}^8\frac{V_{nhk}}{R_{X,Y}} \cos((9-2k)\frac{\pi}{8})\cos(\phi_{nhk})\right)^2\\
    {}&+\left(\sum_{k=1}^8\frac{V_{nhk}}{R_{X,Y}} \cos((9-2k)\frac{\pi}{8})\sin(\phi_{nhk})\right)^2\Bigg]^{1/2},\label{eq:higher-harmonic-rad-amp-x}
\end{split}\\
\begin{split}
    |\delta\mathcal{E}_{Ynh}|={}&\Bigg[\left(\sum_{k=1}^8\frac{V_{nhk}}{R_{X,Y}} \sin((9-2k)\frac{\pi}{8})\cos(\phi_{nhk})\right)^2\\
    {}&+\left(\sum_{k=1}^8\frac{V_{nhk}}{R_{X,Y}} \sin((9-2k)\frac{\pi}{8})\sin(\phi_{nhk})\right)^2\Bigg]^{1/2}.\label{eq:higher-harmonic-rad-amp-y}
\end{split}
\end{align}
The amplifiers driving each of the electrodes are nominally identical and so the harmonic distortion is approximately equal on each electrode. As a first-order approximation, we take the amplitude of the signal on each electrode to be equal, $V_{nhk}=V_{nh}$, and the phases to be locked to the fundamental with a small offset that can be different on each electrode, $\phi_{nhk}=\phi_{nh0}+n(9-2k)\tfrac{\pi}{8}+\delta\phi_{nhk}$. 

We first consider $\delta\phi_{nhk}=0$. In this case all the radial fields from all electrodes cancel one another unless $n$ differs from a multiple of 8 by $\pm1$. When $n$ is one greater than a multiple of 8, the relative phases on neighboring electrodes are the same as for the fundamental and so the harmonic field corotates with $\Erot$. For $n$ one less than a multiple of 8, the relative phases on neighboring electrodes are flipped compared to the fundamental and the resulting field counterrotates. In both cases the amplitude is $4V_{nh}/R_{X,Y}$ and we use this as an estimate of the field size for these values of $n$ in the third column of Table~\ref{tab:higher-harmonic-fields}. For the other harmonics, we can estimate their size by looking at the variance of their amplitude due to the variance of the amplifier phases,
\begin{align}
    \langle|\delta\mathcal{E}_{Xnh}|^2\rangle ={}& \sum_{k=1}^8 \left(\frac{\partial|\delta\mathcal{E}_{Xnh}|}{\partial\delta\phi_{nhk}}\right)^2\langle\delta\phi_{nhk}^2\rangle,\\
    \langle|\delta\mathcal{E}_{Ynh}|^2\rangle ={}& \sum_{k=1}^8 \left(\frac{\partial|\delta\mathcal{E}_{Ynh}|}{\partial\delta\phi_{nhk}}\right)^2\langle\delta\phi_{nhk}^2\rangle,
\end{align}
where the derivatives in brackets are evaluated at $\delta\phi_{nhk}=0$. Since in these cases the phase relationship between the fields is random, we will be interested in their quadrature sum. After some lengthy algebra, we find
\begin{equation}
\begin{split}
    \langle|\delta\mathcal{E}_{X,Ynh}|^2\rangle^{1/2}={}&\left(\langle|\delta\mathcal{E}_{Xnh}|^2\rangle+\langle|\delta\mathcal{E}_{Ynh}|^2\rangle\right)^{1/2} \\
    ={}& \frac{ V_{nh}}{R_{X,Y}}\Bigg[4+(-1)^n\sqrt{2}\cos\left(\frac{n\pi}{4}\right)\\
    &-(-1)^n\sqrt{2}\cos\left(\frac{3 n\pi}{4}\right)\Bigg]^{1/2}\langle\delta\phi_{nh}^2\rangle^{1/2},
\end{split}
\end{equation}
where we have set the variances of the phases on all amplifiers to be equal, $\langle\delta\phi_{nhk}^2\rangle=\langle\delta\phi_{nh}^2\rangle$. From measurements of the signals on the electrodes, we conservatively estimate $\langle\delta\phi_{nh}^2\rangle=(\pi/8)^2$. The third column of Table~\ref{tab:higher-harmonic-fields} shows the estimated size of the radial electric fields from higher harmonics calculated using this expression. Note that, because we shim the 2nd harmonic, we can make no such claims about relative phases on electrodes and so we assume the worst case of all electrodes on one side of the trap in phase and exactly out of phase with the electrodes on the other side, giving $\delta\mathcal{E}_{X,Y2h}=4\sqrt{1+\tfrac{1}{\sqrt{2}}}V_{2h}/R_{X,Y}$.

\begin{table}
	\caption{Constraints on electric fields from higher-harmonic voltages on radial electrodes and endcaps.}
	\begin{ruledtabular}
	\begin{tabular}{@{}c r r r r@{}}
	     & \multicolumn{2}{c}{Amplitudes (\si{\milli\volt})} & \multicolumn{2}{c}{Fields (\si{\milli\volt\per\meter})}\\
	      \cmidrule(){2-3} \cmidrule(){4-5}
	    n & Radial & Axial & $|\delta\mathcal{E}_{X,Ynh}|$& $|\delta\mathcal{E}_{Znh}|$\\
	    \midrule
 2 & 11 & 0.5 & 250 & 1.0 \\
 3 & 310 & 0.3 & 1296 & 2.1 \\
 4 & 190 & 0.1 & 649 & 1.8 \\
 5 & 410 & 0.2 & 1715 & 2.7 \\
 6 & 100 & 0.4 & 341 & 0.8 \\
 7 & 230 & 0.5 & 4000 & 1.7 \\
 8 & 45 & 0.7 & 154 & 3.1 \\
 9 & 170 & 1.0 & 2957 & 1.8 \\
 10 & 37 & 1.4 & 126 & 2.0 \\
 11 & 140 & 1.8 & 586 & 2.7 \\
 12 & 42 & 2.1 & 143 & 3.1 \\
	\end{tabular}
	\label{tab:higher-harmonic-fields}
	\end{ruledtabular}
\end{table}

We can take a similar approach to estimate the possible higher harmonic fields in the $Z$ direction. The most direct way of generating axial electric fields with angular frequency $n\wrot$ is for them to appear directly on the endcaps of our trap where they generate a field $V_{nhZ}/R_Z^{ax}$, with $R_Z^{ax}\simeq\SI{137}{\centi\meter}$. Our measurements of possible higher-harmonic signals onto the endcaps are shown in the third column of Table~\ref{tab:higher-harmonic-fields}; for $n>3$ these are upper limits, constrained by the frequency-dependent noise floor of our FFT measurements. We find that in most cases the fields generated by these voltages are dominated by another contribution, signals on the radial electrodes combined with axial displacement of the ions from the center of the trap. We can write the axial electric field due to $n\wrot$ signals on the radial electrodes at axial displacement $z_0$ as
\begin{equation}
    \delta\mathcal{E}_{Znh} = \sum_{k=1}^8 \frac{V_{nhk} z_0}{R_Z^2}\cos(n\wrot t+\phi_{nhk}),
\end{equation}
with amplitude
\begin{equation}
\begin{split}
    |\delta\mathcal{E}_{Znh}| ={}& \frac{z_0}{R_Z^2}\Bigg(\left(\sum_{k=1}^8 V_{nhk} \cos(\phi_{nhk})\right)^2\\
    {}&+\left(\sum_{k=1}^8 V_{nhk} \sin(\phi_{nhk})\right)^2\Bigg)^{1/2}.
\end{split}
\end{equation}
The constant $R_Z\simeq\SI{27}{\centi\meter}$ is fixed by the trap geometry. We measure the axial displacement by deliberately applying an ellipticity plus a large second harmonic in phase to all radial electrodes and find $z_0=\SI{0.6}{\milli\meter}$. 

Again, we make the substitutions $V_{nhk}=V_{nh}$ and $\phi_{nhk}=\phi_{nh0}+n(9-2k)\tfrac{\pi}{8} + \delta\phi_{nhk}$ and start by assuming $\delta\phi_{nhk}=0$,
\begin{equation}
    \delta\mathcal{E}_{Znh} = \frac{ V_{nh}z_0}{R_Z^2}\frac{\sin(n\pi)}{\sin\left(\frac{n\pi}{8}\right)}\cos(n\wrot t+\phi_{nh0}).
\end{equation}
The contributions of the 8 electrodes cancel out except for when $n$ is a multiple of 8 when they are all in phase with one another and we have $|\delta\mathcal{E}_{Znh}|=8V_{nh}z_0/R_Z^2$. For these values of $n$, we use this as an estimate of the axial field amplitude from the radial electrodes. For the other possible harmonics we can estimate their amplitudes in the same way we did for the radial fields. The variance of the amplitude of the field is given by
\begin{equation}
    \langle|\delta\mathcal{E}_{Znh}|^2\rangle = \sum_{k=1}^8 \left(\frac{\partial|\delta\mathcal{E}_{Znh}|}{\partial\delta\phi_{nhk}}\right)^2\langle\delta\phi_{nhk}^2\rangle,
\end{equation}
where again, the derivative in brackets is evaluated at $\delta\phi_{nhk}=0$. After some more lengthy algebra, we find
\begin{equation}
    \langle|\delta\mathcal{E}_{Znh}|^2\rangle^{1/2} = \frac{V_{nh}z_0}{R_Z^2}\left(4-\frac{\sin(n\pi)}{\sin\left(\frac{n\pi}{4}\right)}\right)^{1/2}\langle\delta\phi_{nh}^2\rangle^{1/2}.
\end{equation}
We use this expression with $\langle\delta\phi_{nh}^2\rangle=(\pi/8)^2$ to calculate the field estimates for all $n$ which are not a multiple of 8. Note that, again due to our shimming of the second harmonic, we can't make any claims about relative phases on the electrodes and so we assume the worst case of all in phase, giving $|\delta\mathcal{E}_{Z2h}|=8V_{2h}z_0/R_Z^2$. The fourth column of Table~\ref{tab:higher-harmonic-fields} shows the sum in quadrature of our estimates for the axial field from the radial electrodes with those from the endcaps. 

In Table~\ref{tab:higher-harmonic-fields-berrys-phases}, we combine these estimates of the radial and axial field amplitudes with Eq.~\ref{eq:higher-harmonic-berrys-phase} to set limits on the Berry's phase from each possible combination of higher-harmonics. Summing all the contributions in quadrature we get an uncertainty on the Berry's phase of \SI{1.4}{\milli\hertz}, corresponding to a systematic uncertainty in $\fBD$ of \SI{3.0}{\micro\hertz} which we include in our systematic uncertainty budget.

\begin{table}
	\caption{Constraints on Berry's phase frequency contributions to $\fB$ or $\fBR$ due to possible combinations of radial and axial field imperfections from harmonic distortion in amplifiers used to drive radial electrodes. All entries are in \si{\micro\hertz}.}
	\begin{ruledtabular}
	\begin{tabularx}{\textwidth}{c *{11}{c}}
	 & \multicolumn{11}{c}{$n_Z$}\\
	\cmidrule(){2-12}
$n_{X,Y}$ & 2 & 3 & 4 & 5 & 6 & 7 & 8 & 9 & 10 & 11 & 12 \\
\midrule
 2 & 0 & 17 & 0 & 0 & 0 & 0 & 0 & 0 & 0 & 0 & 0 \\
 3 & 11 & 0 & 95 & 0 & 0 & 0 & 0 & 0 & 0 & 0 & 0 \\
 4 & 0 & 22 & 0 & 87 & 0 & 0 & 0 & 0 & 0 & 0 & 0 \\
 5 & 0 & 0 & 75 & 0 & 83 & 0 & 0 & 0 & 0 & 0 & 0 \\
 6 & 0 & 0 & 0 & 31 & 0 & 38 & 0 & 0 & 0 & 0 & 0 \\
 7 & 0 & 0 & 0 & 0 & 139 & 0 & 947 & 0 & 0 & 0 & 0 \\
 8 & 0 & 0 & 0 & 0 & 0 & 13 & 0 & 24 & 0 & 0 & 0 \\
 9 & 0 & 0 & 0 & 0 & 0 & 0 & 545 & 0 & 556 & 0 & 0 \\
 10 & 0 & 0 & 0 & 0 & 0 & 0 & 0 & 16 & 0 & 35 & 0 \\
 11 & 0 & 0 & 0 & 0 & 0 & 0 & 0 & 0 & 90 & 0 & 200 \\
 12 & 0 & 0 & 0 & 0 & 0 & 0 & 0 & 0 & 0 & 33 & 0 \\
	\end{tabularx}
	\label{tab:higher-harmonic-fields-berrys-phases}
	\end{ruledtabular}
\end{table}

\subsection{Residual rotation-induced doublet mixing} 
\label{subsec:rotation-induced-mixing-effects}

The systematic effects we have considered so far are all concerned with the second term in Eq.~\ref{eq:freq-expansion}, various sources of $\delta f_0^s$. However, the third and fourth terms in the expression, containing the off-diagonal components of the effective Hamiltonian, can also potentially cause systematic shifts. The third term concerns shifts directly generated by the off-diagonal mixing components, while the fourth term concerns the leaking of frequency shifts into other channels due to this mixing. In this section we discuss possible contributions to each and place constraints on their size during the measurement.


\subsubsection{Leaking of $\fBR$}
\label{sec:fBR_leak}

The largest leak of any frequency channel into $\fBD$ is expected to come from the frequency channel with the largest signal. This is $\fBR$ where a signal of $\sim\SI{200}{\milli\hertz}$ is caused by the charging currents in our electrodes. This could potentially leak into $\fBD$ when combined with a non-zero values of $\delta\Delta^{DR}$ or $\delta\Delta^{R}$. From Eq.~\ref{eq:freq-expansion} we have
\begin{equation}
    \delta \fBD = -\fBR\frac{2\Delta^0\delta\Delta^{DR} + 2\Delta^D\delta\Delta^R}{|f_0^0|^2}.\label{eq:fBR-leak}
\end{equation}
Non-zero values of $\delta\Delta^{DR}$ and $\delta\Delta^{R}$ can come from an axial magnetic field through a 4th-order coupling similar to $\Delta$ and $\Delta^D$ but where one of the rotation matrix elements is replaced with a magnetic-field matrix element. Any effect would also result in a contribution to $\fDR$,
\begin{equation}
    \delta f^{DR} = \frac{2\Delta^0\delta\Delta^{DR} + 2\Delta^D\delta\Delta^R}{|f_0^0|}.
\end{equation}
So we can rewrite Eq.~\ref{eq:fBR-leak} as 
\begin{equation}
    \delta \fBD \simeq -\frac{\fBR}{\fn}\left(\fDR-\frac{\dgF}{g_F}\fR\right).
\end{equation}
Our data places a limit on this contribution to $\fBD$ of $<\SI{170}{\nano\hertz}$ which we include in our systematic uncertainty budget. 

On similar grounds, we could also expect non-zero $\delta\Delta^{DR}$ or $\delta\Delta^{R}$ to cause $\fBR$ to leak into $\fB$, 
\begin{equation}
    \delta \fB = -\fBR\frac{2\Delta^0\delta\Delta^{R} + 2\Delta^D\delta\Delta^{DR}}{|f_0^0|^2}\label{eq:fBR-leak-fB},
\end{equation}
causing a systematic via the correction described in Sec.~\ref{sec:baxgradnr}. Since this expression involves the same terms as Eq.~\ref{eq:fBR-leak}, barring unlikely cancellations, the contribution can be expected to be of similar size. The corresponding systematic in $\fBD$ is suppressed by a further factor of $\dgF/g_F$ and so we include no contribution in our systematic uncertainty budget.

\subsubsection{Axial magnetic fields}
\label{sec:axial_magnetic_fields}

Analogously to how a static axial magnetic field can generate $\delta\Delta^{DR}$ and $\delta\Delta^{R}$, a $\tilB$-odd axial magnetic field $\mathcal{B}^B_Z$ can cause $\delta\Delta^{BDR}$ and $\delta\Delta^{BR}$. Such a field could be generated from the $\vec{\mathcal{B}^0}$ coils if the ions are situated slightly away from their geometric center. The combination of these two effects can cause a shift in $\fBD$ via the third term in Eq.~\ref{eq:freq-expansion},
\begin{equation}
    \delta \fBD = \frac{2\delta\Delta^{DR}\delta\Delta^{BR}+2\delta\Delta^{R}\delta\Delta^{BDR}}{|f^0_0|^2}.\label{eq:axial-B-fields-fDB}
\end{equation}
An order-of-magnitude estimate for these $\tilR$-odd mixing elements can be obtained by taking the measured values of $\Delta$ and $\Delta^D$ and multiplying them by the ratio of the magnetic-field matrix element to the rotational matrix element, $g_F \mu_{\rm B} \mathcal{B}_Z/\hbar\wrot\sim\SI{0.01}{\per\gauss}$. Since $\Delta$ and $\Delta^D$ are $\sim \SI{1}{\hertz}$, we can expect $\delta\Delta^{DR},\delta\Delta^{BR},\delta\Delta^{R},\delta\Delta^{BDR}$ of $\sim\SI{10}{\milli\hertz\per\gauss}$. We took data with large axial magnetic fields $\sim\SI{10}{\gauss}$ and saw shifts in $\fDR$ of \SI{170+-30}{\micro\hertz\per\gauss} for $\fn=\SI{151}{\hertz}$, confirming the order of magnitude of this approximation. Applying these approximations to Eq.~\ref{eq:axial-B-fields-fDB}, we can then expect a systematic shift in $\fBD$ of $\sim \SI{2}{\micro\hertz\per\gauss\squared}\times\mathcal{B}^B_Z\mathcal{B}_Z$. Looking at $\fDR-\tfrac{\dgF}{g_F}\fR$ and $\fBDR-\tfrac{\dgF}{g_F}\fBR$ for our dataset puts limits on the fields of $\mathcal{B}_Z<\SI{0.5}{\gauss}$ and $\mathcal{B}^B_Z<\SI{0.4}{\gauss}$. We note that the same magnetometers used to shim out $\mathcal{B}_Z$ are able to shim $\mathcal{B}_X$ and $\mathcal{B}_Y$ to $\sim\SI{10}{\milli\gauss}$ but since we have not confirmed this for $\mathcal{B}_Z$, we use this more conservative limit. Finally we constrain any systematic shift in $\fBD$ to $<\SI{400}{\nano\hertz}$.
\subsection{Oddities \& miscellany}

\subsubsection{Ion-ion interactions}
\label{ion-ion interactions}

The dominant ion-ion interaction is from the monopole charge on each ion. If we qualitatively model the cloud as 20,000 singly charged ions uniformly distributed across a sphere 1 cm in radius,  we find that the resulting mean-field electric field is much smaller than either $\vecErot$ or the peak $\vec{\mathcal{E}}_{\rm RF}$, but comparable to the time-averaged value of $\vec{\mathcal{E}}_{\rm RF}$. That is to say, the mean-field self-repulsion of the cloud is only modestly smaller than the effective electric fields providing secular confinement.  We see evidence for this, for instance in the frequencies of the breathing modes of the cloud.  The ions therefore experience a net confinement that is somewhat anharmonic. None of the arguments for the limits of the size of systematic errors on $d_e$ hinge on the confinement being particularly harmonic.  Each trapped ion necessarily experiences a total time-averaged axial electric field (whether external or from other ions)  extremely close to zero, and thus to a high precision the rotation of $\vecErot$ causes no Berry’s phase frequency shift.  It is the case, however, that the mean-field repulsion causes the ion cloud to be larger than it otherwise would be at a given temperature. In the presence of various field inhomogeneities, changes in the radius of the cloud can change both the decoherence rate and the average fringe frequency $\fn$. We see shifts in the mean Ramsey frequency $\fn$ that correlate with the number of ions in the trap. These do not appear in the $\tilD$-odd frequency channels. The mean-field coulombic repulsion does not break the various symmetries that keep frequency shifts out of the eEDM channel.

On a microscopic level, the coulomb potential between two nearest-neighbor atoms is typically $10^{-3} kT$, and thus ions are far from the ion-crystal regime \cite{Hansen1973}.  When small-impact-parameter ion-ion collisions occur, the ion-ion electric field can briefly spike to a magnitude which is not infinitesimal compared to $\Erot$ \cite{Leanhardt2011}.  This can cause the tip of the $\vecErot$ to briefly wobble in a way that encloses solid angle, and there can be a resulting random Berry’s phase shift that degrades the coherence between the $m_F =3/2$ and $m_F=-3/2$ states, but does not bias the central value of the Ramsey fringe frequency.  Adiabatic relaxation of the confining potential of the trap dramatically reduces this source of decoherence, at the cost of increasing the decoherence from spatial inhomoegeneity mentioned in this section.  We empirically reoptimize the compromise value of the ramped-down confinement several times during the course of a long data collection run.

Fields from the molecule-frame electric dipole moment are down from the monopole-generated fields by a factor of $10^5$ or more.  

As for magnetic interactions, each ion carries a magnetic dipole moment $(3/2) g_F \mu_{\rm B}$.  Approximating the distribution of 20,000 ions just as described above, the field arising from a uniform  magnetization within the ball of ions could cause frequency shifts of order \si{\pico\hertz}, and is thus neglected.

\subsubsection{Effect of ion-cloud spatial distributions in which the Stark doublets are not perfectly overlapping}
\label{sec:doublet_odd_displacement}

Much of the ultimate accuracy and precision of our measured $\fBD \sim d_e \Eeff$ comes from the fact that we are measuring a resonance in two samples of ions which overlap perfectly in space and in time, but for which $\Eeff$ points in opposite directions. If for some reason the two samples do not perfectly overlap, then spatial variation in $\tilB$-odd frequencies can cause systematic shifts in our eEDM channel. Generically, time-averaged $\tilD$-odd displacements $\langle r_i\rangle^D$ or sizes of the ion cloud $\langle {r_i}^2\rangle^D$ can couple to first- and second-order gradients in $\fB$ to give systematic shifts,
\begin{equation}
    \delta \fBD = \sum_{i=x,y,z}\langle r_i\rangle^D\frac{\partial{\fB}}{\partial r_i}+\langle {r_i}^2\rangle^D\frac{\partial^2{\fB}}{\partial {r_i}^2}.
\end{equation}
By moving the ions around in the trap, we measure typical gradients $\frac{\partial{\fB}}{\partial r_i}\sim \SI{40}{\milli\hertz\per\centi\meter}$ and $\frac{\partial^2{\fB}}{\partial r_i^2}\sim \SI{10}{\milli\hertz\per\centi\meter\squared}$ and so we are potentially interested in $\langle r_i\rangle^D\sim\SI{1e-4}{\centi\meter}$ and $\langle r_i^2\rangle^D\sim\SI{5e-4}{\centi\meter\squared}$. 

We have good a priori reason to believe that the spatial distributions of ions in the two doublets are identical to very high precision. The two Stark doublets are initially populated by optical pumping with \ltrans{} from the $^1\Sigma$ ground state via a $^3\Pi_0$ state, both with $\Omega=0$. To an excellent approximation, both these states have well-defined parity in the modest electric fields used in our experiment and so the pumping process is completely independent of doublet. Any $\tilD$-odd spatial distributions must be imprinted subsequently, either by the lasers used to prepare or readout the states of the molecules, or by $\tilD$-odd forces on the ions.

We first discuss possible effects of the lasers used for state preparation. The next laser to interact with the ions is \lop{} which polarizes the ions by optically pumping them into stretched $m_F$ states. Although this process again proceeds via an excited state with $\Omega=0$, and so each photon scatter is equally likely to populate either doublet, in this case the laser excites out of the $^3\Delta_1$ state, and so can potentially cause a $\tilD$-odd population difference. However, we operate in the strongly saturated regime where all ions have sufficient time to interact with the laser and so any effects of spatial intensity variation are strongly suppressed. Due to this suppression, we expect any effects to be smaller than those discussed in the next paragraph.

The other CW laser addressing molecules in the science state is \ldepl{} used to remove population in unwanted $m_F$ states. \ldepl{} is on for \SI{7}{\milli\second} before the Ramsey sequence, to clean up any population not successfully optically pumped, and for \SI{25}{\milli\second} after the Ramsey sequence to remove population in one stretched state. Using measurements of ion number as a function of \ldepl{} interaction time, we estimate the time taken to remove all molecules in unwanted $m_F$ levels is about \SI{1}{\milli\second} and so both interaction periods are strongly saturated, again greatly suppressing any effects of spatial variation. However for this laser, there is a second relevant time scale; because the electric field is not perfectly aligned with the $k$-vector of the light, molecules in the desired stretched $m_F$ state can also occasionally scatter photons, removing a fraction $\beta_{\rm dep}$ from the state we detect. This effect is largest for the time after the Ramsey sequence, where interaction time is longest and we conservatively estimate that $\beta_{\rm dep}<0.1$. Although this step takes place after the Ramsey sequence, if the ions removed have some $\tilD$-odd spatial dependence, that will result in similar dependence being imprinted onto the remaining ions which we detect. 
The rotating $\vecErot$ causes the ions to move in small circles at speeds of $\sim\SI{1000}{\meter\per\second}$. This motion in the $X$-$Y$ plane causes a sinusoidally varying Doppler shift oscillating at $\wrot$ with amplitude $\Delta_{\rm Dopp}\sim\SI{1000}{\mega\hertz}$. This Doppler shift, combined with the Stark splitting of $\Delta_{\rm St}\sim(\dmf\Erot-\Apar/2)/h\sim\SI{140}{\mega\hertz}$ between the two doublets means that the laser comes into resonance with each at a slightly different time on each rotation. We have identified two possible mechanisms related to this effect which can produce $\tilD$-odd spatial structure in the cloud.

The first is that the interaction with each doublet takes place at slightly different spatial locations, separated by $\delta x_{\rm dep} \sim\SI{60}{\micro\meter}$. This can cause an initial offset between the two clouds, $r_{i0}^D\lesssim\beta_{\rm dep}\delta x_{\rm dep}\sim \SI{6}{\micro\meter}$. Initial offsets in the center position of the two doublets can cause a non-zero $\langle r_i\rangle^D$ but, given that both clouds oscillate about the trap center with trap frequencies $\omega_{\rm sec}\sim2\pi\times\SI{1}{\kilo\hertz}$, the time-averaged effect is heavily suppressed by the long Ramsey times, $t_R\sim\SI{2}{\second}$ used in our experiment, $\langle r_i\rangle^D\sim\langle r_{i0}\rangle\omega_{\rm sec}/2\pi t_R\sim\SI{30}{\nano\meter}$. This effect is $\tilD\tilR$-odd and so produces a systematic when coupled to gradients in $\fBR$. The $\fBR$ gradients we measure in the radial directions are $<\SI{20}{\milli\hertz\per\centi\meter}$, giving a systematic shift of $<\SI{100}{\nano\hertz}$. In reality we expect any effect to be significantly smaller still due to the depletion laser being on for a time which is long compared to one trap period so that any $r_{i0}^D$ is greatly reduced. 

If the laser beam has non-zero displacement from the center of the micromotion, $x_{\rm dep}$, the interaction can selectively remove more of the hotter ions from one doublet than from the other, producing an effective $\langle r_i^2\rangle^D$ in those ions remaining in $\tDo$. In the worst-case limit where the depletion laser is much smaller than the ion cloud, the mean square position of the removed ions is
\begin{equation}
    (x_{\rm dep} +\frac{\delta x_{\rm dep}}{2})^2 - (x_{\rm dep} -\frac{\delta x_{\rm dep}}{2})^2 = 2 x_{\rm dep}\delta x_{\rm dep}.\end{equation}
The ions oscillate backwards and forward in the trap and so, in the limit of no ion-ion collisions, their time-averaged mean square position during the Ramsey time is reduced by a factor of 2. Simulations of more realistic ratios of laser to cloud size reduce this by a further factor of $\sim3$. We estimate $x_{\rm dep}<\SI{2}{\milli\meter}$ and so estimate the difference in mean square position of the remaining ions as $\langle {r_i}^2\rangle^D\sim\beta_{\rm dep} x_{\rm dep}\delta x_{\rm dep}/3\sim\SI{4e-5}{\centi\meter\squared}$. The effect is again $\tilD\tilR$-odd and so can produce a systematic when accompanied by curvature in $\fBR$. The largest such curvature we measure is \SI{70}{\milli\hertz\per\centi\meter\squared}, producing a systematic of $\sim\SI{3}{\micro\hertz}$. In reality, ion-ion collisions redistribute the velocities and positions of ions in the trap so that any initial difference in the size of the clouds is rapidly scrambled. To set an upper limit on the timescale of this thermalization, we note that we measure a curvature in $\fn$ of \SI{1200}{\milli\hertz\per\centi\meter\squared}, caused by the spatial variation of the magnetic field from the $\vec{\mathcal{B}^0}$ coils. For a \SI{1}{\centi\meter} gaussian cloud with no ion-ion collisions, this would cause decoherence of $\sim20\%$ of the contrast in $\sim\SI{50}{\milli\second}$. This is roughly equal to our total loss of contrast over a \SI{3}{\second} Ramsey time and so $\sim\SI{50}{\milli\second}$ sets a rough upper limit for the timescale of thermalization. For the shortest Ramsey times used in the dataset of $\sim\SI{1.5}{\second}$, the possible systematic shift is reduced by a factor of $\SI{50}{\milli\second}/\SI{1.5}{\second}\sim30$. As such we estimate the size of any residual systematic to be $<\SI{100}{\nano\hertz}$ and so do not include its contribution in our systematic error budget.

The second mechanism is that the electric field vector defining the quantization axis is, on average, pointing in slightly different directions when the laser interacts with each doublet. These angles are given by $\theta_{\rm u}=\theta_0/2+\delta\theta$, $\theta_{\rm l}=-\theta_0/2+\delta\theta$ where $\theta_0\sim\Delta_{\rm St}/\Delta_{\rm Dopp}\sim0.14$ is the difference in the angle of $\Erot$ when it addresses each of the two doublets and $\delta\theta\sim\delta_{\rm depl}/\Delta_{\rm Dopp}$ is half the difference in the magnitude of those angles caused by imperfect laser detuning $\delta_{\rm depl}\lesssim\SI{20}{\mega\hertz}$. In the unsaturated limit, the scattering rate with which each doublet interacts with the laser is then different by a factor $\cos\theta_{\rm u}-\cos\theta_{\rm l}\sim-2\Delta_{\rm St}\delta_{\rm depl}/\Delta_{\rm Dopp}^2$. Again, the laser interacts with a fraction $\sim0.1$ of the molecules and so we estimate the possible difference in mean square position as $\langle {r_i}^2\rangle^D\sim\langle r_i^2\rangle 0.1(2\Delta_{\rm St}\delta_{\rm depl}/\Delta_{\rm Dopp}^2)\sim\SI{6e-4}{\centi\meter\squared}$. This effect is $\tilR$-even and so can produce a systematic when accompanied by curvature in $\fB$. The largest such curvature we measure is \SI{10}{\milli\hertz\per\centi\meter\squared}, producing a systematic of $<\SI{6}{\micro\hertz}$. The same thermalization arguments discussed in the previous paragraph reduce this to below \SI{200}{\nano\hertz}. 

Finally, we consider the first dissociation laser. The photodissociation beam is a doubled, pulsed dye laser with a \SI{10}{\nano\second} pulse width. The central frequency of the laser pulse is stabilized to a wavemeter, with the set point chosen to be a compromise between the peaks of the two transitions for dissociating the upper and lower doublets. Over the course of the entire data run we find that on average we see a $\tilD$-odd term of 2\% in the number of Hafnium ions we create.  A residual difference in efficiency, in combination with intensity-driven saturation and spatial structure on the beam can generate a $\tilD$-odd contribution to the mean-square spatial extent of the molecules that participate in the Ramsey fringes. This in turn can combine with curvature in $\fB$, which can be as large as \SI{10}{\milli\hertz\per\centi\meter\squared},  to give a systematic error on $\fBD$.   As with possible $\tilD$-odd effects arising from the optical pumping beams, the size of the resulting systematic error is strongly suppressed by ion-ion thermalization that occurs over the course of the $t_R$ and washes out size differences. The pulsed laser’s width in frequency space is large compared to the transform-limited value for a \SI{10}{\nano\second} pulse width, and frequency structure within the pulse is not well characterized, which makes modeling the effect of the laser beam a little uncertain. 
Just before entering the chamber, however, the dissociation beam passes through an iris with 1 cm diameter. We can make the simplifying and  very pessimistic assumption that the extra 2\% $\tilD$-odd change in Hafnium ions is due entirely to molecules dissociated just at the outer edge of the laser beam. In this way we can set a conservative limit that the size of this effect on $\fBD$ must be less than \SI{3.5}{\micro\hertz}.


In addition to these effects from the lasers, we have also considered possible effects of $\tilD$-odd forces on the ions due to electric field gradients, and $\tilD$-odd heating due to photon scattering. We find each to be significantly smaller than our limits on those from the lasers.

\subsubsection{Effects of frequency drifts}
\label{sec:frequency_drift}
The average frequency probed in our experiment experiences drifts over multiple time scales, driven by multiple physical effects.  The frequency is linear in the magnitude of $\vecErot$, and linear in the magnitude of the reversing $\vec{\mathcal{B}^0}$. The electronics that generate these fields have gains that drift with various thermal time constants. Changes in surface potentials cause the equilibrium position, the center of the trapped ion cloud, to move slightly over the course of hours and days.  In the presence of linear field gradients, this can cause a frequency drift. We reoptimize ion production  every few hours. In between these tune-ups, ion number usually drifts downward. The fringe frequency is modestly affected by ion number.  There is also the possibility of finite `settling time' as e.g. the magnetic field restabilizes after we switch its sign.

The order in which we cycle through the various combinations of the parameters  $\tilB,\tilR,\tilI$ can couple to the various frequency drifts and result in systematic effects in various frequency channels.  The data collection software must periodically perform various overhead functions such as writing data to disk, and this can disrupt the cadence of data collection and cause differential thermal shifts for different switch states.  As described in Section~\ref{sec:exp-switch-states}, every two blocks we reverse the order of switches.  We report data averaged over the frequencies obtained from blocks with the different progression direction, but if we disaggregate the data we see that progression order is associated with shifts in the $\tilD$-even frequency channels of 2 or \SI{3}{\milli\hertz}, in particular in $\fB$, $\fI$ and $\fR$.  We reverse but do not fully randomize the switch progression between blocks, and a simple model for drifts would suggest that $\fRI$, $\fBR$, and $\fBI$ should all be shifted away from zero by switch-progression effects that survive averaging over our reversal of switch progression direction. Indeed we see that both $\fBI$ and $\fRI$ differ from zero by statistically significant amounts of order $\sim \SI{1}{\milli\hertz}$, see Appendix~\ref{app:table-guide}. 
 
The $\tilD$-odd frequency channels, and $\fBD$ in particular are highly insulated from drift effects because the upper and lower doublets are probed simultaneously with each shot of the experiment. We estimate that the degree of temporal overlap is better than a part in $10^4$ of the duration of the Ramsey sequence, which would put any possible leakage from $\tilD$-even channels to $\tilD$-odd channels to well under \SI{1}{\micro\hertz}.

\section{Phase Shifts} 
\label{sec:phase_shifts}

In this section we explore possible systematic effects caused by shifts in the ion's internal quantum phase, or our measurement of that phase. Using trapped ions for our measurement makes us relatively insensitive to these effects when compared to other similar experiments using beams of atoms or molecules. The first reason is that we can vary the free evolution time of our measurement; each block of our dataset consists of data taken at short and long Ramsey times. Most phase shifts caused by state-preparation or measurement are common mode and so cancel out in our eEDM channel which is only sensitive to the differential phase evolution between the early and late time data. The second reason is that any residual shifts---from phase effects which are different for the early- and late-time data---get divided by the difference in free evolution time between the early- and late-time data. The coherence time of the ions in our experiment is roughly three orders of magnitude longer than comparable experiments using beams and so any effects are reduced by a similar factor. In Sec.~\ref{sec:state-prep}, we consider phase errors caused by imperfections in state-preparation, and in Sec.~\ref{sec:internal_state_measurement}, we consider errors caused by imperfections in our measurement of phases.

\subsection{State preparation} 
\label{sec:state-prep}
\subsubsection{phases due to $\pi/2$ pulses} 

During preliminary data runs, we measured a small, and unexpected, contribution to the initial phase (roughly \SI{10}{\milli\radian}) which is odd under $\tilD$, $\tilB$ and $\tilI$. We studied the dependence of this effect on many experimental parameters. In particular we found: (a) the magnitude of the effect depended strongly on the parameters used for the $\pi/2$ pulses--we could greatly suppress the effect by implementing $\pi/2$ pulses with less reduced values of $\Erot$, and compensating by increasing their duration; (b) the sign of the effect changed when we projected into the opposite doublet; and (c) contrary to what we had anticipated for an effect which is odd under $\tilI$, there is no dependence on the side of the phosphor screen that each doublet is imaged onto.

\begin{figure}
    \centering
    \includegraphics[width=\columnwidth]{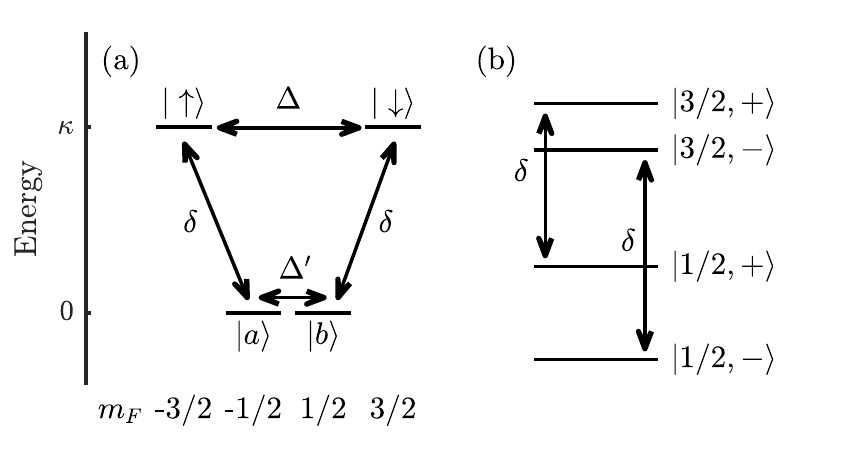}
    \caption{Toy model to explain $\phi^{DBI}$. (a) Uncoupled basis. (b) Coupled basis.}
    \label{fig:pDBI-toy-model}
\end{figure}

We have settled on a probable physical mechanism for the effect which qualitatively matches our observations by including the effects of other states in the $\tDo(v=0,J=1)$ manifold outside the two doublets used for spectroscopy. The important features of this mechanism can be illustrated with the toy model system shown in Fig.~\ref{fig:pDBI-toy-model}(a). The model includes just 4 of the 12 levels in $\tDo(J=1)$: the two $F=3/2, |m_F|=3/2$ states of a given doublet, labeled $\ket{\uparrow}$ and $\ket{\downarrow}$, and the two $F=3/2,|m_F|=1/2$ levels which lie closest to them in energy, labeled $\ket{a}$ and $\ket{b}$. In an electric field, and in the absence of any magnetic field or rotation, the two pairs of states with $|m_F|=3/2$ and $|m_F|=1/2$ are each degenerate. The Stark splitting between the two pairs is $\kappa$. When rotation is included, there is a direct coupling mixing $\ket{\uparrow}$ and $\ket{a}$, and $\ket{\downarrow}$ and $\ket{b}$. In the full 12-level model there is also a second-order coupling between $\ket{a}$ and $\ket{b}$---involving a single matrix element associated with each of rotation and $\Omega$-doubling---and a fourth-order coupling between $\ket{\uparrow}$ and $\ket{\downarrow}$--- involving three matrix elements associated with rotation and a matrix element associated with $\Omega$-doubling. Although these couplings are much smaller than the direct coupling, they are important because they act in degenerate subspaces so we include their effects in our toy model through couplings $\Delta$ and $\Delta'$. We neglect effects of magnetic fields here because we are interested in what happens during a $\pi/2$-pulse where the other effects are all much larger. The Hamiltonian for our toy system in this basis is then
\begin{equation}
    H = \left(\begin{array}{cccc}
         \kappa & \delta & 0 & \Delta \\
         \delta & 0 & \Delta' & 0 \\
         0 & \Delta' & 0 & \delta \\
         \Delta & 0 & \delta & \kappa
    \end{array}\right),
\end{equation}
where the states are in order of $m_F$: $\ket{\uparrow}, \ket{a}, \ket{b}, \ket{\downarrow}$. 

We can simplify our thinking by changing to a different basis. Consider the case where we somehow independently tune $\delta$ to zero; we now have two uncoupled systems, $\ket{\uparrow}$ and $\ket{\downarrow}$, and $\ket{a}$ and $\ket{b}$. We can diagonalize each system to give eigenstates which are fully mixed as shown in Fig.~\ref{fig:pDBI-toy-model}(b): $\ket{3/2,+} = \tfrac{1}{\sqrt{2}}(\ket{\uparrow}+\ket{\downarrow})$ and $\ket{3/2,-} = \tfrac{1}{\sqrt{2}}(\ket{\uparrow}-\ket{\downarrow})$, and $\ket{1/2,+} = \tfrac{1}{\sqrt{2}}(\ket{a}+\ket{b})$ and $\ket{1/2,-} = \tfrac{1}{\sqrt{2}}(\ket{a}-\ket{b})$. Rewriting the full toy-model Hamiltonian in this basis we have 
\begin{equation}
    H = \left(\begin{array}{cccc}
         \Delta' & \delta & 0 & 0 \\
         \delta & \kappa + \Delta & 0 & 0 \\
         0 & 0 & -\Delta' & \delta \\
         0 & 0 & \delta & \kappa - \Delta
    \end{array}\right),\label{eq:pDBI-toy-ham2}
\end{equation}
where the states are in the order $\ket{1/2,+}, \ket{3/2,+}, \ket{1/2,-}, \ket{3/2,-}$. The Hamiltonian is now block diagonal. During the second $\pi/2$ pulse, the mixing between $\ket{\uparrow}$ and $\ket{\downarrow}$ causes population to oscillate between them (as intended), but it also causes a weak oscillation between the $|m_F|=3/2$ states and the $|m_F|=1/2$ states. This oscillation depends on the initial phase difference between the $\ket{\uparrow}$ and $\ket{\downarrow}$ states. The two most important phases to consider are the sides of the fringe; where we are most sensitive to phase shifts, and thus where we take most of our data. On the side of the fringe, the state immediately before the second $\pi/2$ pulse is $\ket{3/2,+}$ or $\ket{3/2,-}$. It can be seen from Eq.~\ref{eq:pDBI-toy-ham2} that when $\delta=0$, these states are eigenstates of the Hamiltonian and the $\pi/2$-pulse does nothing. When $\delta$ is non-zero, the coupling causes population to be transferred from $\ket{3/2,+}$ to $\ket{1/2,+}$ and from $\ket{3/2,-}$ to $\ket{1/2,-}$ with Rabi frequencies $\Omega_+$ and $\Omega_-$ respectively,
\begin{equation}
\begin{split}
    \Omega_+ &= \sqrt{4\delta^2 + (\kappa+\Delta-\Delta')^2}\sim \kappa+\Delta-\Delta',\\
    \Omega_- &= \sqrt{4\delta^2 + (\kappa-\Delta+\Delta')^2}\sim \kappa-\Delta+\Delta'.
    \end{split}
\end{equation}
The two Rabi frequencies are different because the energy gap is different in each case. The amplitudes of the oscillations in each case are then
\begin{equation}
\begin{split}
    A_+ &= \frac{\delta^2}{4\delta^2 + (\kappa+\Delta-\Delta')^2}\sim\frac{\delta^2}{\kappa^2},\\
    A_- &= \frac{\delta^2}{4\delta^2 + (\kappa-\Delta+\Delta' )^2}\sim\frac{\delta^2}{\kappa^2}.
\end{split}
\end{equation}
The approximate expressions in both cases assume $\kappa\gg\delta,\Delta,\Delta'$. This difference in Rabi frequencies results in a different amount of the population being transferred to $\ket{a}$ and $\ket{b}$, dependent on the initial phase. We measure only the population in either $\ket{\uparrow}$ \textit{or} $\ket{\downarrow}$ and never the population in $\ket{a}$ or $\ket{b}$, and so this population transfer appears as an apparent phase shift. The population oscillations are happening very fast---the frequencies are of order the energy gap between the 3/2 states and the 1/2 states, $\kappa\sim\si{\mega\hertz}$---and so the size of the apparent phase shift can depend very sensitively on the exact parameters of the $\pi/2$-pulse. The maximum size of the effect goes as $1/\kappa^2$ and so depends strongly on the size of $\Erot$ during the $\pi/2$ pulse.

We now examine the dependence of this effect on the experimental switch state:
\begin{enumerate}[(i)]
    \item $\tilI$: in our implementation of the $\tilI$ switch, the $m_F$ state we project into, and read out of, is $\tilI$ odd. This corresponds to measuring, for example, $\ket{\uparrow}$ instead of $\ket{\downarrow}$. On the side of the fringe where the population in $\ket{\uparrow}$ is decreasing with Ramsey time, the population in $\ket{\downarrow}$ is increasing with Ramsey time and as a result the apparent phase shift---a relative change in population which has same sign for both states---is $\tilI$ odd.
    \item $\tilB$: the phase shift is completely independent $\vec{\mathcal{B}^0}$ and so is $\tilB$-odd.
    \item $\tilD$: in the opposing doublet, the sign of $\kappa$ is flipped; the $|m_F|=1/2$ states are above the $|m_F|=3/2$ states in energy. This means that the relative rate of transfer between the two sides of the fringe are flipped and the effect is $\tilD$ odd.
    \item $\tilR$: $\delta$, $\Delta$ and $\Delta'$ all involve odd numbers of matrix elements associated with rotation and so flip sign with rotation direction. This flips the sign of the effect so that the relative rates of population transfer out of the two states are reversed. However, the flipping sign of $\Delta$ means that the initial superposition produced by the first $\pi/2$ pulse is $\pi$ out of phase compared with the opposite rotation direction so the phase shift at a given Ramsey time ends up being $\tilR$ even.
\end{enumerate}
In the experiment it is very difficult to control the parameters of the $\pi/2$ pulses at the required level to quantitatively verify this mechanism is at play. In addition, small differences in the electric field experienced by different ions, and coupling to other states in the full Hamiltonian, cause the population oscillations to dephase rapidly, so we were never able to observe them. However, we believe the qualitative agreement with what we observe to be convincing.

We took three steps to mitigate the effects of $\phi^{DBI}$. We reduced the size of the effect dramatically by increasing the value of $\Erot$ that we ramp down to during the $\pi/2$-pulse relative to Ref.~\cite{Zhou2020}. We also added the $\tilP$ chop to our experimental sequence for the eEDM dataset; in every other block, we read out of the opposite state, so that any remaining effect would change sign. Finally, midway through the dataset, we reversed the polarization of all the optical pumping, depletion and dissociation lasers, again flipping the sign of any effect. We can still check for any residual effect by looking at the difference in $\phi^{DBIP}$ before and after the waveplate change; in the eEDM dataset its average value was \SI{1.1+-0.1}{\milli\radian}. Such phase shifts could potentially leak into frequency channels if, for example, the second $\pi/2$ pulse depended on the Ramsey time, perhaps because of heating in the amplifiers or similar. We note that during our exploration of the effect we were able to increase its size to \SI{170}{\milli\radian}, and saw no shift in $\fDBI$ at the $1\sigma$ confidence level. Using this data we can set a limit on any shift caused by the phase at \SI{-3+-4}{\micro\hertz\per\milli\radian}, increasing confidence in our ability to reject systematic effects associated with phases caused by the $\pi/2$ pulses. In the eEDM dataset, none of the other $\tilD$-odd phases (except $\phi^D$) exceed \SI{750}{\micro\radian}.

\subsection{Internal state measurement}
\label{sec:internal_state_measurement}
\subsubsection{Improperly characterized imaging contrast} 
\label{sec:systematic_fDpulling}

\begin{figure}
    \centering
        \includegraphics[width=\columnwidth]{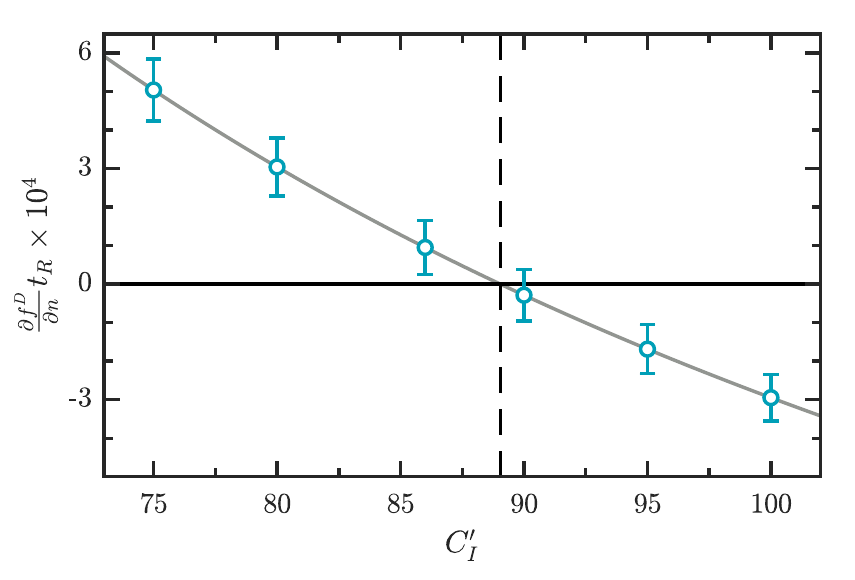}
    \caption{Example data. Plotting the slope of $\fD$ when the early-time data is taken at varying offsets from the \textit{beat}. The $x$-intercept gives $C_{\rm I}'=\num{0.89+-0.01}$, which we then use in our fitting program.}
    \label{fig:fD_pulling}
\end{figure}

\begin{figure}
    \centering
        \includegraphics[width=\columnwidth]{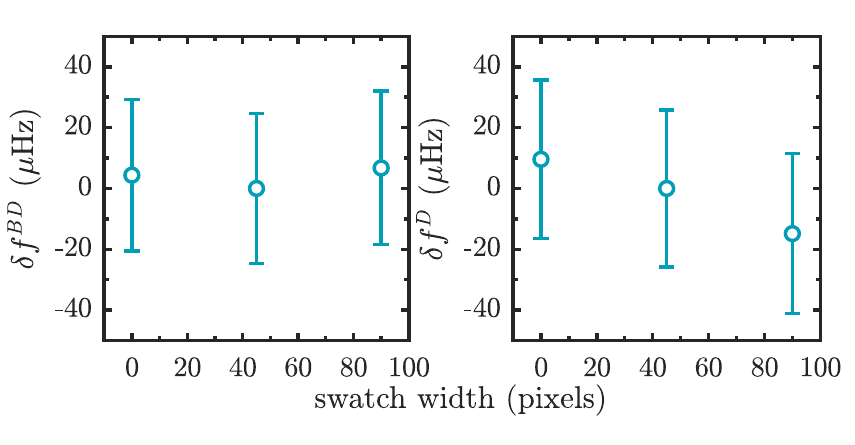}
    \caption{Varying the swatch width while correcting for the doublet contrast. We see no meaningful shifts in our eEDM channel and a small residual shift in $\fD$. In the right panel we have averaged over the $3$ distinct values of $\fD$ that we operated at.}
    \label{fig:fBD_vs_swatchwidth}
\end{figure}

\begin{figure}
    \centering
        \includegraphics[width=\columnwidth]{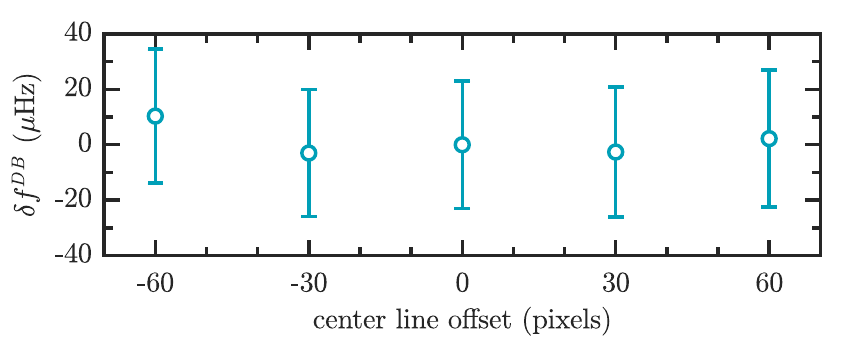}
    \caption{The effect of moving the center of the swatch on our eEDM channel.}
    \label{fig:fBD_vs_swatchposition}
\end{figure}

Our imaging process allows us to differentiate between Hf ions originating from the upper and lower doublets by projecting them onto different sides of the imaging MCP. However, due to the initial size of the cloud and its finite temperature the two clouds are not perfectly separated and they overlap slightly at the center of the screen. This overlap can cause our measurement of the difference frequency to be pulled and, if improperly characterised, can potentially cause a systematic error. To see how this works, consider how the number of ions detected on each side of the MCP, and in each phase, depends on the phase evolution of the upper and lower doublets $\Phi_{\rm u}$ and $\Phi_{\rm l}$,
\begin{equation}
    \begin{split}
        N_{u}^{\rm anti} &= \frac{N}{2}(1-\sin(\Phi_{\rm u}))(1-\epsilon)+\frac{N}{2}(1-\sin(\Phi_{\rm l}))(\epsilon)\\
        N_{u}^{\rm in} &= \frac{N}{2}(1+\sin(\Phi_{\rm u}))(1-\epsilon)+\frac{N}{2}(1+\sin(\Phi_{\rm l}))(\epsilon)\\
        N_{l}^{\rm anti} &= \frac{N}{2}(1-\sin(\Phi_{\rm l}))(1-\epsilon)+\frac{N}{2}(1-\sin(\Phi_{\rm u}))(\epsilon)\\
        N_{l}^{\rm in} &= \frac{N}{2}(1+\sin(\Phi_{\rm l}))(1-\epsilon)+\frac{N}{2}(1+\sin(\Phi_{\rm u}))(\epsilon),
    \end{split}
\end{equation}
where $N$ is the mean number of ions measured on each side of the MCP, $\epsilon$ is a parameter which characterises the amount of leakage of each cloud onto the other side of the MCP screen, and we have assumed perfect contrast. We can now form asymmetries as we do in our analysis,
\begin{equation}
    \begin{split}
        A_{\rm u} &= \frac{N_{\rm u}^{\rm in}-N_{\rm u}^{\rm anti}}{N_{\rm u}^{\rm in}+N_{\rm u}^{\rm anti}}\\
        &= (1-\epsilon)\sin\Phi_{\rm u}+\epsilon\sin\Phi_{\rm l},\\
        A_{\rm l} &= \frac{N_{\rm l}^{\rm in}-N_{\rm l}^{\rm anti}}{N_{\rm l}^{\rm in}+N_{\rm l}^{\rm anti}}\\
        &= (1-\epsilon)\sin\Phi_{\rm u}+\epsilon\sin\Phi_{\rm l}.\label{eq:imaging-contrast-asymmetries}
    \end{split}
\end{equation}
We can simplify our thinking by assuming that we are taking data close to the side of the fringe such that $\Phi_{\rm u}\simeq2 p \pi+\delta_{\rm u}$ and $\Phi_{\rm l}\simeq2 q \pi+\delta_{\rm l}$ for $p$ and $q$ integer. In this limit we have 
\begin{equation}
    \begin{split}
        A_{\rm u} \simeq \delta_{\rm u} -\epsilon(\delta_{\rm u}-\delta_{\rm l})\\
        A_{\rm l} \simeq \delta_{\rm l} +\epsilon(\delta_{\rm u}-\delta_{\rm l}).
    \end{split}
\end{equation}
The apparent phase of each of the doublets are pulled towards each other by an amount that depends on the leakage $\epsilon$ and their difference in true phase $\delta_{\rm u}-\delta_{\rm l}$. We account for this in our analysis by including a parameter $C_{\rm I}$, the imaging contrast. By comparing Eq.~\ref{eq:imaging-contrast-asymmetries} with Eq.~\ref{eq:asymmetries} we can identify $\epsilon=(1-C_{\rm I})/2$. Now suppose we mischaracterize the imaging contrast $C_{\rm I}$, assigning it an incorrect value $C_{\rm I}'$. The resultant fitted values for the phases of the upper and lower doublet $\delta_{\rm u}'$ and $\delta_{\rm l}'$ will then satisfy the following expressions
\begin{equation}
    \begin{split}
        \delta_{\rm u}' -\frac{1-C_{\rm I}'}{2}(\delta_{\rm u}'-\delta_{\rm l}')=\delta_{\rm u} -\frac{1-C_{\rm I}}{2}(\delta_{\rm u}-\delta_{\rm l}),\\
        \delta_{\rm l}' +\frac{1-C_{\rm I}'}{2}(\delta_{\rm u}'-\delta_{\rm l}')=\delta_{\rm l} +\frac{1-C_{\rm I}}{2}(\delta_{\rm u}-\delta_{\rm l}).
    \end{split}\label{eq:pD-pulling-simp}
\end{equation}
We find that the fitted difference phase $\Theta'=\delta_{\rm u}'-\delta_{\rm l}'$ is different from the true difference phase $\Theta=\delta_{\rm u}-\delta_{\rm l}$ by an amount
\begin{equation}
    \Theta'-\Theta = \frac{(C_{\rm I}-C_{\rm I}')\Theta}{C_{\rm I}'}.
\end{equation}
Both the early and late time data in our Ramsey fringes are potentially susceptible to this phase pulling effect and so it is important that we properly characterize and understand $C_{\rm I}$. To do so, we took fringes where the data was collected deliberately offset from the early- or late-time zero crossings of the difference fringe. Taking the late time data removed from the beat by a number of $\fn$ periods $n$ causes the fitted $\fD$ to be pulled by an amount
\begin{equation}
    \delta\fD=\frac{\Theta'-\Theta}{2\pi t_R} = \frac{n}{t_R}\frac{(C_{\rm I}-C_{\rm I}')}{C_{\rm I}'}\frac{\fD}{f}.
\end{equation}
Moving the early time data has the same size effect but with the opposite sign. For a given $C_{\rm I}'$ and swatch size, we can fit the extracted $\fD$ vs the number of periods we are offset from the zero crossing of the difference fringe. Figure~\ref{fig:fD_pulling} shows how this slope varies with the value of $C_{\rm I}'$ for the swatch size of 90 pixels used in our measurement, chosen to maximize $C\sqrt{N}$. The correct value of $C_{\rm I}'$ is the one for which $\frac{\partial\fD}{\partial n}=0$, here $C_{\rm I}'=\num{0.89+-0.01}$, corresponding to $\epsilon=\num{0.050+-0.005}$, which we use for analysis of our whole dataset. We note that, because the points in Fig.~\ref{fig:fD_pulling} correspond to reanalysis of the same dataset, the signal to noise in the shifts is much better than indicated by the error bars. This also applies to Figs.~\ref{fig:fBD_vs_swatchwidth}--\ref{fig:various_vs_swatchposition} and the lower part of Fig.~\ref{fig:imaging_saturation_data}.

Whilst we can measure $C_{\rm I}$ very precisely, its value can wander slightly over time. We have repeated this measurement many times over the course a year and find that the early and late time $C_{\rm I}$ are consistent with one another, and between measurements, to $\pm0.05$. We conservatively estimate the largest possible deviation in $C_{\rm I}$, averaged over the dataset to be $\delta C_{\rm I}\sim0.05$.

For this effect to leak into $\fBD$ and cause a systematic shift in our eEDM signal requires either the amount we miss the beat by to be $\tilB$-odd, or the imaging contrast itself to be $\tilB$-odd. We consider the $\tilB$-odd phase first. Because the mean value of $\phi^D$ is negative, the early time beat happens before zero, and so we are forced to take data with nonzero $\Theta$. In this case a non-zero value of $\phi^{BD}$ will mean that the early time data is taken at a different $\Theta$ depending on the $\tilB$ switch. The value of $\phi^{BD}$ over the dataset is \SI{-30+-140}{\micro\radian}. Combining this with the uncertainty in $C_{\rm I}$ above leads to a systematic uncertainty $\tfrac{\phi^{DB}\delta C_{\rm I}}{2\pi t_R C_{\rm I}}\sim \SI{0.7}{\micro\hertz}$.

A systematic shift can also potentially be caused by $\phi^D$ combined with a $\tilB$-odd imaging contrast $C_{\rm I}^B$. In order for this shift to reach the \SI{5}{\micro\hertz} level, would require $C_{\rm I}^B\sim0.005$, or equivalently $\epsilon^B\sim\num{2.5e-3}$, or about 5\% of the total leakage. We know of no mechanism which can cause such an effect. As a final check, we refit all \numblocks{} blocks in the dataset with $C_I$ deliberately offset from our best estimate and see no concerning shift in our final value of $\fBD$ and so include no contribution in our systematic uncertainty budget.

\subsubsection{Swatch position}
\label{sec:systematic_swatch_position}

\begin{figure}
    \centering
        \includegraphics[width=\columnwidth]{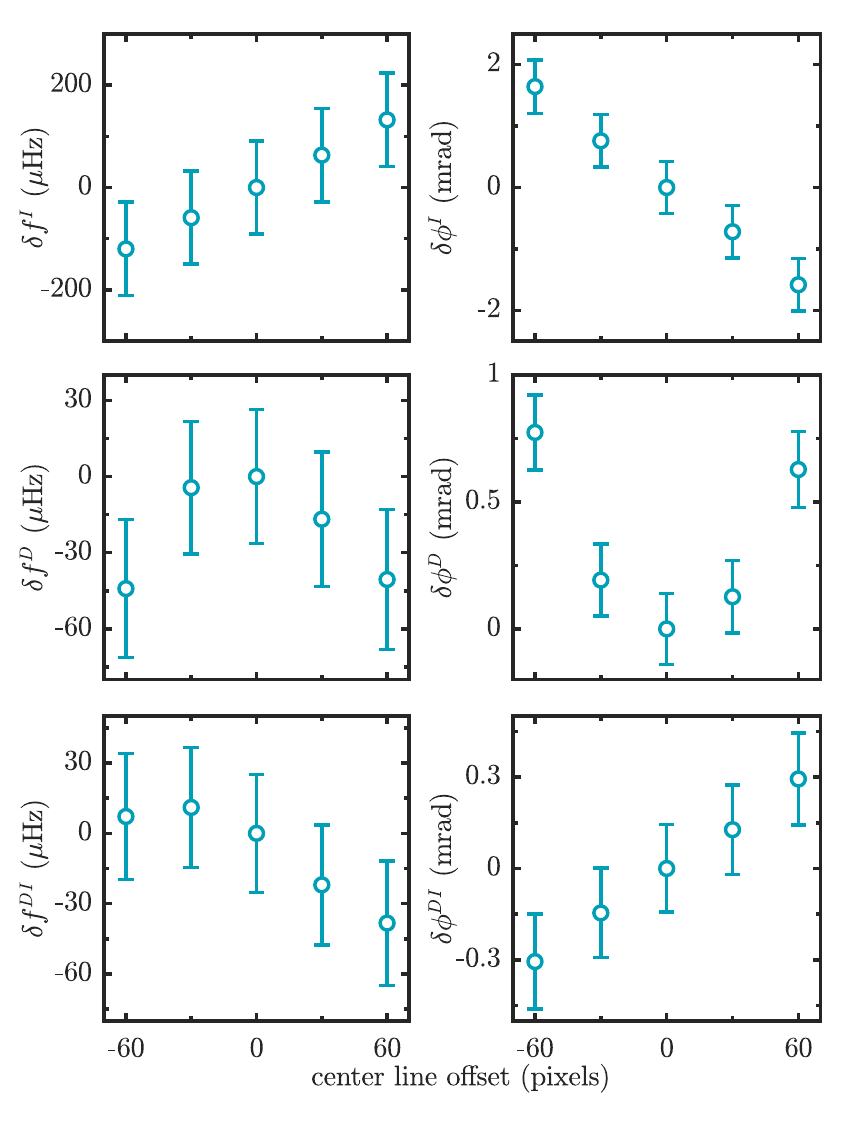}
    \caption{Moving the center of the swatch has a measurable effect in several parity channels. Note all panels show the entire aggregated dataset except for the $\fD$ panel, which shows a subset of measurements performed at a single $\fn$}
    \label{fig:various_vs_swatchposition}
\end{figure}

As described in Section \ref{sec:experimental_overview}, the position of the center line and corresponding swatch for each switch state is determined by an algorithm based on images of the cloud taken without any $\pi/2$-pulses. In this section we explore the consequences of a possible systematic error in the center position dividing line between the two doublets.

We reanalyzed the dataset, with the center line in each image displaced by different number of pixels left and right. The largest observed shifts in the frequency and phase channels are shown in Fig.~\ref{fig:various_vs_swatchposition}. We can explain the observed shifts using ideas from the previous section. Let's use Eq.~\ref{eq:pD-pulling-simp} as before but this time with the phase pulling on each of the doublets separated,
\begin{equation}
\begin{split}
    \delta_{\rm u}'-\delta_{\rm u} &= \frac{(\delta_{\rm u}-\delta_{\rm l})(\epsilon-\epsilon')}{2\epsilon'-1},\\
    \delta_{\rm l}'-\delta_{\rm l} &= -\frac{(\delta_{\rm u}-\delta_{\rm l})(\epsilon-\epsilon')}{2\epsilon'-1}.
\end{split}
\end{equation}
We can combine both of these equations as 
\begin{equation}
    \delta\phi = \tilde{D}\frac{\Theta(\epsilon-\epsilon')}{2\epsilon'-1} \label{eq:sep-pD-pulling}
\end{equation}
where $\delta\phi$ is the phase pulling of a single fringe and $\Theta=\delta_{\rm u}-\delta_{\rm l}$ is the amount we are missing the beat by. We can express the effect of moving the swatch center around by modifying $\epsilon$, the amount each doublet leaks into the other
\begin{equation}
    \epsilon = \epsilon_0 + \kappa \tilde{D} \tilde{I} Y_c - \nu (\tilde{D}\tilde{I} Y_c)^2.\label{eq:mod-doublet-leak}
\end{equation}
This equation expresses the fact that as we move the swatch over to one side, there is more leakage into one of the doublets and less into the other. The direction of the leakage depends on the $I$ switch. The first order effect, $\propto \kappa$, is linear but for larger swatch displacements, the second order effect, $\propto\nu$, starts to become important too. Substituting this back into Eq.~\ref{eq:sep-pD-pulling} and assuming for simplicity that we have done a good job of picking the doublet contrast in the first place so that $\epsilon'=\epsilon_0$, we get
\begin{equation}
    \delta\phi = \frac{\Theta}{2\epsilon'-1}(\tilde{I}\kappa Y_c - \tilde{D}\nu Y_c^2).
\end{equation}
The last factor we need to include to explain the data is to make the approximation that the main place we are missing the beat is at early time; a small amount of differential phase evolution occurs during the $\pi/2$-pulses so that even at $t_R=0$, the two doublets are slightly out of phase. Including the largest (by far) contribution to the $D$-odd phase only, we have $\Theta\sim\phi^D$, and
\begin{equation}
    \delta\phi = \frac{1}{2\epsilon'-1}(\tilde{I}\kappa \phi^D Y_c - \tilde{D}\nu \phi^D Y_c^2).\label{eq:swatch-pos-phi-shift}
\end{equation}
So the principle effects we expect are a linear shift in $\phi^I$ and a smaller quadratic effects in $\phi^D$. Because (almost) all of the pulling is happening at early time, each of these could be expected to have an echo in the corresponding $f$ channel, with opposite sign and scaled by a factor of $2\pi\times T \sim \SI{15}{\second}$. Each of these can be seen in Fig.~\ref{fig:various_vs_swatchposition} from which we infer $\kappa=\num{-1.89+-0.06e-3}$ per pixel and $\nu=\num{-7.0+-0.2e-6}$ per pixel squared. The shifts seen in $\phi^{DI}$ and $\fDI$ (about a factor of 5 smaller than those in $\phi^I$ and $\fI$) could be caused by a $\tilI$-odd systematic error in the swatch position $Y^I$. Including this in Eq.~\ref{eq:swatch-pos-phi-shift} yields an additional contribution $\tfrac{\phi^D}{2\epsilon'-1}(\kappa Y^I-\tilD\nu{Y^I}^2-2\tilD\tilI\nu Y_c Y^I)$. Comparing the term linear in $Y_c$ to the observed gradient in $\phi^{DI}$ gives $Y^I=\num{25+-1}$ pixels. Other possible systematic shifts in swatch position can be constrained by the data in the same way to less than 5 pixels.

The largest gradient in a frequency channel seen in this analysis is $\partial \fI/\partial Y_c \sim\SI{2}{\micro\hertz}$ per pixel. It is important to realize that the swatch displacement that is carried out here is just one of many possible ways of moving the swatch. Because we moved the swatch the same direction in each switch state, $Y_c$ shows up in Eq.~\ref{eq:mod-doublet-leak} with $\tilD\tilI$. There are 7 other possible ways to move the swatch, always $\tilD$-odd but with all other possible switch-state dependence. We can infer the effects of these possible systematic effects from those shown here. We are interested in any possible shifts in $\phi^{DB}$ and $f^{DB}$ which can only show up quadratically in any systematic error in the swatch position. To see an effect requires either missing the beat by a $\tilB$-odd amount, or a combination of two systematic shifts in swatch position, one $\tilB$-odd and one $\tilB$-even. As discussed in the previous section, the former can happen due to a combination of our non-zero $\phi^D$ and a non-zero $\phi^{DB}$. In our dataset, $\phi^{DB}=\SI{-30+-140}{\micro\radian}$ which, combined with $Y^I$ gives a systematic $|\tfrac{\phi^{DB} \nu {Y^I}^2}{2\pi T(1-e\epsilon')}|< \SI{70}{\nano\hertz}$. The largest contribution to the latter is from $Y^I$ combined with $Y^{BI}$, constrained by looking at the gradient of $\phi^{DBI}$ in our analysis to be $\num{1+-2}$ pixels. This gives a systematic of $|\tfrac{2 \phi^D \nu Y^I Y^{BI}}{2\pi T(1-e\epsilon')}|< \SI{1.2}{\micro\hertz}$. We include the quadrature sum of these two in our systematic uncertainty budget.

\subsubsection{Counting saturation}
Our imaging system is subject to ion-counting saturation, and if that were somehow magnetic field and doublet odd it could be a concern. It is hard to conceive of a mechanism for this to be both magnetic field and doublet odd, however, because even if one side of the screen saturated faster than the other, that asymmetry is heavily suppressed by the fact that the $\tilde{I}$ chop swaps which side of the screen we read each doublet out on. 

\begin{figure}
    \centering
        \includegraphics[width=\columnwidth]{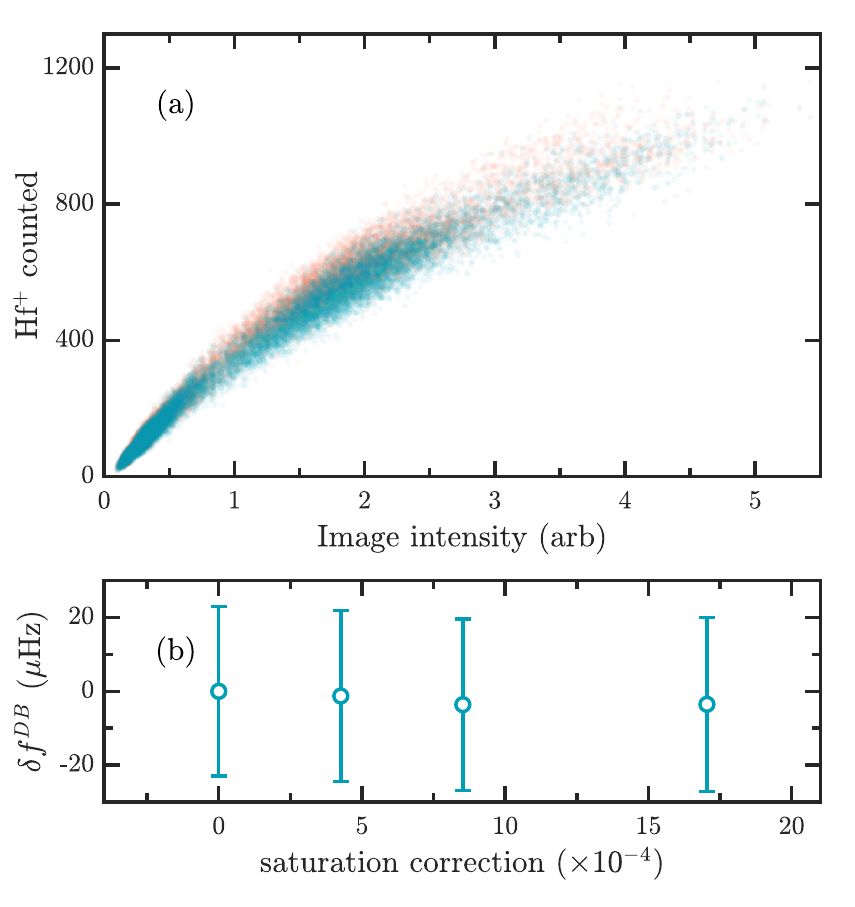}
    \caption{Imaging saturation. (a) Imaging data collated over 25 blocks. The upper doublet data is in pink while the lower doublet data is in blue. (b) Effect of saturation correction on $\fBD$, averaged over eEDM dataset.}
    \label{fig:imaging_saturation_data}
\end{figure}

Collating the imaging data over $25$ blocks  we can see that our imaging does indeed saturate (the integrated image intensity, $I$, does not scale linearly with the number of individually counted ions, see Figure \ref{fig:imaging_saturation_data}).

We can fit the data in Figure \ref{fig:imaging_saturation_data} to the function $\rm{Hf}^+_{\rm counted} = \beta I/(1+I/I_{\rm sat})$. For small numbers of ions, the image intensity should scale linearly with $\rm{Hf}^+_{\rm counted}$. Therefore, we can apply a saturation correction to our data: $\rm{Hf}^+_{\rm real} = \rm{Hf}^+_{\rm counted}(1+\alpha \rm{Hf}^+_{\rm counted})$, where $\alpha=1/(I_{\rm sat}\beta)$ is the saturation correction parameter. We re-analyzed all our data with several saturation corrections applied: none/zero, two incorrect ones, and the \textquote{true} value of $\alpha=8.52\times 10^{-4}$. We see no significant shift in our eEDM channel and so we do not include this effect in our systematic error budget.


\subsubsection{Time-varying offsets}

Our asymmetry fringes are slightly offset from zero due to the necessary compromise in our $\pi/2$-pulse parameters when addressing both doublets simultaneously, discussed in Appendix~\ref{app:table-guide}. The offsets are characterized by the parameters $O_{\rm u/l}$ from Eq.~\ref{eq:decaying-sine-wave}. It is possible that these offsets might decay as a function of Ramsey time, for example if collisions slowly redistribute population between the two states in a doublet. Since we fit our fringes with a single offset for all Ramsey times, this could potentially lead to a systematic shift in the relative phases. To investigate this possibility, we refit our entire dataset with a fixed differential offset between the early and late time asymmetry data points and looked for changes in $\fBD$. We observe no significant shifts and so include no contribution in our systematic uncertainty budget.

\section{Conclusion}
\label{sec:conclusion}

\begin{table*}
	\caption{Systematic error budget}
	\centering
	\begin{ruledtabular}
	\begin{tabular}{l l r r}
	    Effect & Section & Correction ($\mu$Hz) & Uncertainty ($\mu$Hz)\\
	    \midrule
	    \textbf{Magnetic}\\
	    \quad Non-reversing $\vec{\mathcal{B}^0}$ & \ref{sec:baxgradnr} & 0.1 & $<0.1$  \\ 
        \quad Second harmonic of $\Erot$ and transverse magnetic field & \ref{sec:2h+B} & & 2.2  \\ 
        \quad Third harmonic of $\Erot$ and magnetic field gradients & \ref{sec:nh-magnetic} & & 1.5\\
        \quad Higher harmonics of $\Erot$ and higher order magnetic field gradients &\ref{sec:nh-magnetic} & & $<0.1$  \\
        \quad Ellipticity of $\Erot$ and magnetic field gradients & \ref{sec:ellipticity-mag} & & 1.7  \\
	    \textbf{Berry's phase}\\
        \quad Phase modulation due to axial secular motion & \ref{subsec:systematic_zslosh} & & 3.4  \\ 
        \quad Axial 2nd harmonic with ellipticity of $\Erot$ & \ref{sec:axial2f_plus_radial_ellipticity}& & 1.7  \\ 
        \quad Higher harmonics of $\Erot$ & \ref{sec:berrys-phase-higher-harmonics} & & 3.0  \\
	    \textbf{Rotation induced mixing}\\
	    \quad Leaking of $\fBR$ & \ref{sec:fBR_leak} & & $0.2$ \\
        \quad Axial magnetic fields & \ref{sec:axial_magnetic_fields} & & $0.4$  \\
	    \textbf{Other frequency}\\
        \quad Imperfect overlap of spatial distributions & \ref{sec:doublet_odd_displacement}& & $3.5$  \\
	    \textbf{Phase}\\
        \quad Improperly characterized imaging contrast & \ref{sec:systematic_fDpulling}& & 0.7  \\
        \quad Swatch position & \ref{sec:systematic_swatch_position}& & 1.2  \\
        \addlinespace
            \textbf{Total} & & 0.1 & 6.9
	\end{tabular}
	\label{tab:systematic_error_budget}
	\end{ruledtabular}
\end{table*}

Table \ref{tab:systematic_error_budget} presents our error budget. We have demonstrated a significant advance in the characterization of our experiment resulting in a lower systematic uncertainty and a measurement which is statistics dominated. The use of rotating bias fields and trapped molecules is a powerful technique for suppressing systematics; the rotation means that most stationary field-induced systematics average away over a rotation cycle, and the trapped species allows us to measure both early- and late-time phase which eliminates a large class of state-preparation systematics one might otherwise have to worry about. As eEDM sensitivity is pushed into the next decade of accuracy, these advantages may prove essential.


\begin{acknowledgments}
We thank the staff at JILA for making this experiment possible. 
This work was supported by the Marsico foundation, the Sloan foundation, the Gordon and Betty Moore foundation, NIST, and the NSF (Award PHY-1125844). TW acknowledges funding support from NSF GRFP.

\end{acknowledgments}

\bibliography{references}

\appendix
\section{Guide to the tables of fit values}
\label{app:table-guide}

\subsection{Overview of data}

The data blocks can be divided into three main categories, based on the approximate value of $\fn$, corresponding to roughly $77$, $105$, and \SI{151}{\hertz}.  $\fD \simeq \tfrac{\dgF}{g_F} \fn$,  or respectively at $-171$, $-230$ and \SI{-327}{\milli\hertz}, and by choosing late-time Ramsey fringes centered at $1/2\fD$ or respectively roughly 2.9, 2.2 and \SI{1.5}{\second}, the late-time Ramsey fringes were taken at the re-phasing of the fringes for the upper and lower doublets.  
In the tables below, we present averages taken separately over the three ranges of $\fn$, plus an average taken over all the blocks.  In each case the quoted one-$\sigma$ errors are based on combining the estimated error on all the relevant blocks, and then relaxing that value by multiplying by $\sqrt{\chi^2}$ to correct for overscatter. For reference we include the relevant value of $\chi^2$ as calculated before the error bars were relaxed. The $\chi^2$ associated with the relaxed values of $\sigma$ displayed here are all by construction equal to 1.

The values are presented in Tables~\ref{tab:f-parity-channels}--\ref{tab:O-parity-channels}.

\subsection{Overall comments}

The parameters which are even in all switch states (superscript 0) are the noisiest, as they are unprotected from all the drifts in the experimental conditions.  These drifts arise from many sources, for instance: (i) changes in the temperature of the lab, or of the water cooling the power op-amps that drive the ion-trap electrodes; (ii) changes in the ambient magnetic field and its gradients, arising e.g. from a distant freight elevator, a less-distant optical table, or a still-closer weld in an only nominally nonmagnetic UHV chamber; (iii) variation in the number, temperature, and density of the ion cloud; (iv) drifts in the current supply that drives the $\vec{\mathcal{B}^0}$ coils. Cloud size and thus ion density and trap temperature were not well-characterized parameters. Ion number in particular drifts from block to block, and due to mean-field coulomb repulsion within the cloud the ion number is coupled to cloud size, and from there to $\fn$ which is an average over spatial field inhomogeneities. The strength of radial and axial trap confinement were treated as parameters to be tweaked to maximize the precision per block of data. Ion production was frequently re-optimized as well, with resulting jumps in $\fn$ between blocks of tens of \si{\milli\hertz} or more.

All $\tilP$-odd values are in a category by themselves, in that they were not collected as part of a rigorously implemented intra-block chop. Instead, we reverse the direction of the relevant polarizer every other block. On some days, we took an odd number of blocks. In other instances we vetoed an entire block of data which left the next block of data unpaired with a block of opposite P switch.  For these reasons, from our entire run of $\sim1300$ blocks we do not try to generate $\sim650$ sets of $\tilP$-odd fit values based on pairs of  matched, sequentially collected blocks. Instead, for each of the $\fn$ superblocks, we divide the blocks into the $\tilP=1$ and $\tilP=-1$ piles, calculate weighted mean values for each fitting parameter and each of the $\{\tilD,\tilB,\tilR,\tilI\}$ parity channels, with error bars corrected for scatter. Then, element-wise across the large table of values, we take either a sum or difference between the $\tilP=1$ and $\tilP=-1$ to create the overall $\tilP$-even or $\tilP$-odd parity channels. In this method, there is no particular meaning to $\chi^2$ for the $\tilP$-odd parity channels, so no such value presented in the tables below, however we use the $\tilP$-even $\chi^2$ to relax the associated error bars. 

After each block we attempt to servo the value of $\fB$ back to zero for the subsequent block. Thus the mean value of $\fB$ is very low even though $\fB$ channel is susceptible to drifts in ambient conditions. These drifts can be tracked because we record the value of the current in the coils used to servo $\fB$. 

Looking across the 480 parity-channel values associated with the three frequency superblocks, and  the 160 fully aggregated  values, we see the overwhelming majority of the values are quite close to zero. There are a few numbers which are dramatically different from zero, for good reason: 
\begin{enumerate}[(i)]
    \item $\fn$ itself of course, the absolute value of the energy difference between $m_F = \pm 3/2$, which we apply a bias magnetic field to set to a value between 75 and \SI{155}{\hertz}, depending on the superblock.
    \item $\fD$ comes in at \SI{-225}{\milli\hertz},  smaller than the mean value of $\fn$ by a factor of about $-1/450$, which is half the fractional difference in $g$-factor between the upper and lower states, plus a few smaller correction factors.
    \item $\fBR$, at \SI{212}{\milli\hertz}, comes from a magnetic field, rotating at \SI{375}{\kilo\hertz}, generated by oscillating currents that arise from charging the fins to the oscillating voltages that create the rotating electric bias field. 
    \item $\fBDR$, at roughly \SI{410}{\micro\hertz}, is  $\fBR$ echoed in corresponding $\tilD$-odd channel by the same factor of $\dgF/g\sim -1/450$. After a correction for finite $\fBR$ is applied, $\fBDR$ is within $1.7\,\sigma$ of zero. 
    \item $\fBD$ is proportional to the electron’s electric dipole moment $d_e$ and its value in principle could be quite large. Taking at face value the ACME 90\% confidence limit \cite{ACME2018}, $d_e < \SI{1.1e-29}{\electron\centi\meter}$, we would have similar confidence  that our measured value will lie between $\SI{-120}{\micro\hertz}< \fBD<\SI{120}{\micro\hertz}$. For all but the last few weeks of the experimental effort described here,  the blinding procedure built into our analysis software made $\fBD$ appear to have an arbitrary value near \SI{16.1}{\milli\hertz}.
    \item $\gamma^0$, the mean decay rate of coherence is about $\SI{0.1}{\per\second}$.  Arising from ion-ion collisions and from spatial inhomogeneities,  gamma is a noisy number that depends sensitively on number of trapped ions and fine details of the trap shimming.
    \item $\phi^0$, fairly large at tens of milliradians, is measured to be proportional to $\fn$ and is consistent with $\sim\SI{140}{\micro\second}$ offset in where we define $t_R = 0$.  This is not unexpected given that the Rabi frequency varies during the \SI{1}{\milli\second} duration of the $\pi/2$ pulses.
    \item $O^0, O^D, C^D, \phi^D$. These nonzero terms all arise predominantly from the same basic cause. Due to the presence of the $F=1/2$ manifold intermingled with the $F=3/2$ levels, the coupling procedure we use to drive nominal $\pi/2$ pulses is characterized by a slightly different effective Rabi frequency for the upper and lower levels. The duration of our coupling pulses is chosen as a compromise, and results in a state mixing of slightly larger than $\pi/2$ for one doublet and slightly smaller than $\pi/2$ for the other. This is the origin of the fringe offset $O^0$ and this compromise also contributes to the deviation of fringe contrast from unity.  Our compromise $\pi/2$ pulse duration was in retrospect chosen imperfectly, such that the deviation from perfect $\pi/2$ duration was not equal and opposite for the upper and lower levels. This led to finite values of $O^D$ and $C^D$.  With the $\pi/2$ pulses not applied at exactly zero detuning, a $\phi^D$ term also results. 
\end{enumerate}

Beyond these large (and largely understood) nonzero fit values, in a perfect version of the experiment, we would like to see the remainder of the 480 values be very small, and ideally within measurement error of zero.  

\subsection{Frequencies} 

Of the remaining 27 frequency channels, there are five that differ from zero by more than $4 \sigma$, i.e. by more than 4 times their scatter-adjusted estimated error.  

The four largest are $\fR, \fRI,  \fI$, and $\fBI$ with frequencies of $-4.2, -2.2,  -0.8$ and \SI{-0.42}{\milli\hertz} respectively. $\fRI$ and $\fBI$ are discussed in Section~\ref{sec:frequency_drift}.  What $\fR, \fRI$, and  $\fI$  have in common is that they correspond to switches which involve changing the inputs to our direct digital synthesis (DDS) boards. Each fin is driven by a distinct DDS board.  To change the sign of rotation of $\vecErot$, the $\tilR$-switch , we change the relative digital phases we load into the different DDS boards. To effect the $\tilI$ switch, which has to do with the direction of the the $\vecErot$ electric field at the instant we do the photodissociation  pulse, we switch the sign of the amplitude input into all 8 boards.  The digital math performed within the DDS boards must always include at least implicitly a truncation to the least significant bit (LSB)  of the digital-to-analog converter, and this process is repeatable but subject to tiny discontinuities from the round-off error.  Our fringe frequency is directly proportional to the rotating electric field amplitude.  The commanded voltage on any fin is sinusoidal in time. A shift of \SI{4}{\milli\hertz} could result in a change in the magnitude of $\vecErot$ corresponding to much less than 1 LSB of commanded voltage. In this context, our nonzero values of $\fR, \fRI$ and $\fI$ seem unsurprising.  $\fR$ is additionally affected by small drifts in the equilibrium location of the center of the ion cloud, and by drifts in the magnitude of residual second harmonic contamination in $V_{\rm rot}$. 

After the discussion above, we are left with 9 $\tilD$-even and 14 $\tilD$-odd channels which we believe should be quiet, and near zero after a ``$\dgF/g$ correction'' for the nonzero value in the corresponding $\tilD$-even channel, (the correction leaves $\fDR$ at $2.7\,\sigma$ from zero, and $\fBDR$ at $1.6\,\sigma$ from zero).  We have 23 ``quiet'' channels, and an additional channel $\fBD$, which we believe should be quiet except for any nonzero value of $d_e$, the electron’s EDM.  We do not have an independent measure of the correct value of $d_e$, but the 23 channels offer a chance for test of our accuracy and precision claims.  We take each quiet frequency channel, divide it by its corresponding sigma, and then we can ask what is the rms amount by which the items in the ensemble differ from zero? If we omit the 2.7-sigma value, $\fDR$, and look at only the remaining 22 values, we get a very pleasing answer -- the frequencies differ from zero by an rms amount of 1.01. If instead we include $\fDR$, the rms normalized difference from zero is 1.28. This is not a especially large number for an ensemble of 23 independent points; for 23 normally distributed points, there is about a 15\% chance that at least one will be as far away from its ``correct'' value as $2.7\,\sigma$. We have no explanation for why $\fDR$ should be nonzero, and we can’t rule out that an uncharacterized systematic error contributed to its nonzero value.  We put a lot more effort into thinking about and characterizing systematic errors that will move $\fBD$ around compared to errors in $\fDR$, so even if one in 23 measured frequencies does have an uncharacterized systematic error, this is not a very worrisome impeachment of $\fBD$.  In the end we have chosen to comment on this slightly aberrant observation but to make no corresponding relaxation in our error budget for our main measurement. 

\subsection{Contrasts, and contrast decays} 

We've already mentioned $C^0$ and $C^D$.  The fit value $C$, the initial contrast in the Ramsey fringes, is affected by parameters of the ion cloud exactly at the moment of photodissociation. The cloud mean position, its mean velocity, and the direction in which Hafnium ions are ejected after photodissociation, all impact where the atomic Hafnium ions eventually impact on the imaging ion detector. Different regions of the multi-channel ion detector have different sensitivity and different propensity to saturate.  We have a procedure which region of the detector receives ions ejecting from upper doublet and which from lower doublet, but this procedure can be subject to biases.  All this means that changes in contrast can appear in multiple channels. In particular $C^{DI}$ corresponds to a different contrast detected on the right or on the left side of the detector. This is a particularly large term. $C^{RI}$  corresponds to the direction of the \hffp{} velocity, along the direction towards the detector, at the moment of photodissociation. This in turn affects the mean time-of-flight for the \hfp{} ions and the \hffp{} ions. We impose a hardware gate to cause the image of ions to display only the \hfp{} ions and not the heavier, later-arriving \hffp{} ions. The timing of this gate is designed to maximize count rate from \hfp{} ions but suppress the contrast-destroying \hffp{} ions.  Changing the molecule velocity at the moment of dissociation changes the optimum time for the gate, but we do not make any adjustment to the detection circuitry. Therefore a substantial value of $C^{RI}$ comes as no surprise. $C^R, C^{DR}$ and $C^{DRI}$ likely arise from similar but smaller effects. The entire point of including the contrast $C$ as a fitting term is to prevent the nonlinear fitting routine from interpreting changes in contrasts as changes in frequencies.  Frequencies are further isolated from misinterpreted changes in contrast (as for instance, if the detector loses more sensitivity due to large-count rate saturation) by ensuring we choose Ramsey times separately for data each switch state so that we take points as close as possible to the exactly halfway  up the sides of fringes.


What we call $\gamma$ is really just a measure of the ratio of the contrast in the early-time fringe to that in the late-time fringe. We don't routinely collect over the intermediate times between the first, ``early-time'' sinusoidal cycle of the Ramsey fringe and the last ``last-time'' fringe. On those occasions when we do fill in some of the intermediate times, the resulting full fringe does not fit well to a pure exponential decay in contrast.  The various effects that limit contrast cannot be combined in a purely multiplicative way, and thus our approach to fitting (Eq.~\ref{eq:asymmetry-simultaneous}) means that any channel with a distinctly nonzero contrast, say $C^{RI}$,  will have a nonzero decay rate, i.e. $\gamma^{RI}$.

\subsection{Phases}

The largest observed nonzero channels in phase and offset ($O$, $O^D$, $\phi$, $\phi^D$) are well understood. Phase errors typically arise from small imperfections in the $\pi/2$ pulses.  The observed value of $\phi^{BR}$ is for instance consistent with the presence of an otherwise imperceptibly small axial gradient in the magnitude of $\vecErot$.  Other channels of $\phi$ differ from zero by statistically significant amounts, but all less than \SI{3}{\milli\radian}.  With exception of $\phi^D$, no $\tilD$-odd channel differs from zero by more than \SI{1}{\milli\radian}, and the all important $\phi^{DB}$ is measured to be \SI{0.03+-0.14}{\milli\radian}. The frequencies we measure are isolated from phase errors by the early fringe/late fringe chop. Our best effort at deliberately creating a nonzero fringe phase resulted in a \SI{170}{\milli\radian} phase,  which is 1000 times larger than the $1\,\sigma$ limit on our measured $\phi^{DB}$,  but even in that case we could resolve no frequency shift in the corresponding frequency channel.  

\subsection{Offsets}

Offsets in our Ramsey fringes can occur due to imperfections in $\pi/2$ pulses, as occurs in $O^0, O^D$ and presumably $O^{BR}$. There are scenarios where offsets in some channels can occur due to minor spatial variations in the polarization of laser beams, such as could happen from a dusty waveplate. This may account for the anomalous value of $O^{DIP}$.  Other than $O^0$ and $O^D$, the magnitudes of offset channels are less than one thousandth of the average fringe amplitude.   Even if we did not allow the fitting routine to relax around a nonzero offset, the spacing of our Ramsey time points is such that it would strongly suppress coupling of offsets to frequencies in our fits.  
\begin{turnpage}
\renewcommand{\arraystretch}{1} 

\begin{table}
\centering
\begin{ruledtabular} \begin{tabular}{@{}l*{15}{r}@{}}
&\multicolumn{4}{c}{$f\sim\SI{77}{\hertz}$} &\multicolumn{4}{c}{$f\sim\SI{105}{\hertz}$} &\multicolumn{4}{c}{$f\sim\SI{151}{\hertz}$} &\multicolumn{3}{c}{Average}\\
\cmidrule(l){2-5} \cmidrule(l){6-9}\cmidrule(l){10-13} \cmidrule(l){14-16}
 & \multicolumn{1}{c}{$\langle f \rangle$} & \multicolumn{1}{c}{$\sigma$} & \multicolumn{1}{c}{$\chi^2$} & \multicolumn{1}{c}{$\langle f \rangle/\sigma$} & \multicolumn{1}{c}{$\langle f \rangle$} & \multicolumn{1}{c}{$\sigma$} & \multicolumn{1}{c}{$\chi^2$} & \multicolumn{1}{c}{$\langle f \rangle/\sigma$} & \multicolumn{1}{c}{$\langle f \rangle$} & \multicolumn{1}{c}{$\sigma$} & \multicolumn{1}{c}{$\chi^2$} & \multicolumn{1}{c}{$\langle f \rangle/\sigma$} & \multicolumn{1}{c}{$\langle f \rangle$} & \multicolumn{1}{c}{$\sigma$} & \multicolumn{1}{c}{$\langle f \rangle/\sigma$} \\
\midrule
\addlinespace
$f^{0}$ & 76991.98 & 1.06 & 167.63 & 72848.89 & 105043.75 & 1.35 & 263.04 & 77837.69 & 151225.70 & 4.65 & 944.01 & 32487.03 & 89624.37 & 0.82 & 109419.08 \\
$f^{P}$ & -0.15 & 1.06 &  & -0.14 & -0.05 & 1.35 &  & -0.04 & -3.19 & 4.65 &  & -0.69 & -0.21 & 0.82 & -0.25 \\
$f^{B}$ & -0.11 & 0.14 & 3.04 & -0.79 & 0.00 & 0.15 & 3.07 & 0.03 & 0.04 & 0.25 & 2.76 & 0.14 & -0.04 & 0.09 & -0.45 \\
$f^{BP}$ & 0.05 & 0.14 &  & 0.33 & 0.07 & 0.15 &  & 0.46 & 0.30 & 0.25 &  & 1.20 & 0.09 & 0.09 & 0.97 \\
\addlinespace
$f^{R}$ & -1.62 & 0.13 & 2.42 & -12.73 & -4.76 & 0.14 & 2.74 & -34.55 & -15.00 & 0.28 & 3.50 & -52.97 & -4.23 & 0.09 & -47.73 \\
$f^{RP}$ & 0.22 & 0.13 &  & 1.71 & 0.03 & 0.14 &  & 0.25 & 0.06 & 0.28 &  & 0.21 & 0.13 & 0.09 & 1.42 \\
$f^{I}$ & -0.66 & 0.10 & 1.49 & -6.63 & -0.96 & 0.12 & 2.03 & -8.06 & -1.18 & 0.23 & 2.20 & -5.23 & -0.82 & 0.07 & -11.39 \\
$f^{IP}$ & -0.07 & 0.10 &  & -0.67 & 0.01 & 0.12 &  & 0.07 & -0.25 & 0.23 &  & -1.09 & -0.06 & 0.07 & -0.79 \\
\addlinespace
$f^{BR}$ & 213.28 & 0.09 & 1.29 & 2297.42 & 210.86 & 0.10 & 1.57 & 2023.15 & 209.14 & 0.18 & 1.46 & 1141.01 & 211.82 & 0.06 & 3266.90 \\
$f^{BRP}$ & 0.10 & 0.09 &  & 1.09 & -0.15 & 0.10 &  & -1.48 & -0.41 & 0.18 &  & -2.24 & -0.06 & 0.06 & -0.95 \\
$f^{BI}$ & -0.27 & 0.09 & 1.13 & -3.08 & -0.53 & 0.09 & 1.27 & -5.62 & -0.62 & 0.17 & 1.25 & -3.70 & -0.42 & 0.06 & -6.99 \\
$f^{BIP}$ & -0.06 & 0.09 &  & -0.68 & 0.04 & 0.09 &  & 0.43 & -0.04 & 0.17 &  & -0.26 & -0.02 & 0.06 & -0.28 \\
\addlinespace
$f^{RI}$ & -1.72 & 0.09 & 1.23 & -19.05 & -2.50 & 0.10 & 1.45 & -25.01 & -3.34 & 0.18 & 1.39 & -18.73 & -2.23 & 0.06 & -35.53 \\
$f^{RIP}$ & 0.07 & 0.09 &  & 0.80 & -0.01 & 0.10 &  & -0.05 & -0.09 & 0.18 &  & -0.48 & 0.02 & 0.06 & 0.35 \\
$f^{BRI}$ & 0.10 & 0.09 & 1.30 & 1.11 & -0.05 & 0.09 & 1.25 & -0.58 & -0.10 & 0.16 & 1.12 & -0.61 & 0.01 & 0.06 & 0.11 \\
$f^{BRIP}$ & -0.10 & 0.09 &  & -1.11 & -0.03 & 0.09 &  & -0.28 & -0.18 & 0.16 &  & -1.14 & -0.08 & 0.06 & -1.34 \\
\addlinespace
$f^{D}$ & -171.27 & 0.04 & 1.07 & -4426.93 & -230.12 & 0.03 & 1.06 & -7070.22 & -327.98 & 0.06 & 1.19 & -5496.53 & -223.87 & 0.02 & -9740.04 \\
$f^{DP}$ & 0.07 & 0.04 &  & 1.73 & -0.03 & 0.03 &  & -1.07 & -0.07 & 0.06 &  & -1.18 & 0.00 & 0.02 & -0.18 \\
$f^{DB}$ & -0.04\footnotemark[1] & 0.04 & 1.03 & -1.13\footnotemark[1] & 0.00\footnotemark[1] & 0.03 & 1.09 & 0.15\footnotemark[1] & 0.00\footnotemark[1] & 0.06 & 1.06 & -0.08\footnotemark[1] & -0.01\footnotemark[1] & 0.02 & -0.61\footnotemark[1] \\
$f^{DBP}$ & -0.13 & 0.04 &  & -3.37 & 0.00 & 0.03 &  & 0.15 & 0.01 & 0.06 &  & 0.20 & -0.04 & 0.02 & -1.83 \\
\addlinespace
$f^{DR}$ & 0.04 & 0.04 & 1.17 & 1.09 & 0.08 & 0.03 & 1.05 & 2.36 & 0.11 & 0.05 & 0.87 & 2.19 & 0.07 & 0.02 & 3.23 \\
$f^{DRP}$ & -0.01 & 0.04 &  & -0.36 & 0.00 & 0.03 &  & -0.03 & -0.10 & 0.05 &  & -2.02 & -0.03 & 0.02 & -1.12 \\
$f^{DI}$ & -0.01 & 0.04 & 1.03 & -0.16 & -0.09 & 0.03 & 1.09 & -2.80 & 0.02 & 0.06 & 1.07 & 0.28 & -0.04 & 0.02 & -1.91 \\
$f^{DIP}$ & 0.02 & 0.04 &  & 0.49 & -0.02 & 0.03 &  & -0.60 & -0.04 & 0.06 &  & -0.70 & -0.01 & 0.02 & -0.40 \\
\addlinespace
$f^{DBR}$ & -0.37 & 0.04 & 1.15 & -9.20 & -0.45 & 0.03 & 0.99 & -14.42 & -0.35 & 0.06 & 1.03 & -6.23 & -0.41 & 0.02 & -18.09 \\
$f^{DBRP}$ & -0.04 & 0.04 &  & -1.10 & 0.05 & 0.03 &  & 1.48 & 0.07 & 0.06 &  & 1.19 & 0.02 & 0.02 & 0.93 \\
$f^{DBI}$ & -0.01 & 0.04 & 1.04 & -0.18 & 0.05 & 0.03 & 1.07 & 1.58 & 0.05 & 0.05 & 0.96 & 0.84 & 0.03 & 0.02 & 1.34 \\
$f^{DBIP}$ & -0.05 & 0.04 &  & -1.27 & -0.03 & 0.03 &  & -0.79 & 0.05 & 0.05 &  & 0.88 & -0.02 & 0.02 & -0.93 \\
\addlinespace
$f^{DRI}$ & 0.03 & 0.04 & 1.11 & 0.83 & 0.00 & 0.03 & 0.99 & -0.06 & 0.06 & 0.06 & 1.11 & 1.08 & 0.02 & 0.02 & 0.85 \\
$f^{DRIP}$ & -0.03 & 0.04 &  & -0.83 & 0.03 & 0.03 &  & 0.97 & -0.01 & 0.06 &  & -0.20 & 0.00 & 0.02 & 0.15 \\
$f^{DBRI}$ & 0.00 & 0.04 & 1.05 & 0.02 & 0.01 & 0.03 & 1.10 & 0.42 & -0.02 & 0.06 & 1.02 & -0.32 & 0.00 & 0.02 & 0.17 \\
$f^{DBRIP}$ & 0.04 & 0.04 &  & 0.97 & 0.01 & 0.03 &  & 0.36 & -0.03 & 0.06 &  & -0.49 & 0.01 & 0.02 & 0.63 \\
\addlinespace
\end{tabular} \end{ruledtabular}
\footnotetext[1]{Mean values of $\fBD$ have had blind subtracted for clarity but all analysis was done before blind was removed.}
\caption{Table of $f$ parity channels. All values of $\langle f\rangle$ and $\sigma$ are given in \si{\milli\hertz}.}
\label{tab:f-parity-channels}
\end{table}
\begin{table}
\centering
\begin{ruledtabular} \begin{tabular}{@{}l*{15}{r}@{}}
&\multicolumn{4}{c}{$f\sim\SI{77}{\hertz}$} &\multicolumn{4}{c}{$f\sim\SI{105}{\hertz}$} &\multicolumn{4}{c}{$f\sim\SI{151}{\hertz}$} &\multicolumn{3}{c}{Average}\\
\cmidrule(l){2-5} \cmidrule(l){6-9}\cmidrule(l){10-13} \cmidrule(l){14-16}
 & \multicolumn{1}{c}{$\langle \phi \rangle$} & \multicolumn{1}{c}{$\sigma$} & \multicolumn{1}{c}{$\chi^2$} & \multicolumn{1}{c}{$\langle \phi \rangle/\sigma$} & \multicolumn{1}{c}{$\langle \phi \rangle$} & \multicolumn{1}{c}{$\sigma$} & \multicolumn{1}{c}{$\chi^2$} & \multicolumn{1}{c}{$\langle \phi \rangle/\sigma$} & \multicolumn{1}{c}{$\langle \phi \rangle$} & \multicolumn{1}{c}{$\sigma$} & \multicolumn{1}{c}{$\chi^2$} & \multicolumn{1}{c}{$\langle \phi \rangle/\sigma$} & \multicolumn{1}{c}{$\langle \phi \rangle$} & \multicolumn{1}{c}{$\sigma$} & \multicolumn{1}{c}{$\langle \phi \rangle/\sigma$} \\
\midrule
\addlinespace
$\phi^{0}$ & -3074.16 & 0.88 & 1.09 & -3509.83 & -3045.96 & 0.64 & 1.14 & -4782.22 & -3017.22 & 0.80 & 1.18 & -3750.96 & -3044.51 & 0.43 & -7018.26 \\
$\phi^{P}$ & 0.37 & 0.88 &  & 0.42 & -0.87 & 0.64 &  & -1.37 & 0.94 & 0.80 &  & 1.17 & -0.04 & 0.43 & -0.09 \\
$\phi^{B}$ & 0.86 & 0.87 & 1.08 & 0.99 & -0.95 & 0.66 & 1.21 & -1.44 & -3.10 & 0.83 & 1.25 & -3.74 & -1.10 & 0.44 & -2.47 \\
$\phi^{BP}$ & 0.61 & 0.87 &  & 0.70 & -0.37 & 0.66 &  & -0.56 & 0.33 & 0.83 &  & 0.40 & 0.08 & 0.44 & 0.19 \\
\addlinespace
$\phi^{R}$ & -1.06 & 0.93 & 1.21 & -1.14 & 0.42 & 0.62 & 1.06 & 0.68 & 0.03 & 0.77 & 1.09 & 0.03 & -0.02 & 0.43 & -0.04 \\
$\phi^{RP}$ & -1.26 & 0.93 &  & -1.36 & 0.21 & 0.62 &  & 0.34 & -0.15 & 0.77 &  & -0.19 & -0.21 & 0.43 & -0.50 \\
$\phi^{I}$ & 0.27 & 0.81 & 0.92 & 0.33 & -0.18 & 0.61 & 1.04 & -0.29 & -0.97 & 0.78 & 1.12 & -1.23 & -0.28 & 0.41 & -0.67 \\
$\phi^{IP}$ & 1.14 & 0.81 &  & 1.42 & -0.51 & 0.61 &  & -0.84 & -0.27 & 0.78 &  & -0.34 & -0.01 & 0.41 & -0.02 \\
\addlinespace
$\phi^{BR}$ & 5.21 & 0.87 & 1.07 & 6.00 & 6.00 & 0.67 & 1.26 & 8.95 & 7.27 & 0.80 & 1.15 & 9.13 & 6.19 & 0.44 & 14.01 \\
$\phi^{BRP}$ & 0.99 & 0.87 &  & 1.14 & 1.15 & 0.67 &  & 1.72 & 1.75 & 0.80 &  & 2.20 & 1.29 & 0.44 & 2.93 \\
$\phi^{BI}$ & -1.98 & 0.90 & 1.16 & -2.19 & -1.47 & 0.65 & 1.18 & -2.27 & -4.53 & 0.78 & 1.12 & -5.77 & -2.54 & 0.44 & -5.81 \\
$\phi^{BIP}$ & -0.21 & 0.90 &  & -0.23 & 0.69 & 0.65 &  & 1.07 & -0.38 & 0.78 &  & -0.49 & 0.15 & 0.44 & 0.34 \\
\addlinespace
$\phi^{RI}$ & -0.76 & 0.88 & 1.09 & -0.87 & -0.84 & 0.61 & 1.04 & -1.38 & 0.21 & 0.74 & 1.00 & 0.28 & -0.49 & 0.41 & -1.19 \\
$\phi^{RIP}$ & -1.29 & 0.88 &  & -1.47 & -0.15 & 0.61 &  & -0.25 & 0.53 & 0.74 &  & 0.72 & -0.19 & 0.41 & -0.46 \\
$\phi^{BRI}$ & 0.19 & 0.87 & 1.08 & 0.22 & 2.11 & 0.62 & 1.08 & 3.40 & 3.23 & 0.77 & 1.08 & 4.19 & 2.00 & 0.42 & 4.72 \\
$\phi^{BRIP}$ & 1.06 & 0.87 &  & 1.21 & 0.77 & 0.62 &  & 1.23 & 1.45 & 0.77 &  & 1.88 & 1.04 & 0.42 & 2.46 \\
\addlinespace
$\phi^{D}$ & -8.77 & 0.26 & 1.06 & -33.68 & -15.68 & 0.20 & 1.12 & -79.30 & -11.44 & 0.26 & 1.03 & -43.74 & -12.69 & 0.13 & -94.12 \\
$\phi^{DP}$ & -0.29 & 0.26 &  & -1.11 & -0.07 & 0.20 &  & -0.36 & -0.30 & 0.26 &  & -1.16 & -0.19 & 0.13 & -1.41 \\
$\phi^{DB}$ & -0.21 & 0.27 & 1.15 & -0.79 & 0.17 & 0.20 & 1.14 & 0.83 & -0.19 & 0.27 & 1.11 & -0.72 & -0.03 & 0.14 & -0.19 \\
$\phi^{DBP}$ & 0.19 & 0.27 &  & 0.71 & 0.26 & 0.20 &  & 1.32 & -0.41 & 0.27 &  & -1.51 & 0.07 & 0.14 & 0.50 \\
\addlinespace
$\phi^{DR}$ & -0.02 & 0.27 & 1.16 & -0.06 & -0.27 & 0.19 & 1.02 & -1.44 & 0.11 & 0.28 & 1.15 & 0.41 & -0.12 & 0.14 & -0.86 \\
$\phi^{DRP}$ & 0.03 & 0.27 &  & 0.11 & 0.07 & 0.19 &  & 0.36 & 0.55 & 0.28 &  & 1.99 & 0.17 & 0.14 & 1.28 \\
$\phi^{DI}$ & -0.19 & 0.26 & 1.03 & -0.74 & 0.86 & 0.20 & 1.19 & 4.23 & 0.17 & 0.28 & 1.19 & 0.61 & 0.39 & 0.14 & 2.78 \\
$\phi^{DIP}$ & 0.14 & 0.26 &  & 0.54 & -0.26 & 0.20 &  & -1.27 & 0.48 & 0.28 &  & 1.72 & 0.04 & 0.14 & 0.28 \\
\addlinespace
$\phi^{DBR}$ & -0.58 & 0.27 & 1.14 & -2.15 & -0.57 & 0.19 & 1.02 & -3.02 & -1.21 & 0.27 & 1.11 & -4.47 & -0.73 & 0.13 & -5.43 \\
$\phi^{DBRP}$ & 0.45 & 0.27 &  & 1.65 & -0.33 & 0.19 &  & -1.77 & -0.16 & 0.27 &  & -0.58 & -0.10 & 0.13 & -0.73 \\
$\phi^{DBI}$ & 0.14 & 0.25 & 0.96 & 0.57 & 0.20 & 0.21 & 1.21 & 0.95 & 0.48 & 0.26 & 1.03 & 1.85 & 0.26 & 0.14 & 1.90 \\
$\phi^{DBIP}$ & -1.04 & 0.25 &  & -4.21 & 0.13 & 0.21 &  & 0.65 & 1.08 & 0.26 &  & 4.15 & 0.04 & 0.14 & 0.27 \\
\addlinespace
$\phi^{DRI}$ & -0.20 & 0.27 & 1.13 & -0.73 & 0.12 & 0.20 & 1.19 & 0.60 & -0.42 & 0.27 & 1.08 & -1.57 & -0.11 & 0.14 & -0.78 \\
$\phi^{DRIP}$ & -0.01 & 0.27 &  & -0.02 & -0.04 & 0.20 &  & -0.22 & 0.23 & 0.27 &  & 0.87 & 0.04 & 0.14 & 0.29 \\
$\phi^{DBRI}$ & 0.01 & 0.26 & 1.09 & 0.03 & -0.74 & 0.20 & 1.13 & -3.75 & -0.59 & 0.26 & 1.04 & -2.25 & -0.50 & 0.14 & -3.71 \\
$\phi^{DBRIP}$ & -0.28 & 0.26 &  & -1.08 & 0.02 & 0.20 &  & 0.11 & 0.55 & 0.26 &  & 2.11 & 0.08 & 0.14 & 0.62 \\
\addlinespace
\end{tabular} \end{ruledtabular}
\caption{Table of $\phi$ parity channels. All values of $\langle\phi\rangle$ and $\sigma$ are given in \si{\milli\radian}.}
\label{tab:p-parity-channels}
\end{table}
\begin{table}
\centering
\begin{ruledtabular} \begin{tabular}{@{}l*{15}{r}@{}}
&\multicolumn{4}{c}{$f\sim\SI{77}{\hertz}$} &\multicolumn{4}{c}{$f\sim\SI{105}{\hertz}$} &\multicolumn{4}{c}{$f\sim\SI{151}{\hertz}$} &\multicolumn{3}{c}{Average}\\
\cmidrule(l){2-5} \cmidrule(l){6-9}\cmidrule(l){10-13} \cmidrule(l){14-16}
 & \multicolumn{1}{c}{$\langle C \rangle$} & \multicolumn{1}{c}{$\sigma$} & \multicolumn{1}{c}{$\chi^2$} & \multicolumn{1}{c}{$\langle C \rangle/\sigma$} & \multicolumn{1}{c}{$\langle C \rangle$} & \multicolumn{1}{c}{$\sigma$} & \multicolumn{1}{c}{$\chi^2$} & \multicolumn{1}{c}{$\langle C \rangle/\sigma$} & \multicolumn{1}{c}{$\langle C \rangle$} & \multicolumn{1}{c}{$\sigma$} & \multicolumn{1}{c}{$\chi^2$} & \multicolumn{1}{c}{$\langle C \rangle/\sigma$} & \multicolumn{1}{c}{$\langle C \rangle$} & \multicolumn{1}{c}{$\sigma$} & \multicolumn{1}{c}{$\langle C \rangle/\sigma$} \\
\midrule
\addlinespace
$C^{0}$ & 552.40 & 1.86 & 14.18 & 296.95 & 541.88 & 1.49 & 16.06 & 364.77 & 545.94 & 1.05 & 5.54 & 519.51 & 545.96 & 0.78 & 700.79 \\
$C^{P}$ & -2.01 & 1.86 &  & -1.08 & -1.03 & 1.49 &  & -0.69 & -1.74 & 1.05 &  & -1.65 & -1.59 & 0.78 & -2.04 \\
$C^{B}$ & -0.64 & 0.54 & 1.18 & -1.19 & -0.23 & 0.41 & 1.24 & -0.55 & -0.26 & 0.48 & 1.16 & -0.54 & -0.34 & 0.27 & -1.26 \\
$C^{BP}$ & -0.35 & 0.54 &  & -0.64 & 0.01 & 0.41 &  & 0.01 & 0.07 & 0.48 &  & 0.15 & -0.06 & 0.27 & -0.23 \\
\addlinespace
$C^{R}$ & -2.07 & 0.57 & 1.32 & -3.64 & -1.62 & 0.41 & 1.25 & -3.91 & -3.81 & 0.49 & 1.18 & -7.85 & -2.43 & 0.28 & -8.82 \\
$C^{RP}$ & -0.55 & 0.57 &  & -0.97 & -0.04 & 0.41 &  & -0.09 & 0.35 & 0.49 &  & 0.72 & -0.03 & 0.28 & -0.12 \\
$C^{I}$ & -1.12 & 0.57 & 1.34 & -1.96 & 1.19 & 0.42 & 1.30 & 2.82 & -1.60 & 0.53 & 1.39 & -3.03 & -0.20 & 0.29 & -0.72 \\
$C^{IP}$ & -0.96 & 0.57 &  & -1.68 & 0.02 & 0.42 &  & 0.04 & -0.57 & 0.53 &  & -1.08 & -0.40 & 0.29 & -1.40 \\
\addlinespace
$C^{BR}$ & -0.70 & 0.53 & 1.15 & -1.32 & 1.01 & 0.42 & 1.26 & 2.43 & 0.56 & 0.53 & 1.40 & 1.06 & 0.41 & 0.28 & 1.48 \\
$C^{BRP}$ & -0.11 & 0.53 &  & -0.20 & 0.12 & 0.42 &  & 0.28 & -0.89 & 0.53 &  & -1.68 & -0.22 & 0.28 & -0.80 \\
$C^{BI}$ & -0.55 & 0.57 & 1.34 & -0.96 & 0.88 & 0.39 & 1.13 & 2.22 & 0.30 & 0.50 & 1.26 & 0.61 & 0.38 & 0.27 & 1.40 \\
$C^{BIP}$ & 0.15 & 0.57 &  & 0.26 & 0.08 & 0.39 &  & 0.20 & -0.31 & 0.50 &  & -0.62 & -0.02 & 0.27 & -0.07 \\
\addlinespace
$C^{RI}$ & 3.79 & 0.56 & 1.30 & 6.74 & 5.30 & 0.44 & 1.40 & 12.08 & 6.44 & 0.51 & 1.28 & 12.72 & 5.27 & 0.29 & 18.47 \\
$C^{RIP}$ & 0.94 & 0.56 &  & 1.66 & 0.77 & 0.44 &  & 1.75 & -0.57 & 0.51 &  & -1.13 & 0.39 & 0.29 & 1.35 \\
$C^{BRI}$ & 0.16 & 0.55 & 1.25 & 0.28 & -0.71 & 0.42 & 1.30 & -1.69 & -1.11 & 0.53 & 1.43 & -2.08 & -0.60 & 0.28 & -2.09 \\
$C^{BRIP}$ & -0.99 & 0.55 &  & -1.79 & -0.75 & 0.42 &  & -1.77 & 0.56 & 0.53 &  & 1.04 & -0.44 & 0.28 & -1.56 \\
\addlinespace
$C^{D}$ & 11.87 & 0.43 & 4.89 & 27.48 & 16.72 & 0.24 & 3.49 & 70.15 & 24.36 & 0.39 & 5.48 & 62.66 & 17.55 & 0.18 & 95.46 \\
$C^{DP}$ & 0.25 & 0.43 &  & 0.58 & -0.22 & 0.24 &  & -0.94 & 0.17 & 0.39 &  & 0.44 & -0.05 & 0.18 & -0.27 \\
$C^{DB}$ & 0.04 & 0.22 & 1.22 & 0.19 & -0.13 & 0.14 & 1.13 & -0.96 & -0.08 & 0.19 & 1.26 & -0.45 & -0.08 & 0.10 & -0.84 \\
$C^{DBP}$ & -0.46 & 0.22 &  & -2.11 & -0.01 & 0.14 &  & -0.10 & 0.14 & 0.19 &  & 0.77 & -0.06 & 0.10 & -0.63 \\
\addlinespace
$C^{DR}$ & -1.20 & 0.21 & 1.19 & -5.65 & -0.32 & 0.14 & 1.16 & -2.31 & -0.92 & 0.18 & 1.24 & -4.99 & -0.67 & 0.10 & -6.89 \\
$C^{DRP}$ & -0.35 & 0.21 &  & -1.64 & 0.07 & 0.14 &  & 0.52 & -0.01 & 0.18 &  & -0.08 & -0.04 & 0.10 & -0.43 \\
$C^{DI}$ & 6.28 & 0.24 & 1.52 & 26.14 & 6.95 & 0.14 & 1.28 & 48.34 & 8.74 & 0.18 & 1.18 & 48.49 & 7.40 & 0.10 & 72.70 \\
$C^{DIP}$ & 0.01 & 0.24 &  & 0.03 & 0.17 & 0.14 &  & 1.17 & 0.19 & 0.18 &  & 1.08 & 0.15 & 0.10 & 1.45 \\
\addlinespace
$C^{DBR}$ & 0.11 & 0.22 & 1.31 & 0.47 & -0.32 & 0.13 & 1.09 & -2.44 & -0.61 & 0.17 & 1.05 & -3.58 & -0.34 & 0.09 & -3.54 \\
$C^{DBRP}$ & -0.02 & 0.22 &  & -0.11 & 0.00 & 0.13 &  & 0.02 & -0.28 & 0.17 &  & -1.63 & -0.09 & 0.09 & -0.94 \\
$C^{DBI}$ & 0.01 & 0.20 & 1.08 & 0.04 & -0.02 & 0.13 & 1.10 & -0.13 & -0.19 & 0.19 & 1.28 & -1.04 & -0.06 & 0.10 & -0.60 \\
$C^{DBIP}$ & 0.13 & 0.20 &  & 0.62 & -0.11 & 0.13 &  & -0.80 & -0.20 & 0.19 &  & -1.06 & -0.08 & 0.10 & -0.82 \\
\addlinespace
$C^{DRI}$ & 0.31 & 0.22 & 1.33 & 1.38 & 0.44 & 0.15 & 1.32 & 3.02 & 0.54 & 0.19 & 1.30 & 2.86 & 0.44 & 0.10 & 4.31 \\
$C^{DRIP}$ & -0.05 & 0.22 &  & -0.22 & 0.14 & 0.15 &  & 0.97 & 0.25 & 0.19 &  & 1.30 & 0.13 & 0.10 & 1.29 \\
$C^{DBRI}$ & -0.18 & 0.24 & 1.45 & -0.76 & 0.00 & 0.14 & 1.19 & 0.03 & -0.13 & 0.19 & 1.28 & -0.69 & -0.07 & 0.10 & -0.68 \\
$C^{DBRIP}$ & -0.07 & 0.24 &  & -0.31 & -0.02 & 0.14 &  & -0.15 & 0.14 & 0.19 &  & 0.73 & 0.02 & 0.10 & 0.15 \\
\addlinespace
\end{tabular} \end{ruledtabular}
\caption{Table of $C$ parity channels. All values of $\langle C\rangle$ and $\sigma$ are multiplied by 1000.}
\label{tab:C0-parity-channels}
\end{table}
\begin{table}
\centering
\begin{ruledtabular} \begin{tabular}{@{}l*{15}{r}@{}}
&\multicolumn{4}{c}{$f\sim\SI{77}{\hertz}$} &\multicolumn{4}{c}{$f\sim\SI{105}{\hertz}$} &\multicolumn{4}{c}{$f\sim\SI{151}{\hertz}$} &\multicolumn{3}{c}{Average}\\
\cmidrule(l){2-5} \cmidrule(l){6-9}\cmidrule(l){10-13} \cmidrule(l){14-16}
 & \multicolumn{1}{c}{$\langle \gamma \rangle$} & \multicolumn{1}{c}{$\sigma$} & \multicolumn{1}{c}{$\chi^2$} & \multicolumn{1}{c}{$\langle \gamma \rangle/\sigma$} & \multicolumn{1}{c}{$\langle \gamma \rangle$} & \multicolumn{1}{c}{$\sigma$} & \multicolumn{1}{c}{$\chi^2$} & \multicolumn{1}{c}{$\langle \gamma \rangle/\sigma$} & \multicolumn{1}{c}{$\langle \gamma \rangle$} & \multicolumn{1}{c}{$\sigma$} & \multicolumn{1}{c}{$\chi^2$} & \multicolumn{1}{c}{$\langle \gamma \rangle/\sigma$} & \multicolumn{1}{c}{$\langle \gamma \rangle$} & \multicolumn{1}{c}{$\sigma$} & \multicolumn{1}{c}{$\langle \gamma \rangle/\sigma$} \\
\midrule
\addlinespace
$\gamma^{0}$ & 75.89 & 1.70 & 9.62 & 44.75 & 87.82 & 1.57 & 8.30 & 56.10 & 121.71 & 2.41 & 7.52 & 50.53 & 89.64 & 1.04 & 86.36 \\
$\gamma^{P}$ & -2.28 & 1.70 &  & -1.34 & -0.97 & 1.57 &  & -0.62 & 0.18 & 2.41 &  & 0.08 & -1.25 & 1.04 & -1.20 \\
$\gamma^{B}$ & -1.18 & 0.59 & 1.16 & -2.01 & 0.54 & 0.59 & 1.18 & 0.91 & 0.34 & 1.00 & 1.29 & 0.34 & -0.23 & 0.39 & -0.59 \\
$\gamma^{BP}$ & -0.75 & 0.59 &  & -1.28 & 0.17 & 0.59 &  & 0.28 & -0.40 & 1.00 &  & -0.40 & -0.31 & 0.39 & -0.81 \\
\addlinespace
$\gamma^{R}$ & 2.65 & 0.61 & 1.23 & 4.37 & 0.91 & 0.59 & 1.19 & 1.54 & -11.28 & 1.00 & 1.27 & -11.34 & -0.24 & 0.39 & -0.62 \\
$\gamma^{RP}$ & -0.08 & 0.61 &  & -0.13 & -0.54 & 0.59 &  & -0.92 & 0.13 & 1.00 &  & 0.13 & -0.25 & 0.39 & -0.64 \\
$\gamma^{I}$ & -0.30 & 0.58 & 1.14 & -0.51 & 1.77 & 0.60 & 1.24 & 2.93 & 2.85 & 0.99 & 1.28 & 2.87 & 1.02 & 0.39 & 2.65 \\
$\gamma^{IP}$ & -1.23 & 0.58 &  & -2.12 & -0.10 & 0.60 &  & -0.16 & -0.64 & 0.99 &  & -0.65 & -0.68 & 0.39 & -1.76 \\
\addlinespace
$\gamma^{BR}$ & -1.47 & 0.55 & 1.02 & -2.65 & -0.33 & 0.61 & 1.25 & -0.54 & 0.48 & 1.00 & 1.28 & 0.49 & -0.75 & 0.38 & -1.97 \\
$\gamma^{BRP}$ & 0.47 & 0.55 &  & 0.86 & 0.45 & 0.61 &  & 0.75 & -0.73 & 1.00 &  & -0.74 & 0.29 & 0.38 & 0.77 \\
$\gamma^{BI}$ & -0.42 & 0.62 & 1.30 & -0.68 & 0.35 & 0.58 & 1.14 & 0.61 & -0.98 & 1.00 & 1.30 & -0.98 & -0.16 & 0.39 & -0.40 \\
$\gamma^{BIP}$ & -0.26 & 0.62 &  & -0.42 & -0.04 & 0.58 &  & -0.06 & -0.43 & 1.00 &  & -0.43 & -0.19 & 0.39 & -0.48 \\
\addlinespace
$\gamma^{RI}$ & 5.39 & 0.58 & 1.12 & 9.33 & 6.41 & 0.63 & 1.36 & 10.10 & 8.42 & 0.96 & 1.19 & 8.76 & 6.28 & 0.39 & 16.07 \\
$\gamma^{RIP}$ & 0.59 & 0.58 &  & 1.02 & 0.09 & 0.63 &  & 0.14 & -1.37 & 0.96 &  & -1.43 & 0.08 & 0.39 & 0.19 \\
$\gamma^{BRI}$ & 0.70 & 0.57 & 1.09 & 1.23 & -0.58 & 0.58 & 1.16 & -0.99 & -0.76 & 0.98 & 1.25 & -0.78 & -0.05 & 0.38 & -0.13 \\
$\gamma^{BRIP}$ & -1.05 & 0.57 &  & -1.83 & -1.41 & 0.58 &  & -2.41 & 0.87 & 0.98 &  & 0.88 & -0.91 & 0.38 & -2.42 \\
\addlinespace
$\gamma^{D}$ & -2.14 & 0.43 & 1.76 & -5.02 & -6.16 & 0.38 & 2.11 & -16.04 & -6.27 & 0.59 & 1.71 & -10.72 & -4.72 & 0.26 & -18.42 \\
$\gamma^{DP}$ & -0.29 & 0.43 &  & -0.69 & 0.04 & 0.38 &  & 0.10 & 0.26 & 0.59 &  & 0.44 & -0.04 & 0.26 & -0.16 \\
$\gamma^{DB}$ & 0.06 & 0.34 & 1.14 & 0.16 & 0.00 & 0.29 & 1.19 & 0.01 & 0.00 & 0.48 & 1.16 & 0.01 & 0.02 & 0.20 & 0.11 \\
$\gamma^{DBP}$ & -0.07 & 0.34 &  & -0.20 & -0.27 & 0.29 &  & -0.95 & 0.05 & 0.48 &  & 0.10 & -0.15 & 0.20 & -0.74 \\
\addlinespace
$\gamma^{DR}$ & -1.15 & 0.37 & 1.31 & -3.14 & -0.41 & 0.30 & 1.33 & -1.35 & -0.36 & 0.48 & 1.16 & -0.75 & -0.65 & 0.21 & -3.07 \\
$\gamma^{DRP}$ & -0.22 & 0.37 &  & -0.59 & -0.16 & 0.30 &  & -0.52 & -0.02 & 0.48 &  & -0.05 & -0.15 & 0.21 & -0.72 \\
$\gamma^{DI}$ & 2.61 & 0.37 & 1.35 & 7.03 & 4.15 & 0.31 & 1.33 & 13.60 & 4.99 & 0.49 & 1.21 & 10.11 & 3.80 & 0.21 & 17.87 \\
$\gamma^{DIP}$ & 0.02 & 0.37 &  & 0.06 & -0.35 & 0.31 &  & -1.15 & 0.19 & 0.49 &  & 0.38 & -0.13 & 0.21 & -0.61 \\
\addlinespace
$\gamma^{DBR}$ & -0.04 & 0.36 & 1.28 & -0.12 & -0.14 & 0.29 & 1.21 & -0.47 & -0.39 & 0.47 & 1.11 & -0.83 & -0.15 & 0.20 & -0.76 \\
$\gamma^{DBRP}$ & -0.24 & 0.36 &  & -0.66 & -0.02 & 0.29 &  & -0.08 & -0.52 & 0.47 &  & -1.10 & -0.18 & 0.20 & -0.90 \\
$\gamma^{DBI}$ & 0.35 & 0.35 & 1.19 & 1.00 & 0.07 & 0.29 & 1.17 & 0.25 & -0.41 & 0.50 & 1.23 & -0.82 & 0.09 & 0.20 & 0.43 \\
$\gamma^{DBIP}$ & 0.29 & 0.35 &  & 0.83 & 0.28 & 0.29 &  & 0.99 & 0.18 & 0.50 &  & 0.37 & 0.27 & 0.20 & 1.33 \\
\addlinespace
$\gamma^{DRI}$ & -0.34 & 0.35 & 1.22 & -0.95 & -0.55 & 0.30 & 1.33 & -1.81 & -0.51 & 0.49 & 1.22 & -1.04 & -0.47 & 0.21 & -2.25 \\
$\gamma^{DRIP}$ & -0.32 & 0.35 &  & -0.91 & 0.66 & 0.30 &  & 2.17 & 0.62 & 0.49 &  & 1.25 & 0.31 & 0.21 & 1.48 \\
$\gamma^{DBRI}$ & -0.61 & 0.39 & 1.45 & -1.59 & -0.42 & 0.28 & 1.14 & -1.49 & -0.59 & 0.48 & 1.15 & -1.22 & -0.51 & 0.21 & -2.46 \\
$\gamma^{DBRIP}$ & -0.47 & 0.39 &  & -1.23 & 0.62 & 0.28 &  & 2.20 & -0.02 & 0.48 &  & -0.05 & 0.19 & 0.21 & 0.93 \\
\addlinespace
\end{tabular} \end{ruledtabular}
\caption{Table of $\gamma$ parity channels. All values of $\langle\gamma\rangle$ and $\sigma$ are in $10^{-3}\,\si{\per\second}$.}
\label{tab:g-parity-channels}
\end{table}
\begin{table}
\centering
\begin{ruledtabular} \begin{tabular}{@{}l*{15}{r}@{}}
&\multicolumn{4}{c}{$f\sim\SI{77}{\hertz}$} &\multicolumn{4}{c}{$f\sim\SI{105}{\hertz}$} &\multicolumn{4}{c}{$f\sim\SI{151}{\hertz}$} &\multicolumn{3}{c}{Average}\\
\cmidrule(l){2-5} \cmidrule(l){6-9}\cmidrule(l){10-13} \cmidrule(l){14-16}
 & \multicolumn{1}{c}{$\langle O \rangle$} & \multicolumn{1}{c}{$\sigma$} & \multicolumn{1}{c}{$\chi^2$} & \multicolumn{1}{c}{$\langle O \rangle/\sigma$} & \multicolumn{1}{c}{$\langle O \rangle$} & \multicolumn{1}{c}{$\sigma$} & \multicolumn{1}{c}{$\chi^2$} & \multicolumn{1}{c}{$\langle O \rangle/\sigma$} & \multicolumn{1}{c}{$\langle O \rangle$} & \multicolumn{1}{c}{$\sigma$} & \multicolumn{1}{c}{$\chi^2$} & \multicolumn{1}{c}{$\langle O \rangle/\sigma$} & \multicolumn{1}{c}{$\langle O \rangle$} & \multicolumn{1}{c}{$\sigma$} & \multicolumn{1}{c}{$\langle O \rangle/\sigma$} \\
\midrule
\addlinespace
$O^{0}$ & 26.08 & 0.34 & 1.86 & 76.73 & 26.11 & 0.23 & 1.60 & 112.94 & 24.39 & 0.30 & 1.76 & 81.92 & 25.60 & 0.16 & 159.15 \\
$O^{P}$ & 0.67 & 0.34 &  & 1.97 & -0.35 & 0.23 &  & -1.50 & 0.00 & 0.30 &  & 0.00 & -0.02 & 0.16 & -0.11 \\
$O^{B}$ & -0.25 & 0.31 & 1.54 & -0.82 & -0.10 & 0.23 & 1.52 & -0.46 & -0.10 & 0.26 & 1.39 & -0.38 & -0.14 & 0.15 & -0.92 \\
$O^{BP}$ & -0.13 & 0.31 &  & -0.42 & -0.28 & 0.23 &  & -1.22 & -0.35 & 0.26 &  & -1.34 & -0.27 & 0.15 & -1.78 \\
\addlinespace
$O^{R}$ & 0.20 & 0.31 & 1.55 & 0.66 & 0.05 & 0.22 & 1.39 & 0.23 & -0.03 & 0.24 & 1.16 & -0.14 & 0.05 & 0.14 & 0.37 \\
$O^{RP}$ & -0.17 & 0.31 &  & -0.54 & -0.18 & 0.22 &  & -0.85 & 0.22 & 0.24 &  & 0.92 & -0.04 & 0.14 & -0.27 \\
$O^{I}$ & 0.12 & 0.29 & 1.36 & 0.43 & -0.13 & 0.21 & 1.28 & -0.62 & -0.23 & 0.28 & 1.52 & -0.84 & -0.09 & 0.14 & -0.66 \\
$O^{IP}$ & -0.52 & 0.29 &  & -1.79 & 0.05 & 0.21 &  & 0.25 & 0.18 & 0.28 &  & 0.66 & -0.05 & 0.14 & -0.37 \\
\addlinespace
$O^{BR}$ & 0.17 & 0.27 & 1.22 & 0.63 & 0.21 & 0.22 & 1.50 & 0.92 & 0.67 & 0.25 & 1.24 & 2.66 & 0.35 & 0.14 & 2.43 \\
$O^{BRP}$ & -0.27 & 0.27 &  & -0.96 & 0.19 & 0.22 &  & 0.85 & -0.05 & 0.25 &  & -0.19 & -0.01 & 0.14 & -0.06 \\
$O^{BI}$ & -0.15 & 0.28 & 1.24 & -0.55 & 0.31 & 0.22 & 1.47 & 1.38 & -0.51 & 0.27 & 1.42 & -1.93 & -0.06 & 0.15 & -0.43 \\
$O^{BIP}$ & 0.46 & 0.28 &  & 1.67 & -0.12 & 0.22 &  & -0.56 & -0.11 & 0.27 &  & -0.42 & 0.04 & 0.15 & 0.28 \\
\addlinespace
$O^{RI}$ & -0.16 & 0.32 & 1.68 & -0.49 & -0.21 & 0.23 & 1.52 & -0.92 & -0.29 & 0.27 & 1.43 & -1.08 & -0.22 & 0.15 & -1.46 \\
$O^{RIP}$ & -0.03 & 0.32 &  & -0.11 & 0.03 & 0.23 &  & 0.12 & -0.05 & 0.27 &  & -0.19 & -0.01 & 0.15 & -0.07 \\
$O^{BRI}$ & 0.02 & 0.28 & 1.27 & 0.07 & -0.60 & 0.22 & 1.45 & -2.73 & 0.02 & 0.26 & 1.35 & 0.08 & -0.25 & 0.14 & -1.71 \\
$O^{BRIP}$ & -0.26 & 0.28 &  & -0.92 & -0.48 & 0.22 &  & -2.19 & -0.34 & 0.26 &  & -1.30 & -0.38 & 0.14 & -2.62 \\
\addlinespace
$O^{D}$ & 10.38 & 0.15 & 2.34 & 67.88 & 10.85 & 0.11 & 2.59 & 100.75 & 1.39 & 0.34 & 14.97 & 4.10 & 10.10 & 0.09 & 118.57 \\
$O^{DP}$ & 0.11 & 0.15 &  & 0.72 & -0.08 & 0.11 &  & -0.76 & -0.18 & 0.34 &  & -0.54 & -0.03 & 0.09 & -0.34 \\
$O^{DB}$ & 0.14 & 0.11 & 1.12 & 1.31 & -0.12 & 0.07 & 1.16 & -1.61 & -0.15 & 0.10 & 1.28 & -1.48 & -0.06 & 0.05 & -1.27 \\
$O^{DBP}$ & 0.04 & 0.11 &  & 0.36 & 0.04 & 0.07 &  & 0.51 & -0.04 & 0.10 &  & -0.45 & 0.02 & 0.05 & 0.30 \\
\addlinespace
$O^{DR}$ & 0.03 & 0.11 & 1.13 & 0.27 & -0.04 & 0.08 & 1.33 & -0.53 & 0.00 & 0.09 & 1.18 & -0.01 & -0.01 & 0.05 & -0.23 \\
$O^{DRP}$ & 0.02 & 0.11 &  & 0.17 & 0.06 & 0.08 &  & 0.75 & -0.05 & 0.09 &  & -0.55 & 0.01 & 0.05 & 0.28 \\
$O^{DI}$ & 0.05 & 0.11 & 1.19 & 0.50 & 0.00 & 0.07 & 1.13 & 0.05 & 0.14 & 0.10 & 1.24 & 1.46 & 0.05 & 0.05 & 1.03 \\
$O^{DIP}$ & 0.68 & 0.11 &  & 6.25 & 0.51 & 0.07 &  & 7.18 & 0.56 & 0.10 &  & 5.72 & 0.56 & 0.05 & 11.03 \\
\addlinespace
$O^{DBR}$ & 0.30 & 0.12 & 1.44 & 2.48 & 0.28 & 0.07 & 1.16 & 3.90 & 0.36 & 0.10 & 1.20 & 3.78 & 0.31 & 0.05 & 5.93 \\
$O^{DBRP}$ & 0.05 & 0.12 &  & 0.44 & -0.01 & 0.07 &  & -0.13 & 0.03 & 0.10 &  & 0.33 & 0.01 & 0.05 & 0.28 \\
$O^{DBI}$ & -0.04 & 0.11 & 1.13 & -0.36 & -0.07 & 0.07 & 1.15 & -0.96 & -0.03 & 0.09 & 1.17 & -0.29 & -0.05 & 0.05 & -1.00 \\
$O^{DBIP}$ & -0.04 & 0.11 &  & -0.38 & 0.10 & 0.07 &  & 1.35 & -0.02 & 0.09 &  & -0.19 & 0.03 & 0.05 & 0.67 \\
\addlinespace
$O^{DRI}$ & -0.17 & 0.11 & 1.21 & -1.51 & -0.58 & 0.08 & 1.26 & -7.76 & -0.46 & 0.10 & 1.26 & -4.69 & -0.45 & 0.05 & -8.64 \\
$O^{DRIP}$ & -0.18 & 0.11 &  & -1.67 & -0.13 & 0.08 &  & -1.76 & 0.07 & 0.10 &  & 0.68 & -0.09 & 0.05 & -1.66 \\
$O^{DBRI}$ & -0.08 & 0.11 & 1.26 & -0.69 & -0.02 & 0.07 & 1.22 & -0.28 & 0.05 & 0.10 & 1.19 & 0.48 & -0.01 & 0.05 & -0.25 \\
$O^{DBRIP}$ & -0.06 & 0.11 &  & -0.55 & 0.10 & 0.07 &  & 1.29 & 0.08 & 0.10 &  & 0.84 & 0.06 & 0.05 & 1.11 \\
\addlinespace
\end{tabular} \end{ruledtabular}
\caption{Table of $O$ parity channels. All values of $\langle O\rangle$ and $\sigma$ are multiplied by 1000.}
\label{tab:O-parity-channels}
\end{table}

\end{turnpage}
\end{document}